\documentclass[aps, epsf, superscriptaddress, two column, pra]{revtex4-1}
\usepackage[utf8]{inputenc}
\usepackage{amsmath}
\usepackage{amsfonts}
\usepackage{subfig}
\usepackage{amssymb}
\usepackage{mathrsfs}
\usepackage{graphicx, psfrag}
\usepackage{float}
\usepackage{xcolor}
\usepackage[user,titleref]{zref}
\usepackage{tikz}
\usepackage{pgfplots}
\usepackage{ulem}
\usepackage[colorlinks=true, citecolor=blue, urlcolor = blue, linkcolor= red, bookmarks=true]{hyperref}
\usepackage{epstopdf}
\usepackage{blindtext}
\graphicspath{{images/}}

\begin{document}

\title{Effect of Andreev Processes on the Goos-H\"anchen (GH) shift in the Graphene-Superconductor-Graphene (GSG) junctions}
\author{Shahrukh Salim}
\affiliation{Department of Physics, Indian Institute of Technology Delhi, Hauz Khas 110 016, New Delhi, INDIA} 
\author{Rahul Marathe}\email{maratherahul@physics.iitd.ac.in} \affiliation{Department of Physics, Indian Institute of Technology Delhi, Hauz Khas 110 016, New Delhi, INDIA} 
\author{Sankalpa Ghosh}\email{sankalpa@physics.iitd.ac.in} \affiliation{Department of Physics, Indian Institute of Technology Delhi, Hauz Khas 110 016, New Delhi, INDIA} 

\begin{abstract}
In this article, we study the transport properties of Graphene-Superconductor-Graphene (GSG) heterojunction where the superconducting region is created in the middle of a graphene sheet, as contrasted to widely studied transport properties through 
a Superconductor-Graphene-Superconductor (SGS) type of Josephson junction. We particularly analyse in detail the Goos-H\"anchen shift of 
the electron and the hole at the GS interface in such a junction, due to normal as well as Andreev reflection, using a transfer matrix-based approach. Additionally, we evaluate the normalised differential conductance as a function of bias voltage that characterises the transport through such junction and point out how they are influenced by Andreev and normal reflection. In the subsequent parts of the article we demonstrate how the GH shift for both electron and hole changes with the width of the superconducting region. The behavior of the differential conductance in such junctions as a function of the bias voltage in the region, dominated by Andreev and normal reflection, is also presented and analysed.
\end{abstract} 
\maketitle
\section{Introduction}
\zlabel{sec:intro}
A normal metallic conductor placed in between two superconducting regions forms superconductor-normal-superconductor (SNS) Josephson junction (JJ) and exhibits unusual electronic properties due to the formation of Andreev bound states (ABS) at excitation energies within the superconducting gap \cite{Andreev1966}. 
The role of ABS in Josephson tunnelling through such junctions has been extensively studied for SNS JJs \cite{Josephson1964, Tinkham1996, Blonder1982, Furusaki1991}. C. W. Beenakker  \cite{Beenakker2006} showed that in a superconductor-graphene (SG) interface, the Andreev process shows additional features under suitable conditions in the form of Specular Andreev reflection (SAR) in addition to Retro Andreev reflection (RAR) as compared to an NS interface \cite{Sauls2018}  where Andreev processes are only restricted to RAR. This leads to a significant body of work both in theory, \cite{Titov2006, Bhattacharjee2006, Maiti2007, Efetov2016, Soori2018}, as well as in experiments \cite{Heersche2007, Du2008, Popinciuc2012, Mizuno2013, Bretheau2013, Efetov2016specular, Bretheau2017, Borzenets2016, Sahu2016, Feng1, Feng2, Feng3, Feng4}. Other interesting phenomena like the crossed Andreev reflection (CAR) and elastic cotunnelling (EC) were also explored \cite{Byers, Guy, Cayssol}. In comparison to the study of such SGS Josephson junctions, another variant where the superconducting region can be placed somewhere in the middle of graphene to introduce proximity-induced superconductivity, leading to the formation of a GSG (graphene-superconductor-graphene) received much less attention. Understanding the nature of transport in such a junction is important to finally consider the transport in an array of GS interfaces and the transport processes in the same.

In this work, we discuss in detail the transport through such GSG junctions. Our study is primarily motivated by the following two aspects of such junctions. These junctions are expected to show an electronic analogue of a well-studied optical phenomenon Goos-H\"anchen (GH) shift \cite{Goos1947, Puri1986, Berman2002, Beenakker2009, Sharma2011, Wu2011, Chen2013}. A recent work \cite{Liu2018} also found that there is an electronic analogue of GH shift in a GS interface both for the normal and Andreev reflection and such GH shift carries distinctive features due to the presence of sub-lattice degrees of freedom in the graphene as compared to a prototype NS interface. 

Another study pointed out that such GH shift in the interface between a non-topological and topological material and quantised circulation of the anomalous shift vector can reveal the topological features of one of the media \cite{Liu2020}. In yet another recent work, GH shift in the surface states of Topological Insulators were also theoretically investigated \cite{Ghadiri22}. However, in most cases, the amount of such shift is of the order of the Fermi-wave vector of the associated material, which makes its direct observation very difficult, even though it may impact the transport \cite{Beenakker2009}. The other motivation of our study is to measure the differential conductance through such GSG interface and see the impact of normal and Andreev reflection on the transport in these junctions.

Accordingly, the rest of the manuscript is organised in the following way. In section \ref{method}, after introducing the basic Dirac Bogoliubov de Gennes (DBdG) Hamiltonian for such junction, we provide the detailed theoretical modelling by evaluating the stationary solutions in different regions, introduce the transfer matrices, and finally calculate the reflection and transmission coefficients of electron and holes through such GSG junctions. We also provide the conditions under which retro and specular Andreev reflection will occur in such junctions. In the next section \ref{RD}, we provide our main results. In the first part of this section \ref{GHshiftsec}, we provide in detail the GH sift in such GSG junctions. In the second part of this section \ref{conductsec} we provide the results of the differential conductance through such junctions under different biasing conditions. Finally, we conclude. We also provide details of some of the expressions used in the main text in the Appendix \ref{append}.

\section{Quantum transport through a GSG junction and their modelling}\label{method}
\zlabel{model}
\begin{figure}[!htb]
\subfloat[]{\includegraphics[width=\columnwidth]{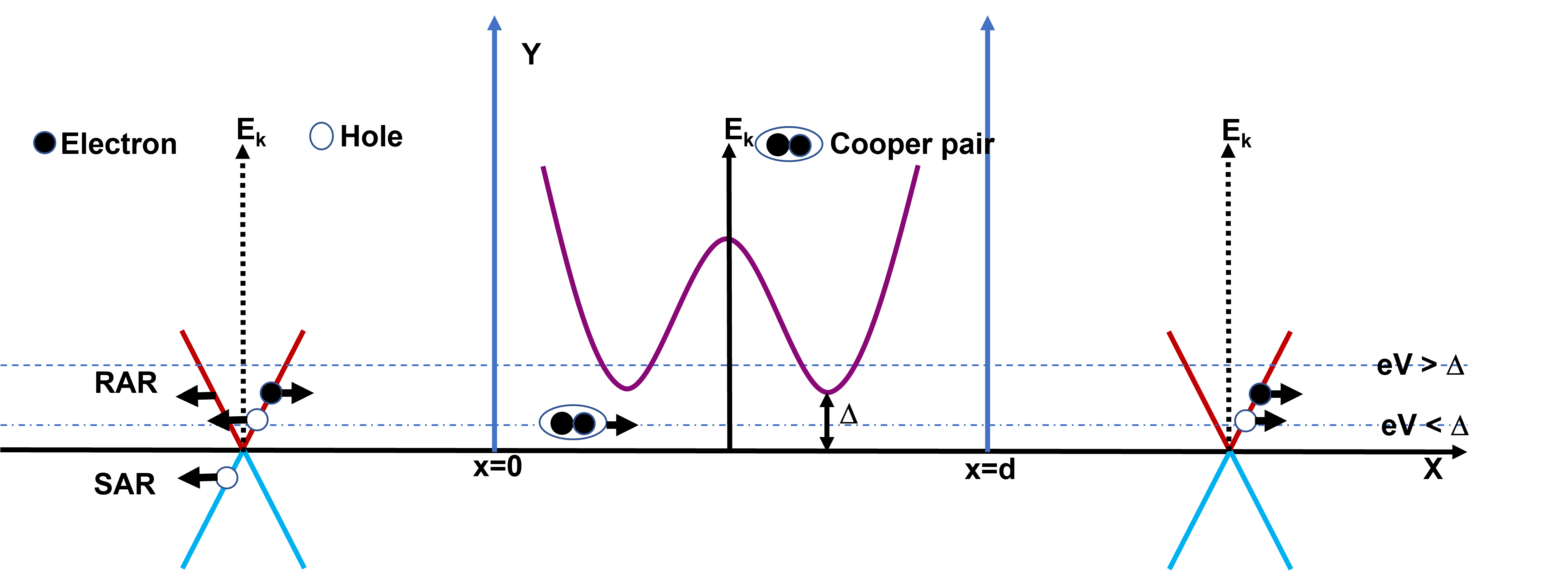}}\\
\subfloat[]{\includegraphics[width=\columnwidth]{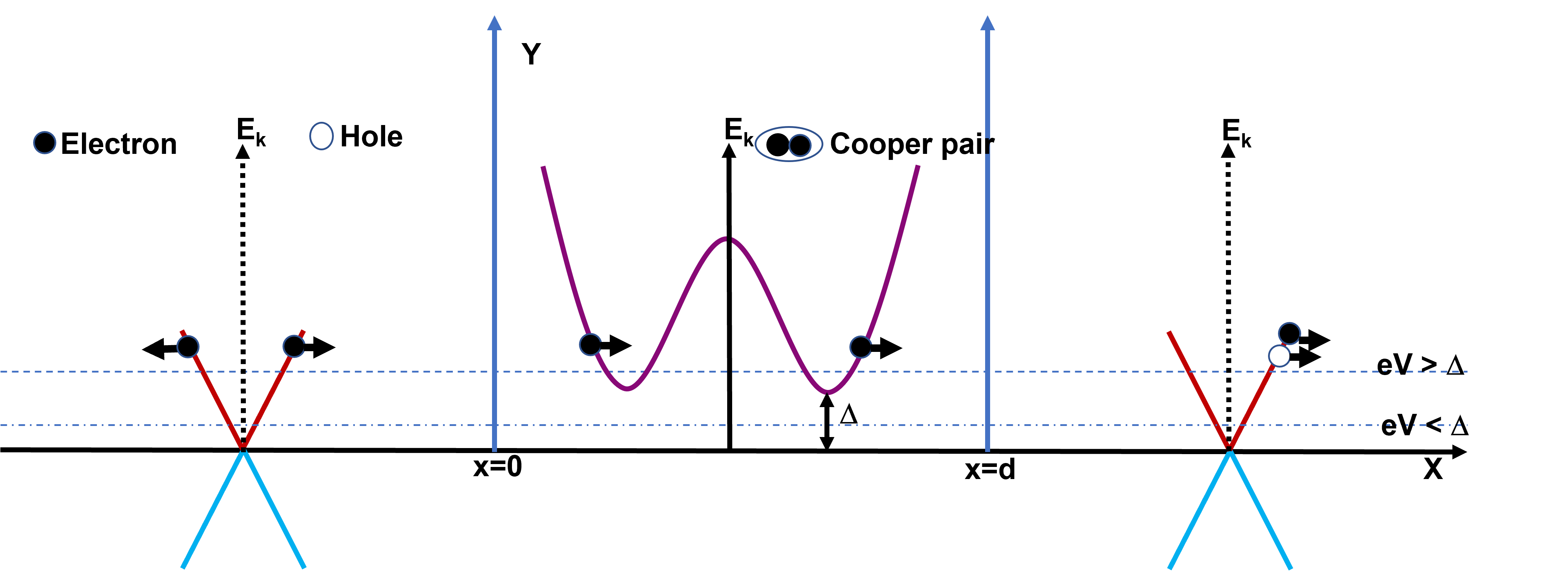}}\\
\subfloat[]{\includegraphics[width=\columnwidth]{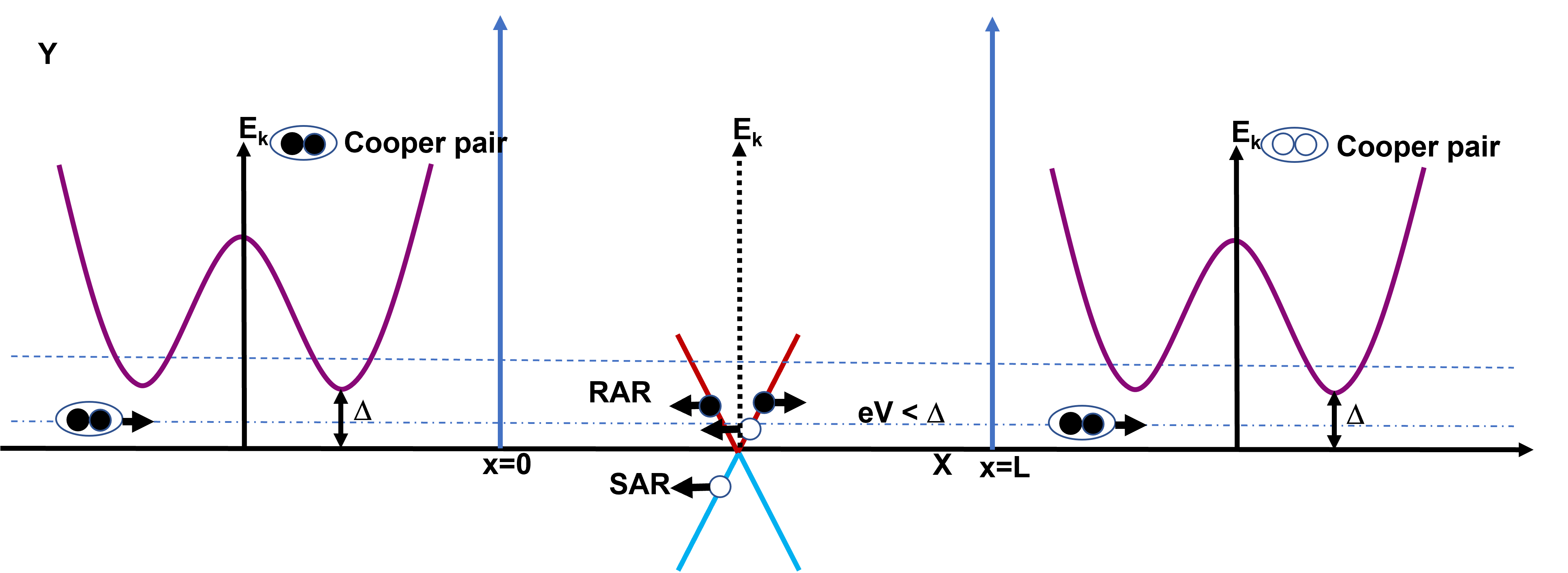}}
\caption{\label{fig:gsg}\textit{(color online)} (a) Andreev processes in a GSG junction. The \textit{black} circle denotes an electron, and a \textit{white} circle denotes a hole. RAR is an intraband process, so the incident particle and the reflected particle lie in the same band (\textit{red} color). For SAR, which is an interband process, the incident particle and the reflected hole lie in opposite bands (\textit{cyan} color). Since the energy of the particle is less than the gap, Cooper pairs are formed in the superconducting region. (b) A normal reflection process for a GSG junction, where the energy of the particle is greater than the superconducting gap. We have single-particle states in the superconducting region. (c) Andreev processes for SGS junction denoting RAR (intraband) and SAR (interband) process at the junction for comparison.}
\zlabel{schm}
\end{figure}

 We begin with a brief discussion on the possible processes that can take place in the GSG heterostructure as compared to the more popular SGS junction. The main difference is that the GSG junction is not a Josephson junction (JJ), unlike the SGS. The details of different processes that are involved in the GSG junction are depicted in Fig.~\ref{fig:gsg}. As shown in Fig.~\ref{fig:gsg}(a), when an electron (denoted by a \textit{black} dot) with energy less than the superconducting gap, $\Delta$, is incident on the GS interface, it is reflected as a hole (denoted by a \textit{white} circle) and a Cooper pair (denoted by a pair of black dots) is transferred in the superconducting region. This process is called the Andreev reflection. The Andreev processes, in turn, are of three types. The first process is an intra-band process where the incident and the reflected particle lie in the same band. This is known as the RAR. The second process is an inter-band process where the incident electron and the reflected hole lie in the opposite bands, known as the SAR. In Fig.~\ref{fig:gsg}(a), one can also see that the incident electron(hole) can be transferred as an electron (hole) or a hole(electron). The former process is called EC, while the latter is termed as CAR \cite{Byers, Guy, Cayssol}. The third process is the normal reflection (see Fig.~\ref{fig:gsg}(b)), where the energy of an incident electron is greater than the superconducting gap, and we only obtain single-particle states in the superconducting region \cite{beenakker2008}. In Fig.~\ref{fig:gsg}(c), we show RAR and SAR processes for the SGS junction. In either case, these processes are extremely important for meaningful modelling of the system and calculation of the GH shift as well as the transport properties across the GSG heterostructure.

\begin{figure*}
\subfloat[]{\includegraphics[width=0.53\columnwidth]{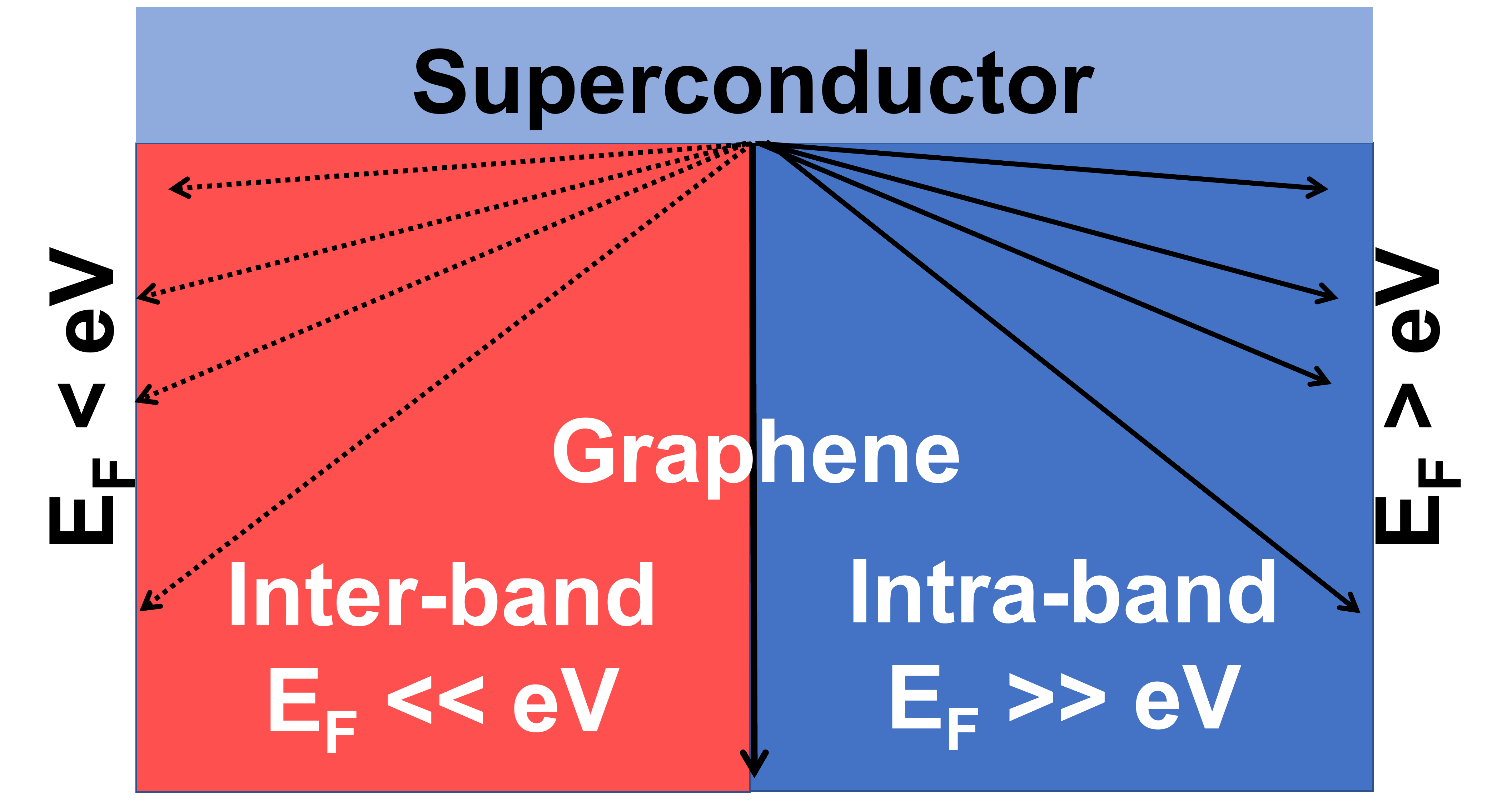}}\hspace*{0.25cm} 
\subfloat[]{\includegraphics[width=0.45\columnwidth]{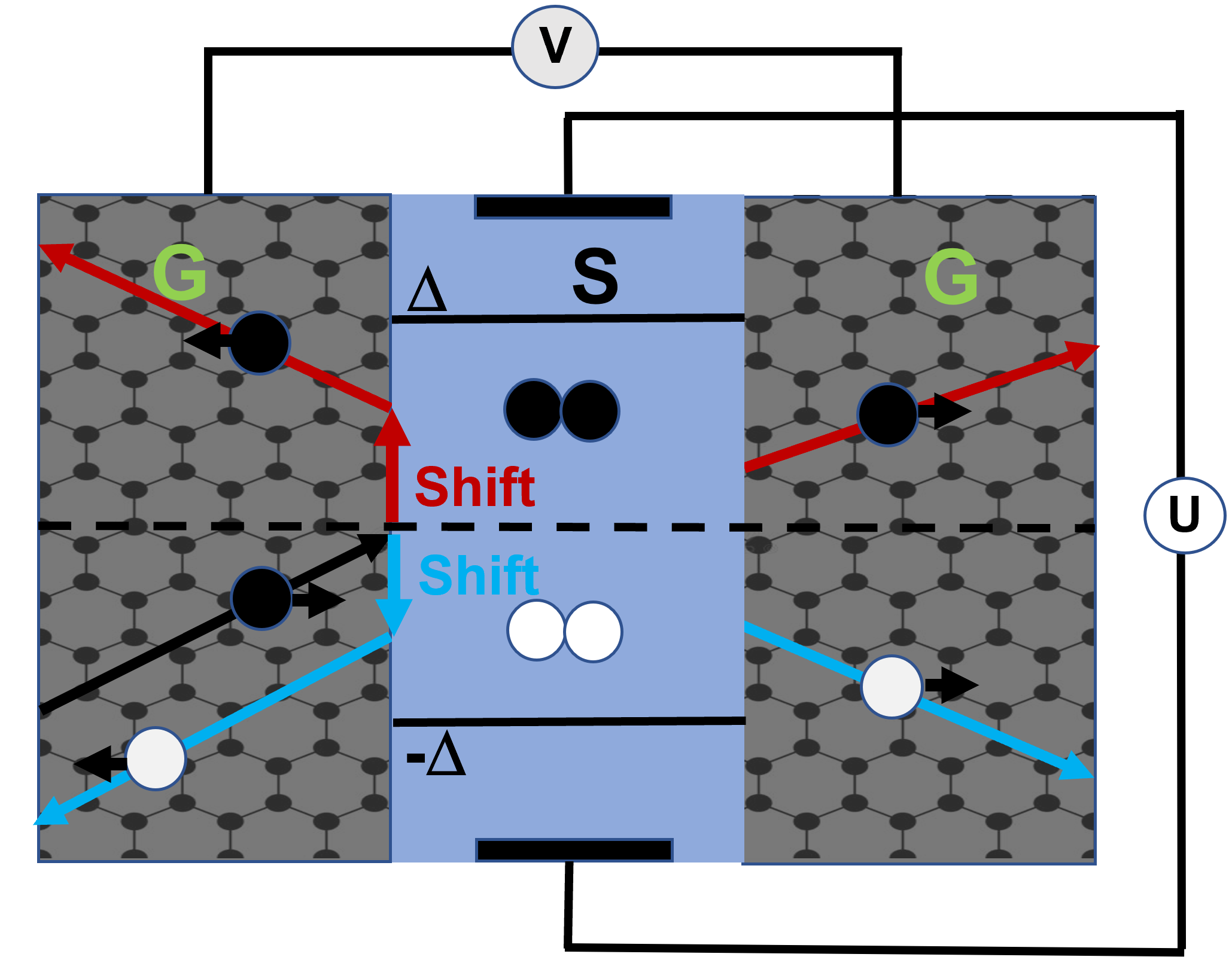}}
\hspace*{0.25cm}
\subfloat[]{\includegraphics[width=0.45\columnwidth]{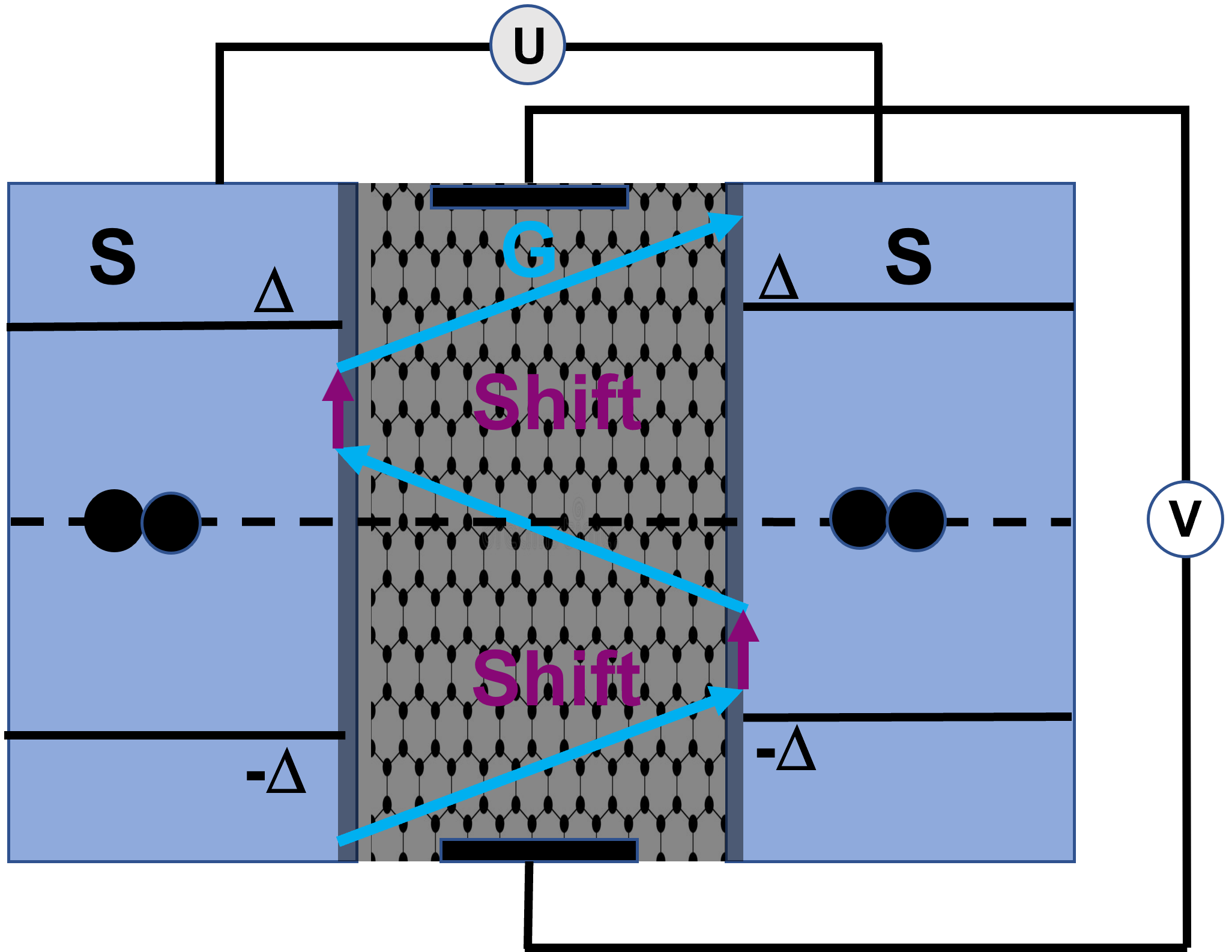}}\hspace*{0.25cm} \subfloat[]{\includegraphics[width=0.55\columnwidth]{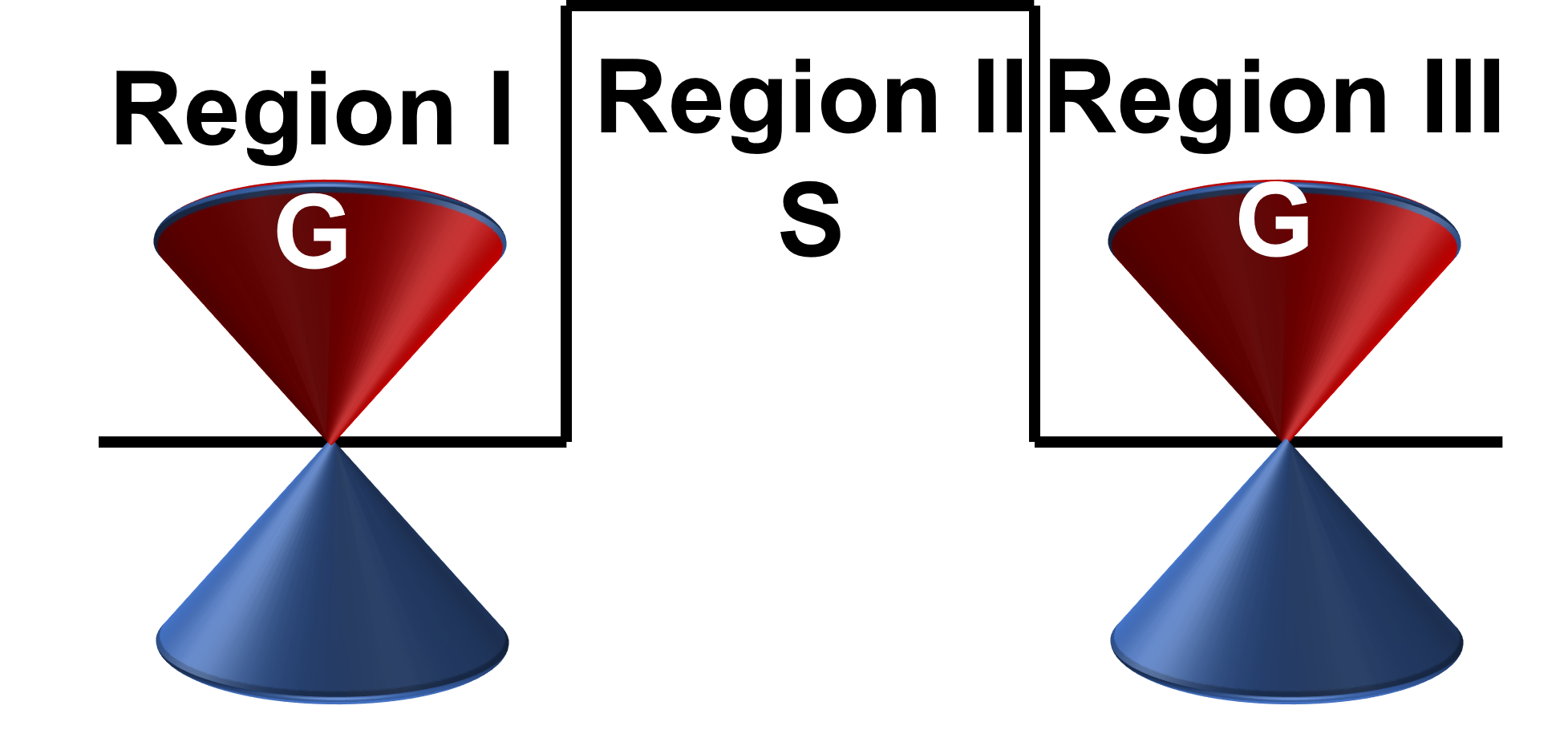}}
\caption {\textit{(color online)} (a) Energy consideration for various Andreev processes happening along the boundary wall for GSG and SGS junctions. (b) Reflection and transmission processes for a GSG junction; $black$ represents an incoming electron, $red$ denotes the reflected electron (Normal reflection) and $blue$ is the reflected hole (Andreev reflection). $V$ is the potential difference across the two graphene regions, while $U$ is the potential applied to the superconducting region. The Goos-H\"anchen shift for electrons(\textit{red}) and holes(\textit{blue}) are denoted by arrows along the boundary wall (GS). (c) Reflection processes for an SGS junction; the Goos-H\"anchen shift is denoted along the SG and GS interfaces in \textit{magenta} color. $U$ is the potential difference between the two superconducting regions, while $V$ is the potential applied to the graphene region. (d) Schematic describing a typical GSG junction and the corresponding potential profile. Region I and III are graphene, and region II is a superconductor. $E$ is the energy of the particle and $\Delta$ is the superconducting energy gap.}\label{fig:AR} 
\end{figure*}

We also show the relevant energy considerations for various Andreev processes that occur along the boundary wall for GSG and SGS junctions in Fig.~\ref{fig:AR}(a). For RAR, the Fermi energy, $E_F$, should be greater than the energy of the incident particle, while for SAR, $E_F$ is smaller than the incident particle energy. As compared to an SGS heterostructure that has been studied widely \cite{Titov2006, Bhattacharjee2006, Heersche2007, Borzenets2016}, the transport across a GSG heterostructure does not take place through the formation of ABS, even though Andreev reflection does occur in such heterojunctions. However, the GH shift, which can be measured if we can trace the path of the reflected particle (see Fig.~\ref{fig:gsg} and Fig.~\ref{fig:AR}(b,c)), plays an important role in determining its transport properties.

 We also note that the GH shift occurs in both SGS and GSG heterostructures. However, experimentally, this can be observed only in the GSG type of junctions where the \textit{left} and the \textit{right} graphene regions support the existence of free particles (electrons and holes; see Fig.~\ref{fig:AR}(b)), as opposed to the SGS junctions where the \textit{left} and the \textit{right} superconducting regions only have bound states (Cooper pairs; see Fig.~\ref{fig:AR}(c)), thus, making the extraction of the paths of the reflected particles difficult. This motivates us to study the GSG junction in detail regarding the GH shift, in terms of the differences in the observed features compared to the earlier studied SG, SGS or SNS junctions, and the transport properties of this junction \cite{Bretheau2017, Beenakker2009, Liu2018}. 
 
In the next section, we describe in detail the theoretical modelling of the system under consideration. 
 
\subsection{Theoretical modelling: Hamiltonian, stationary solutions, and the transfer matrices for GSG junctions} \label{hamwaven}
We consider a GSG junction in a graphene sheet of length \textit{d} in the \textit{x}-\textit{y} plane with the superconducting region extending from \textit{x}=0 to \textit{x}=d  as shown in Fig.~\ref{fig:AR}(d). The graphene leads are held at a potential difference \textit{V} while the voltage applied at the superconducting region is \textit{U$(\vec{x})$}. The chemical potential in the three regions can be modulated using several electrostatic gates. This model also allows us to consider the asymmetric junction where the potential in the \textit{left} and \textit{right} graphene regions can be different. The symmetric junction can then be obtained as a limiting case of the asymmetric junction. The electrons and holes in the graphene region are described by the Bogoliubov-de-Gennes (BdG) equations \cite{Gennes}
\begin{equation}\label{geq}
 \begin{pmatrix}
 H^{\pm}+U(\vec{x})-E_{F}^{L,R} & 0\\
 0 & E_{F}^{L,R}-H^{\pm}-U(\vec{x})
 \end{pmatrix}
\Psi
 =E
\Psi
 \end{equation}\\
 where $H^{\pm}  = \sigma_{x}\partial_{x} \pm \sigma_{y}\partial_{y}$ is the single particle Hamiltonian and $\sigma_{x,y}$ are the Pauli matrices.   
The quasi-electrons and quasi-holes in the superconducting region are described by the Dirac Bogoliubov-de-Gennes (DBdG) equations \cite{Gennes}
 \begin{equation}\label{dbdg}
 \begin{pmatrix}
 H^{\pm}+U(\vec{x})-E_{F}^{L,R} & \Delta\\
 \Delta^{*} & E_{F}^{L,R}-H^{\pm}-U(\vec{x})
 \end{pmatrix}
\Psi
 =E
\Psi
 \end{equation}\\
 where $\Psi$ represents the electron or the hole wave function, $E$ is the excitation energy relative to the Fermi energy, $U(\vec{x})$ is the external potential applied in the superconducting region and  $\Delta$ is the superconducting pair potential  which is given by,
 \begin{equation}
 \Delta = \Delta_{0}\sqrt{1 - \left(\frac{T}{T_{c}}\right)^{2}}
\end{equation}  
where $T_{c}$ is the critical temperature and $\Delta_0$ is the pair potential at $T=0$ K.
The Fermi energy for the $left$ and the $right$ graphene region is given as $E_{F}^{L,R} = E_{F}\pm V/2$. As the electrostatic potentials in the three regions can be independently fixed, we take the case of zero mismatch in the Fermi levels of the three regions \cite{Yeyati}.

The total wave function obtained by solving Eq.~(\ref{geq}) and (\ref{dbdg}) is;
\begin{subequations}
\begin{align}
&\Psi_{G} = \psi^{+}_{Ge}+r_{e}\psi_{Ge}^{-}+r_{h}\psi_{Gh}^{-},~~~	x<0 \\
&\Psi_{S} = A\psi_{Se}^{+}+B\psi_{Se}^{-}+C\psi_{Sh}^{+}+D\psi_{Sh}^{-},~	0<x<d\\
&\Psi_{G'} = t_{e}\psi^{+}_{Ge}+t_{h}\psi^{+}_{Gh},~~~x>d,
\end{align}
\end{subequations}

 where $\Psi_G, \Psi_S$ and $\Psi_{G'}$ are the wave functions in the \textit{left} graphene, superconducting and the \textit{right} graphene regions respectively. They are written as a linear combination of the wave functions for reflected and/or transmitted electrons(e) and/or holes(h) in the different regions. We denote them by the subscripts $(Ge, Gh)$ for the graphene and $(Se, Sh)$ for the superconductor. The $\pm$ superscripts indicate which particle is reflected or transmitted at the two interfaces (at \textit{x}=0 and \textit{x}=d). The symbols \textit{A, B, C, D} denote the amplitudes of the various quasi-particles in the superconducting region. The symbol $r_{e}$ denotes the reflection amplitude for the normal process where an electron is reflected as an electron, $r_{h}$ denotes the reflection amplitude for the Andreev reflection, where an electron is reflected as a hole and $t_{e, h}$ are the transmission amplitudes of electrons and holes respectively.

The explicit form of the wave functions for the particles in the \textit{S} and \textit{G} regions are respectively given as
\begin{subequations}
\begin{equation}\label{eq:superel}
\psi_{Se}^{\pm}=e^{i q y \pm i k_{0} x \mp \kappa x}
\begin{pmatrix}
e^{i\beta}\\
\pm e^{\pm i(\pm\gamma+\beta)}\\
e^{-i\phi}\\
\pm e^{i(\gamma-\phi)}
\end{pmatrix}
\end{equation}
\begin{equation}\label{eq:superhl}
\psi_{Sh}^{\pm}=e^{i q y \mp i k_{0} x \mp \kappa x}
\begin{pmatrix}
e^{-i\beta}\\
\mp e^{\mp i(\mp\gamma-\beta)}\\
e^{-i\phi}\\
\mp e^{i(-\gamma-\phi)}
\end{pmatrix}
\end{equation} 
\begin{equation}\label{eq:grel}
    \psi_{Ge}^{\pm} (x,y) = \frac{e^{i({\pm}k_e^{L,R} x+k_y y)}}{\sqrt{2\cos\alpha^{L,R}}}
\begin{pmatrix}
e^{\mp i\alpha^{L,R}/2}\\
\pm e^{\pm i\alpha^{L,R}/2}\\
0\\
0
\end{pmatrix}
\end{equation}
\begin{equation}\label{eq:grhl}
\psi_{Gh}^{\pm}(x,y) = \frac{e^{i({\pm}k_h^{L,R} x+k_y y)}}{\sqrt{2\cos\alpha'^{L,R}}}
\begin{pmatrix}
0\\
0\\
e^{\mp i\alpha'^{L,R}/2}\\
\mp e^{\pm i\alpha'^{L,R}/2}
\end{pmatrix}
\end{equation}
\end{subequations}
where the superscript $(L,R)$ denotes the \textit{left} and \textit{right} graphene regions respectively. The definitions for various parameters are as follows -
\begin{subequations}\label{param}
\begin{align}
& \alpha^{L,R} = \sin ^{-1}\left(\frac{\hbar v_{F} q}{E+E_{F}^{L,R}}\right)\\
& \alpha'^{L,R} = \sin ^{-1}\left(\frac{\hbar v_{F} q}{E-E_{F}^{L,R}}\right)\\
& k_{e}^{L,R} = \frac{E+ E_{F}^{L,R}}{\hbar v_{F}}\cos\alpha^{L,R}\\
& k_{h}^{L,R} = \frac{E- E_{F}^{L,R}}{\hbar v_{F}}\cos\alpha'^{L,R}\\
& \alpha_{c}^{L} = \sin ^{-1}\left(\frac{E-E_{F}^{L}}{E+E_{F}^{L}}\right)\\
& k_{0} = \sqrt{\frac{(E_{F}-U)^{2}}{(\hbar v_{F})^{2}}- q^{2}}\\
& \kappa = \frac{(E_{F}-U)}{(\hbar v_{F})^{2}k_{0}}\Delta_{0}\sin\beta\\
&\beta  = 
\begin{cases}
\cos^{-1}\left(\frac{E}{\Delta}\right), \text{ if }E<\Delta \\
-i\cosh^{-1} \left(\frac{E}{\Delta}\right), \text{ if }E>\Delta
\end{cases}\\
&\gamma=\sin^{-1}\left(\frac{\hbar v_{F}q}{|E_{F}-U|}\right)
\end{align}
\end{subequations}
Here, $\alpha$ and $\alpha'$ denote the incident angles for electrons and holes respectively, $k_{e}$ and $k_{h}$ are the wave vectors for the electron and the holes in the graphene region respectively, $\alpha_{c}$ is the critical angle, $k_{0}$ and $\kappa$ are the real and the imaginary part of the superconducting wave vector, $q$ is the transverse wave vector, $\beta$ is the ratio of the incident energy of the particle to the superconducting gap and $\gamma$ is the angle for the quasiparticles in the superconducting region. For $\alpha>\alpha_c$, $k_h$ becomes imaginary, and the hole becomes an evanescent wave. So, the hole excitation is impossible for $\alpha>\alpha_c$ and hence the electron is totally reflected. This is the total internal reflection for an electron which will be discussed in the later sections dealing with GH shift.

We now match the wave functions given in Eqs.~(\ref{eq:superel})-(\ref{eq:grhl}) at the interfaces to get the following equations
\begin{eqnarray}
M_{1}(x=0) G_{x=0} &=& M_{2}(x=0) S_{x=0}\\
M_{3}(x=d) S_{x=d} &=& M_{4}(x=d) G_{x=d}
\end{eqnarray}
where the explicit form of the transfer matrices $M_{1,2,3,4}$ are given in Appendix \ref{append} Eq. (\ref{Mmat}). Here, $G$'s and $S$'s are the column vectors containing the amplitudes for all the components of the wave function in the graphene and the superconducting regions and are given in Appendix \ref{append} Eq. (\ref{Gmat}).
The above equations can be combined to yield;
\begin{eqnarray}
G(x=0) &=&  M G(x=d)\nonumber\\
\text{with},~M  &=& M_{1}^{-1}M_{2}M_{3}^{-1}M_{4}
\label{eq:M}
\end{eqnarray}
and $\det{M}=1$.  This condition of the unit determinant guarantees unitary evolution and conservation of probabilities. For finding various transmission and reflection coefficients, we need to use the Eq. (\ref{eq:M})

\begin{eqnarray}\label{TR}
M\begin{bmatrix}
t_{e}\\0\\0\\t_{h}
\end{bmatrix}=
\begin{bmatrix}
1\\r_{e}\\r_{h}\\0
\end{bmatrix}
\end{eqnarray}

Denoting the elements of the matrix $M$ by $a_{ij}$ (detailed expressions for the required $a_{ij}$'s, $r_{e,h}$, and $t_{e,h}$, are given in Appendix \ref{append} (see Eq. (\ref{acof1}) and Eq. (\ref{acof2})), we get the following equations
\begin{subequations}\label{aijs}
\begin{eqnarray}
a_{11}t_{e}+a_{14}t_{h} &=& 1\\
a_{21}t_{e}+a_{24}t_{h} &=& r_{e}\\
a_{31}t_{e}+a_{34}t_{h} &=& r_{h}\\
a_{41}t_{e}+a_{44}t_{h} &=& 0
\end{eqnarray}
\end{subequations}
\begin{figure*}[t]
\subfloat[]{\includegraphics[width=.3\textwidth]{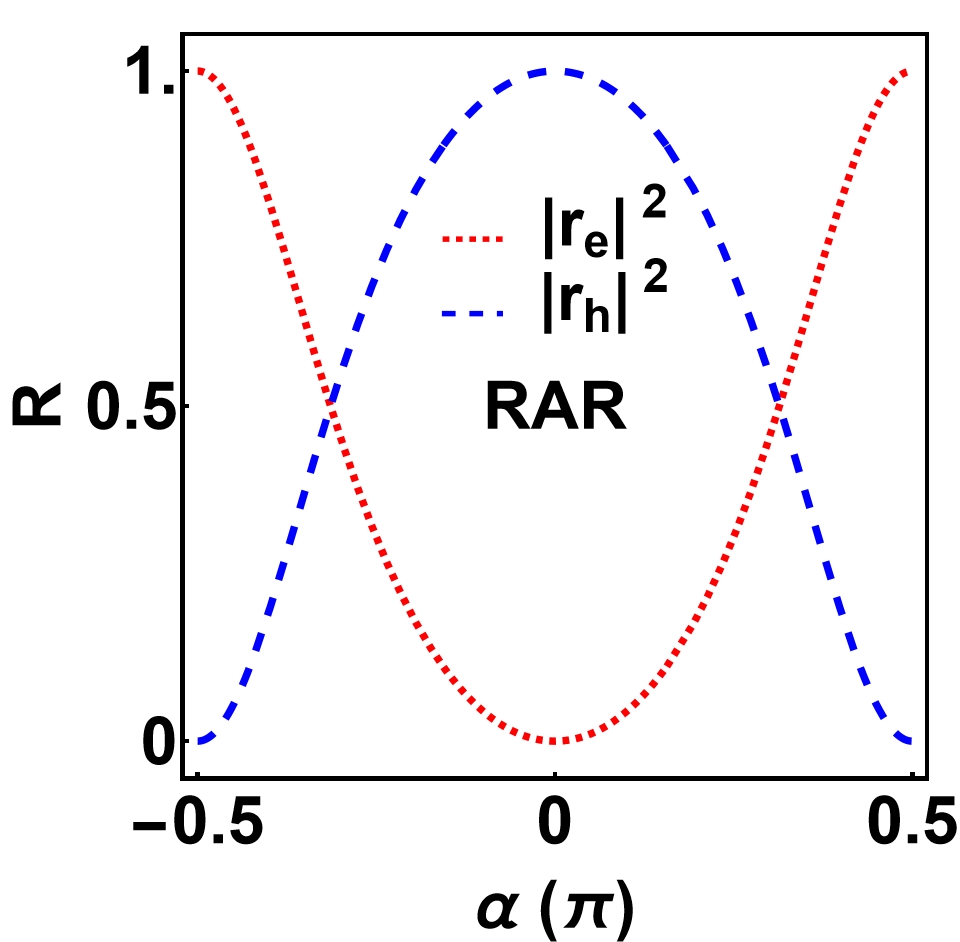}}
\subfloat[]{\includegraphics[width=.3\textwidth]{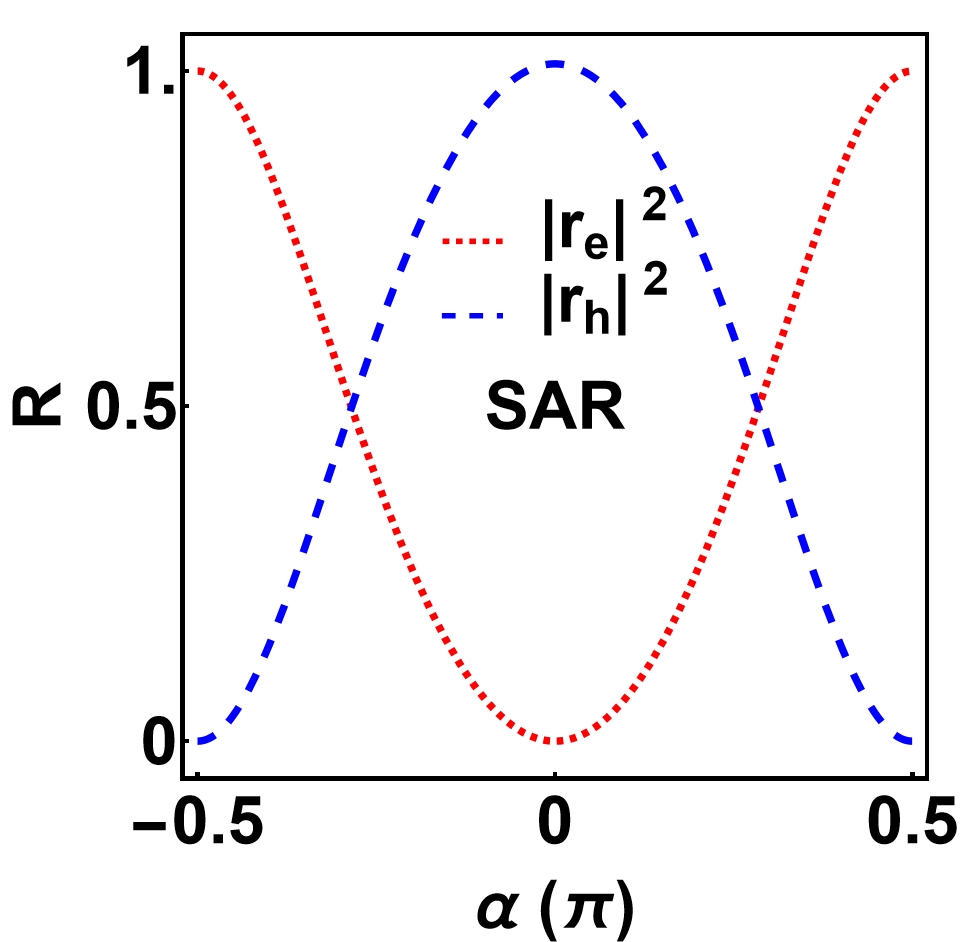}}
\subfloat[]{\includegraphics[width=.31\textwidth]{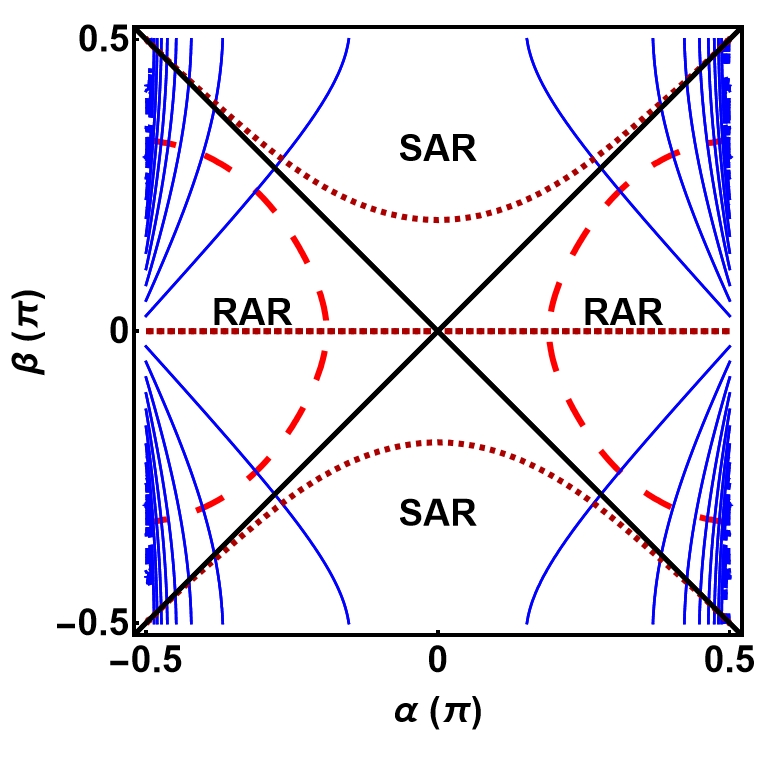}} 
\caption{\textit{(color online)} Symmetric processes - (a-b) Normalisation condition for the conservation of current for the RAR and the SAR processes for the energy of the particle less than the superconducting gap.  (c) Variation of energy-gap ratio, $\beta$, as a function of the incident angle $\alpha$.  The solutions are the intersection of the dashed curves in $magenta$ (RAR) and $brown$ (SAR) with the solid $blue$ lines. $\beta$ is a measure of the energy of the particle, and the angle $\alpha$ is a function of the wave vector $q$. All the plots correspond to the dispersions obtained for the symmetric cases in Eqs. \eqref{symmeq}(a)-(b). The width of the junction is 200 nm. The value of the superconducting gap 1.2 meV.}\label{norm}
\end{figure*}

For a consistent solution, we require det $M=1$, and this, in turn, implies the current conservation in the whole process given by the normalisation condition 
\begin{equation}\label{eq:normal}
|r_{e}|^2+|r_{h}|^2+|t_{e}|^2+|t_{h}|^2 = 1.
\end{equation}
We get analytical expressions for all the transmission and reflection coefficients with this. The explicit mathematical forms of these coefficients are given in Appendix \ref{append} Eq. (\ref{reftranseq}). Using these, we obtain the conditions to be satisfied by the incident energy and the incident angle for RAR and SAR to take place. The incident angles for electron and hole are opposite for RAR and equal for the case of SAR. This leads to the energy constraint for RAR, which demands that the Fermi energy $E_{F}$ should be much greater than the energy of the incident particle for the incident and the reflected particle to lie in the same band. However, for SAR, the incident particle energy should be greater than $E_F$ for the incident and the reflected particle to lie in different bands. These conditions are tabulated in Table \ref{tab:table1}.
\begin{table}[h!]
\fontsize{13}{10}
\centering
\begin{tabular}{|p{2.5cm}|p{2.5cm}|}
\hline
\textbf{RAR}& \textbf{SAR}\\
\hline
$E<E_{F}$ &$E>E_{F}$\\
$\alpha ' \approx -\alpha$ & $\alpha ' \approx  \alpha$\\
$k_{h} \approx  - k_{e} $& $k_{h} \approx  k_{e}$\\
\hline
\end{tabular}
\caption{Conditions for the two types of Andreev reflections namely the RAR and the SAR.}
\label{tab:table1}
\end{table}

We also note that the bias between the \textit{left} and the \textit{right} graphene regions can be adjusted by tuning the voltage $V$. Under these conditions, we get two cases: symmetric ($E_F^L=E_F^R, ~V=0$) and asymmetric ($E_F^L\neq E_F^R, ~V\neq0$). In the symmetric case, we get $|r_{e}|^2+|r_{h}|^2 \approx 1$, as the transmission is very small because transmission below the superconducting gap in the superconducting region is forbidden and most of the incident particles are reflected back, for both RAR and SAR. In the asymmetric case, the potential difference in the $left$ and the $right$ graphene regions leads to finite transmission, and all the four components $(r_e,r_h,t_e,t_h)$ contribute towards the normalisation of the current.

 To elucidate these points, we plot the normalisation for RAR and SAR for the symmetric $(V=0)$ case first and for energies less than the superconducting gap, in Fig.~\ref{norm}(a,b). We see that $|r_e|^2+|r_h|^2\approx 1$ and the reflection amplitudes for electrons and holes are complementary to each other.

Using Eq.~(\ref{TR}), we obtain the dispersion relation for the symmetric case, where we only get the RAR and the SAR.  In this case, the expressions for the dispersion relations are given by
\begin{widetext}
\begin{subequations}\label{symmeq}
\begin{align}
&  \textbf{RAR}\nonumber \\ 
& -4+\cosh 2d\kappa((3+\cos 2\alpha)\cos 2\beta +2\sin^{2}\alpha)
+4\cos 2dk_{0}\sin^{2}\alpha\sin^{2}\beta+4i\cos\alpha\sin 2\beta\sinh 2d\kappa =0\\ \nonumber \\
& \textbf{SAR}\nonumber \\ 
& 4\cos^{2}\alpha - ((3+\cos 2\alpha)\cos 2\beta -2 \sin ^{2}\alpha )\cosh 2d\kappa
 -4\cosh 2dk_{0}\sin^{2}\alpha\sin^{2}\beta -4i\cos\alpha\sin 2\beta\sinh 2d\kappa=0
\end{align} 
\label{dispsym}
\end{subequations}
\end{widetext}

The above equations can be written symbolically as
\begin{equation}\label{eq:symb_sym}
A_{r(s)}\cosh 2d\kappa + B_{r(s)}\sinh 2d\kappa + C_{r(s)}\cos 2dk_{0} - D_{r(s)} = 0
\end{equation}
where the subscript $r(s)$ denotes whether the process is retro(specular). The definitions of $A_{r(s)}$, $B_{r(s)}$, $C_{r(s)}$ and $D_{r(s)}$ are given in Table \ref{tab:table2}.
\begin{widetext}
\begin{center}
\begin{table}[H]
\fontsize{11}{9}
\centering
    \begin{tabular}{|p{7.5cm}|p{7.5cm}|}
    \hline
	\textbf{RAR} & \textbf{SAR}\\
	\hline
	$A_{r} = -(\cos 2\beta(3+\cos 2\alpha)-2 \sin ^{2}\alpha)$  & $A_{s} =  (\cos 2\beta(3+\cos 2\alpha)+2 \sin ^{2}\alpha)$  \\
	$B_{r} = -4i \cos\alpha\sin 2\beta$ & $B_{s} = 4i \cos\alpha\sin 2\beta$\\
	$C_{r} = -4 \sin ^{2}\alpha \sin ^{2} \beta$ & $C_{s} = 4 \sin ^{2}\alpha \sin ^{2} \beta$\\
	$D_{r} = -4\cos ^{2}\alpha$ & $D_{s} = 4$  \\
	\hline  
    \end{tabular}
    \caption{Definitions of the coefficients for retro and specular processes for the symmetric case}
    \label{tab:table2}
\end{table}
\end{center}
\end{widetext}
The dispersion relation obtained in Eq.~(\ref{eq:symb_sym}) is complex in nature, for which the solutions can be obtained only when both the real and complex part of the dispersion is zero. The bound states obtained in this case are rather trivial despite the complex dispersion relation, and the resulting bound states are just the Cooper pairs in the superconducting region of the junction. Therefore, a GSG heterojunction has no Andreev Bound States(ABS).

In Fig.~\ref{norm}(c), we plot the variation of energy-gap ratio, $\beta$, as a function of the incident angle $\alpha$, which is a function of the wave vector \textit{q}. This is obtained from the solution of the complex transcendental dispersion relations discussed above; see Eq.~(\ref{symmeq}). This plot of angular variation helps us understand the dispersion indirectly as it corresponds to the $E-q$ relation indirectly through $\beta$ and $\alpha$. This angular variation plot clearly separates the regions of RAR and SAR. 

\begin{center}
\begin{figure*}
\textbf{RAR}\\
\subfloat[]{\includegraphics[width=.23\textwidth]{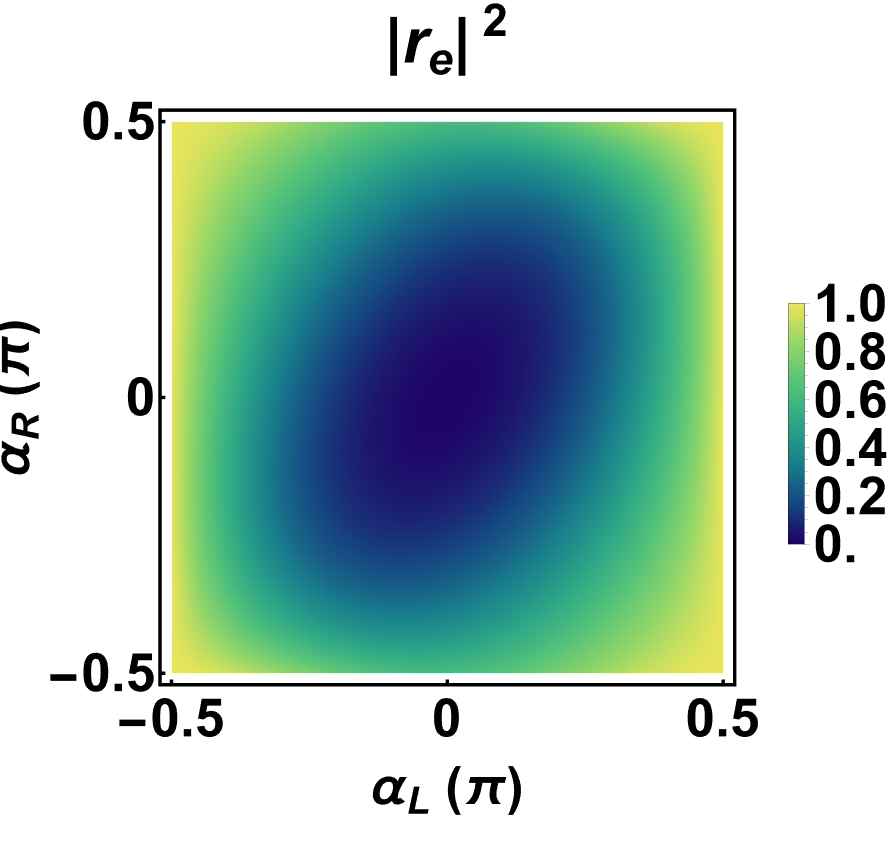}} 
\subfloat[]{\includegraphics[width=.23\textwidth]{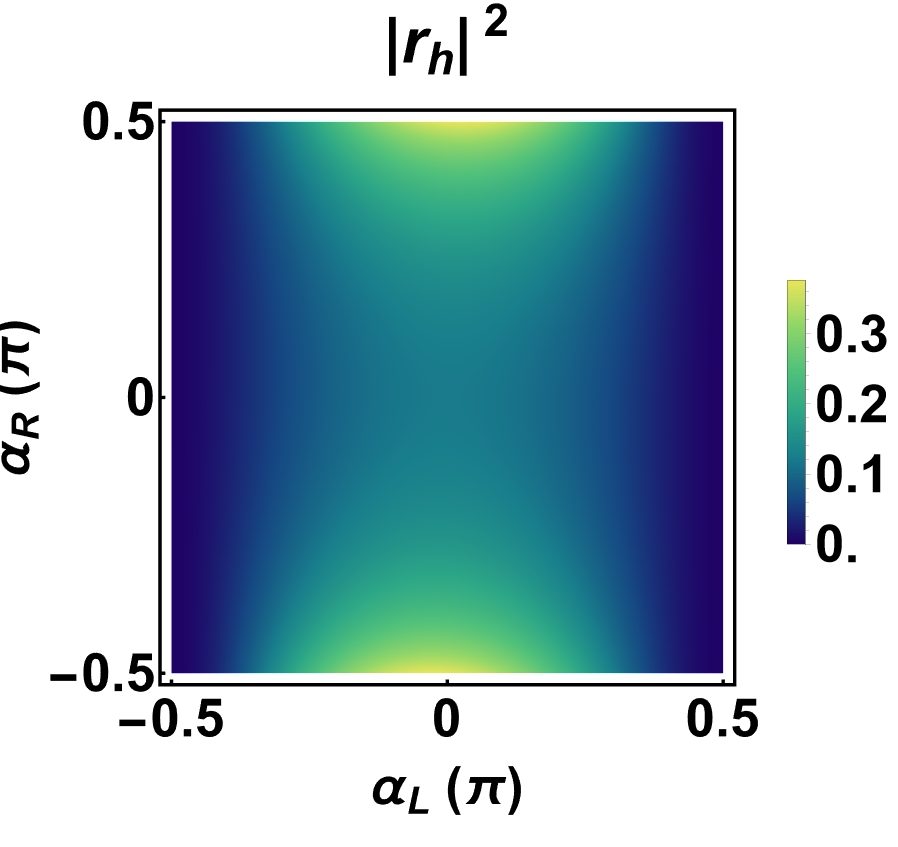}}
\subfloat[]{\includegraphics[width=.23\textwidth]{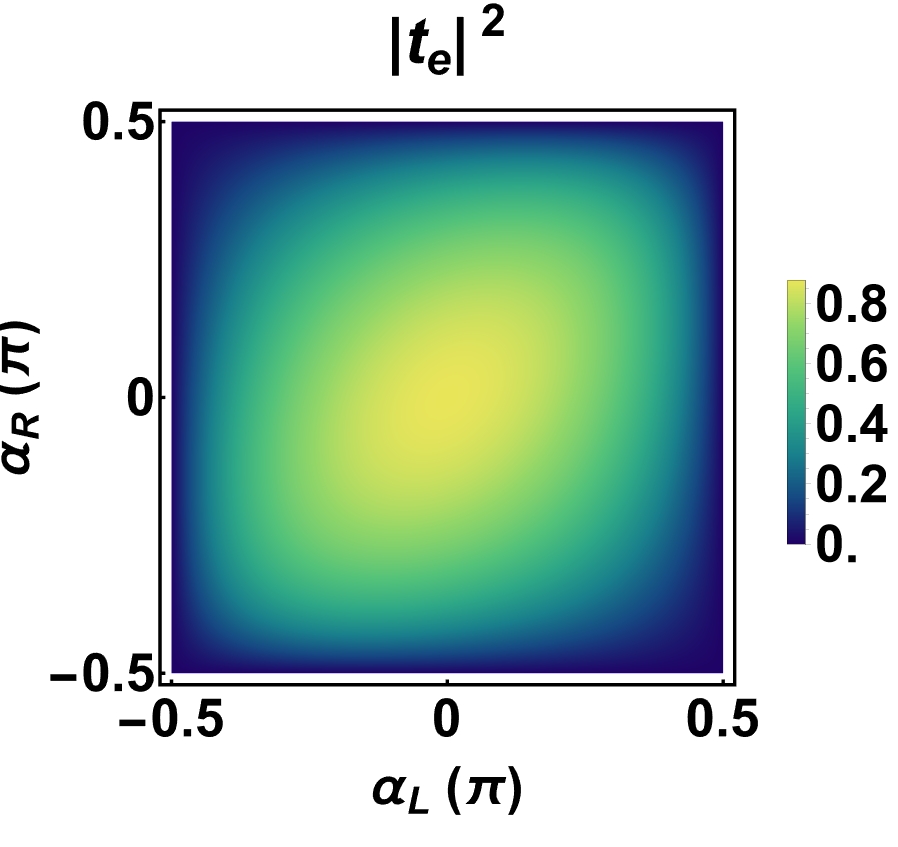}}
\subfloat[]{\includegraphics[width=.25\textwidth]{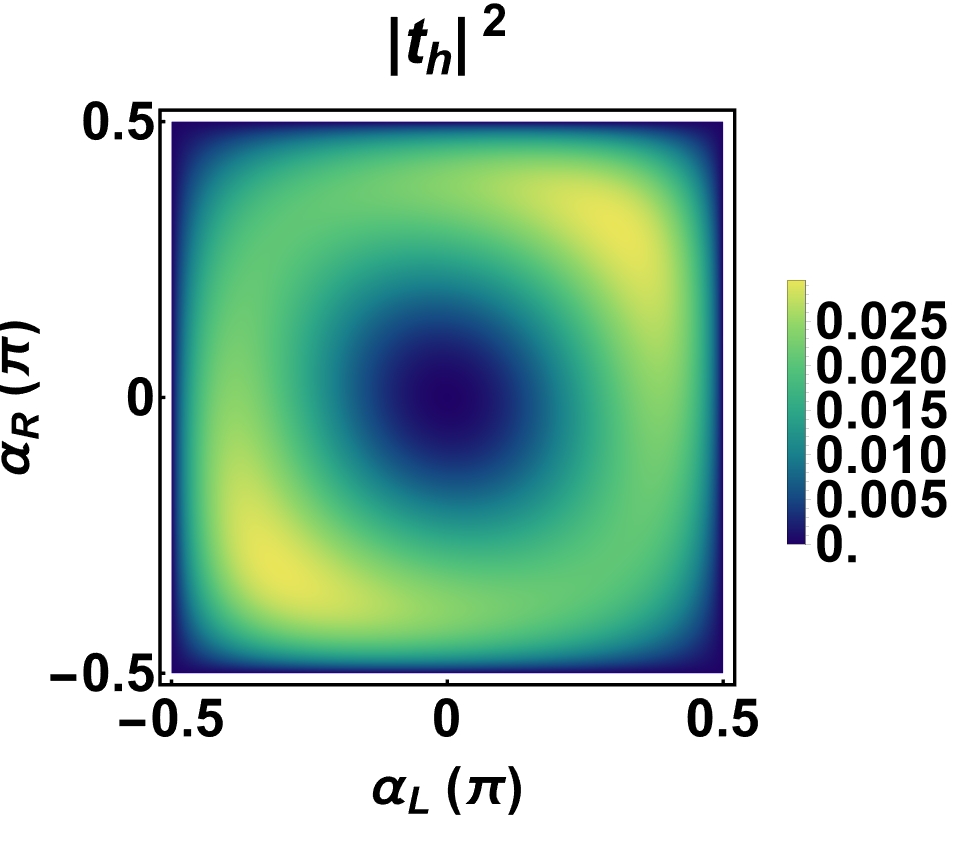}} \\
\textbf{SAR}\\
\subfloat[]{\includegraphics[width=.23\textwidth]{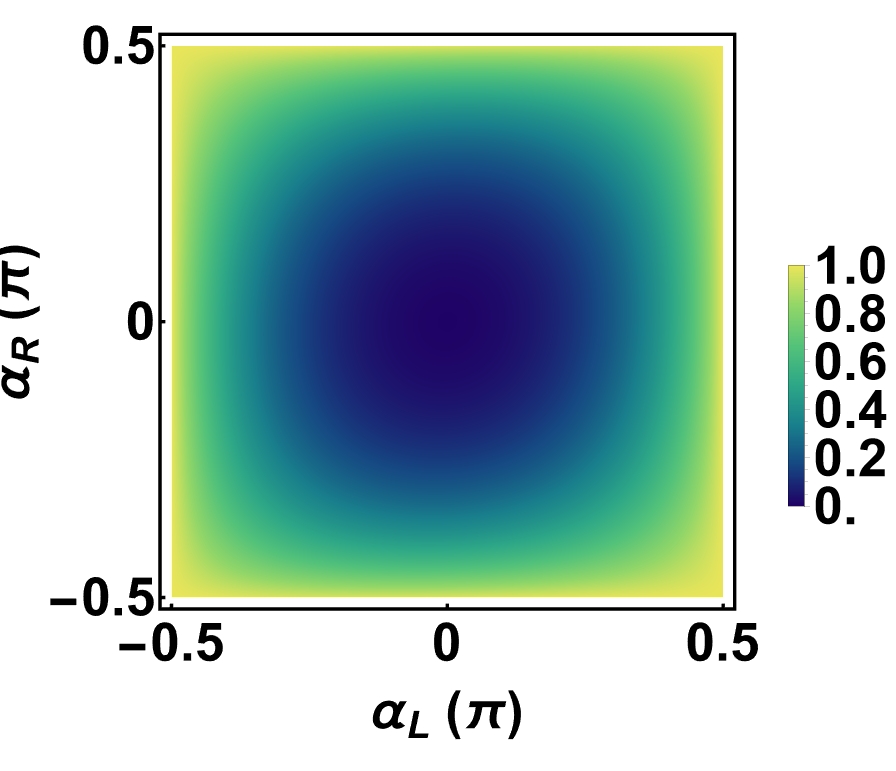}} 
\subfloat[]{\includegraphics[width=.25\textwidth]{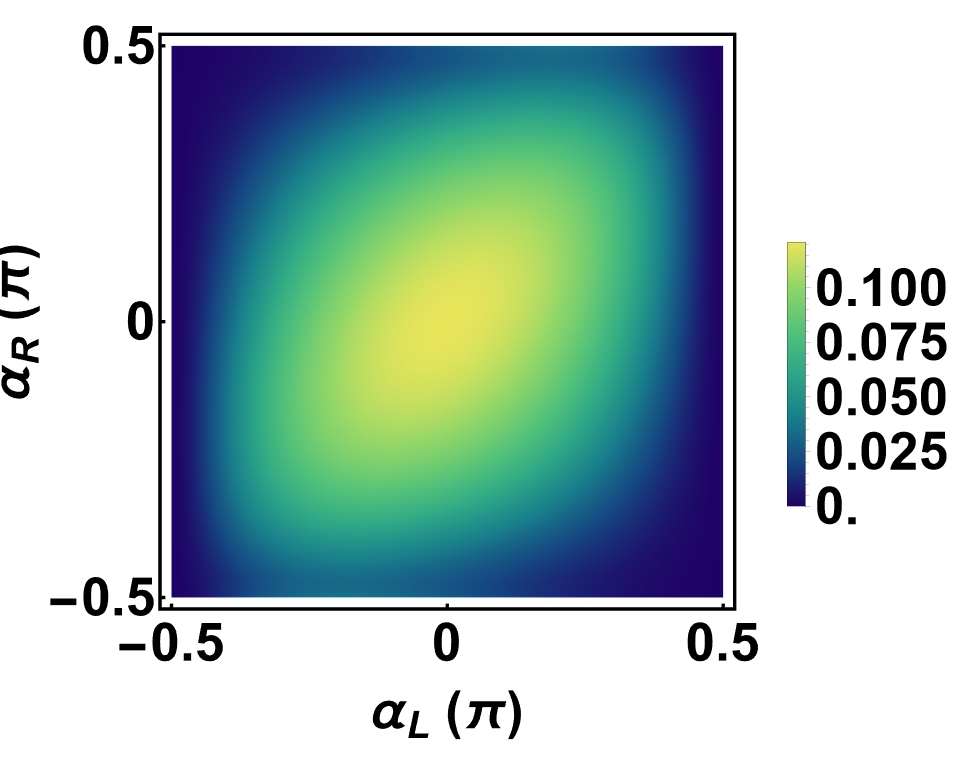}}
\subfloat[]{\includegraphics[width=.23\textwidth]{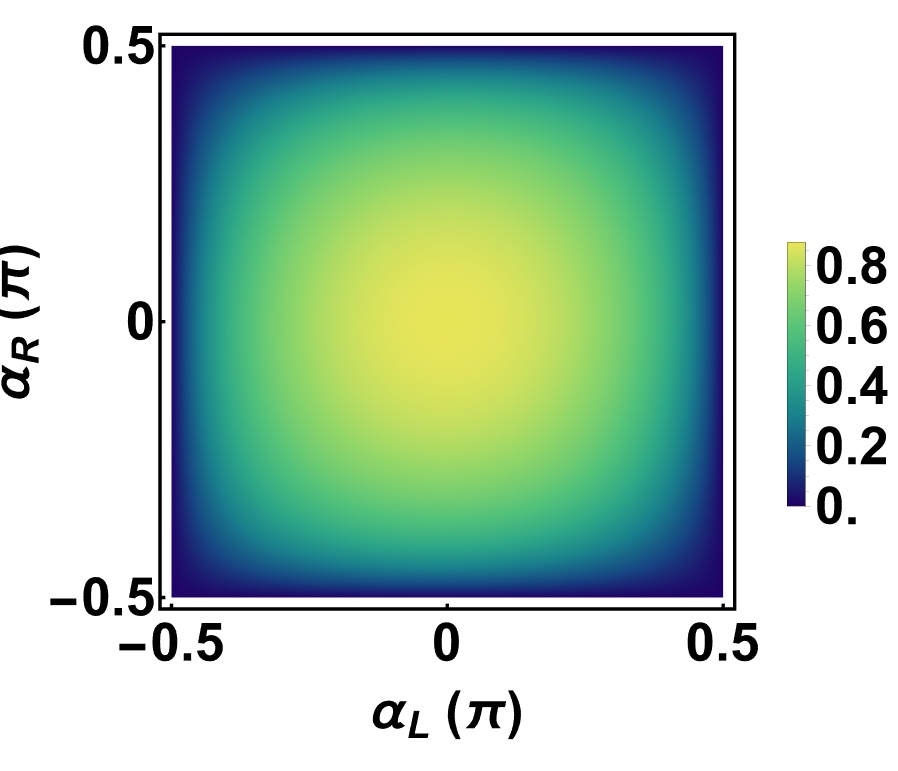}}
\subfloat[]{\includegraphics[width=.24\textwidth]{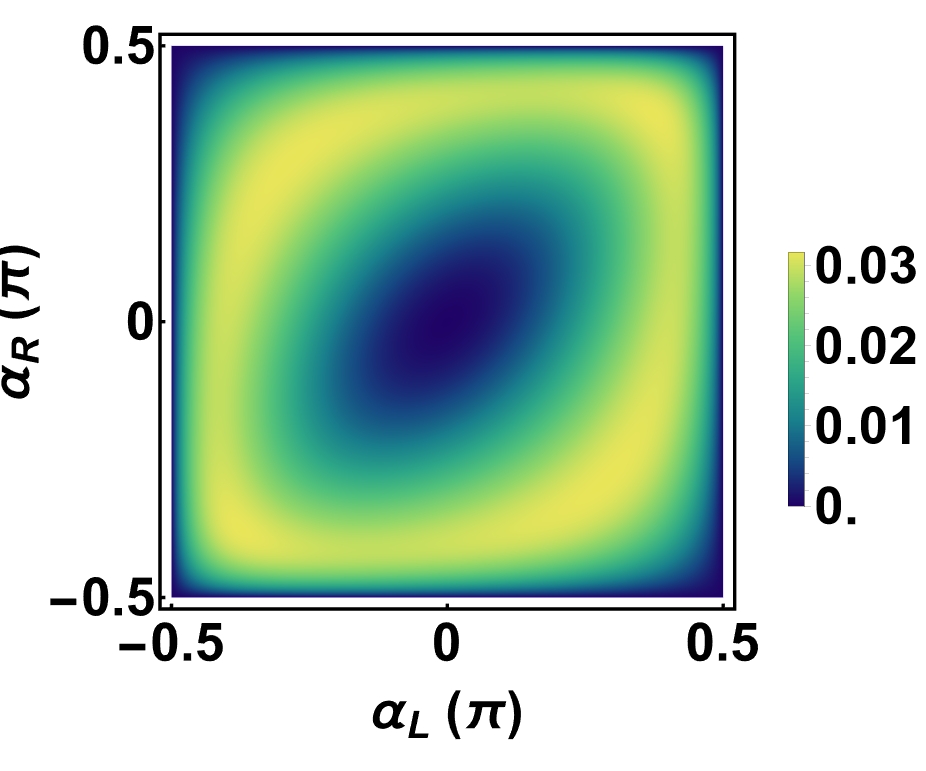}} \\
\textbf{RSP}\\
\subfloat[]{\includegraphics[width=.23\textwidth]{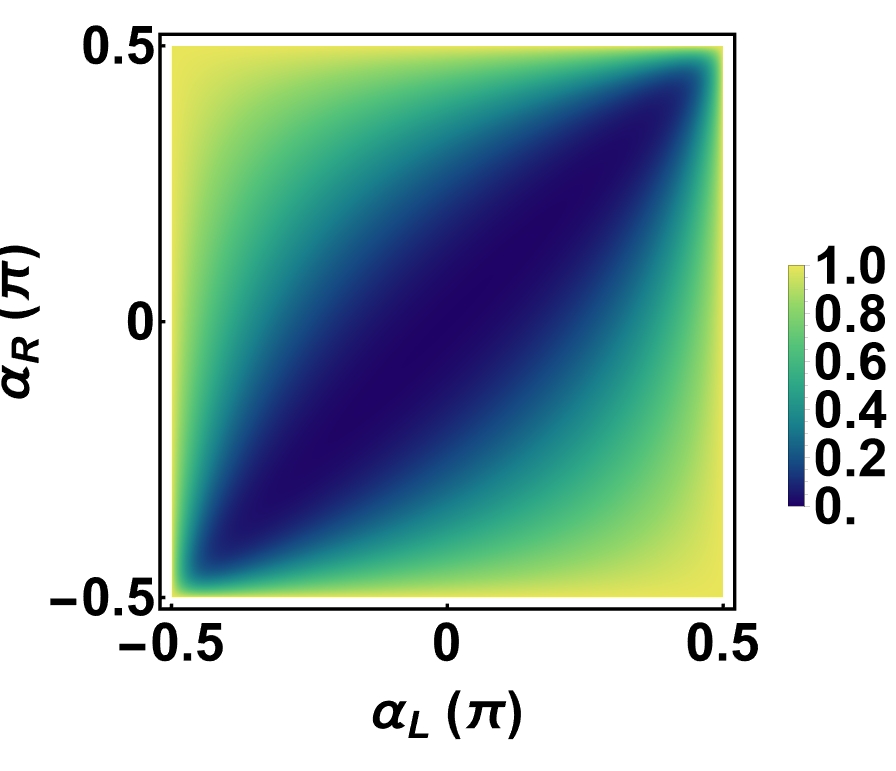}} 
\subfloat[]{\includegraphics[width=.25\textwidth]{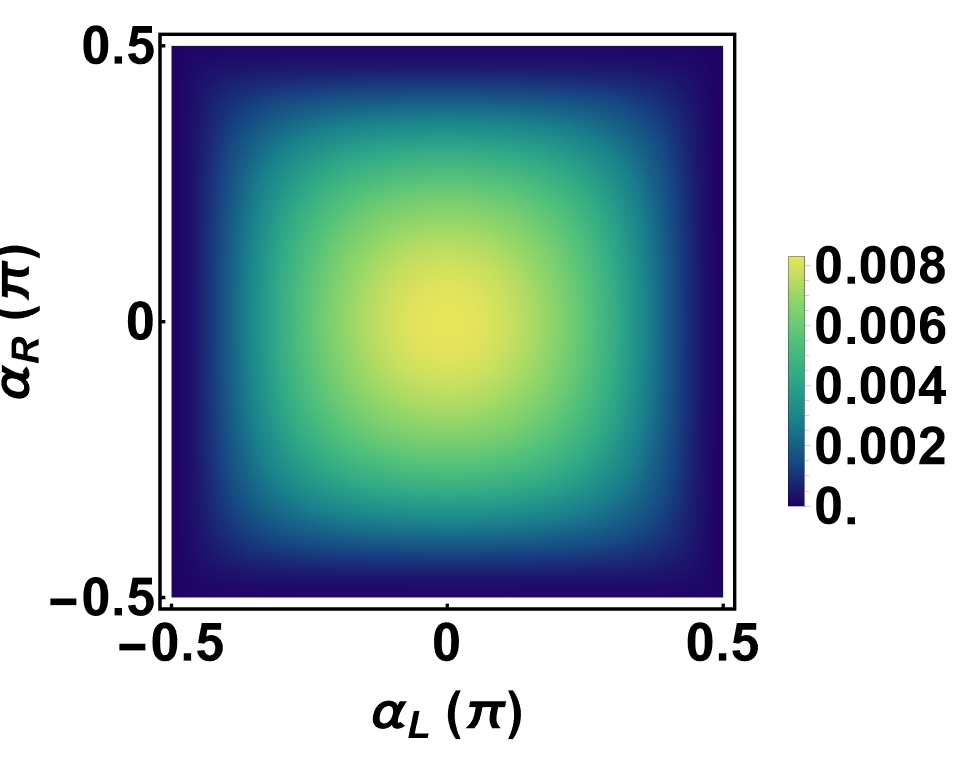}}
\subfloat[]{\includegraphics[width=.23\textwidth]{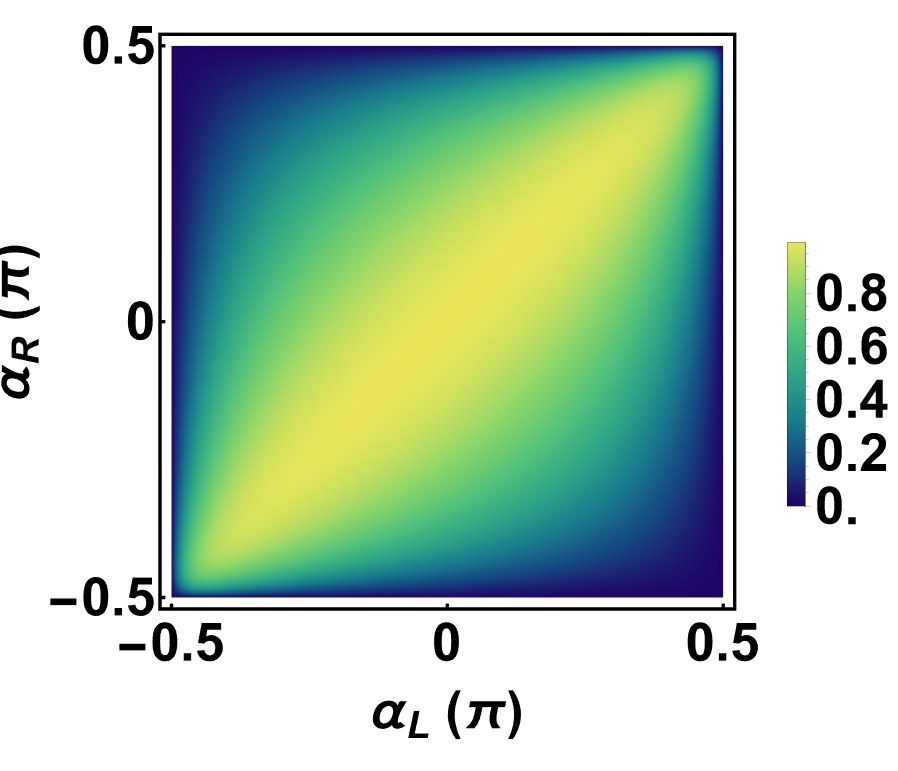}}
\subfloat[]{\includegraphics[width=.24\textwidth]{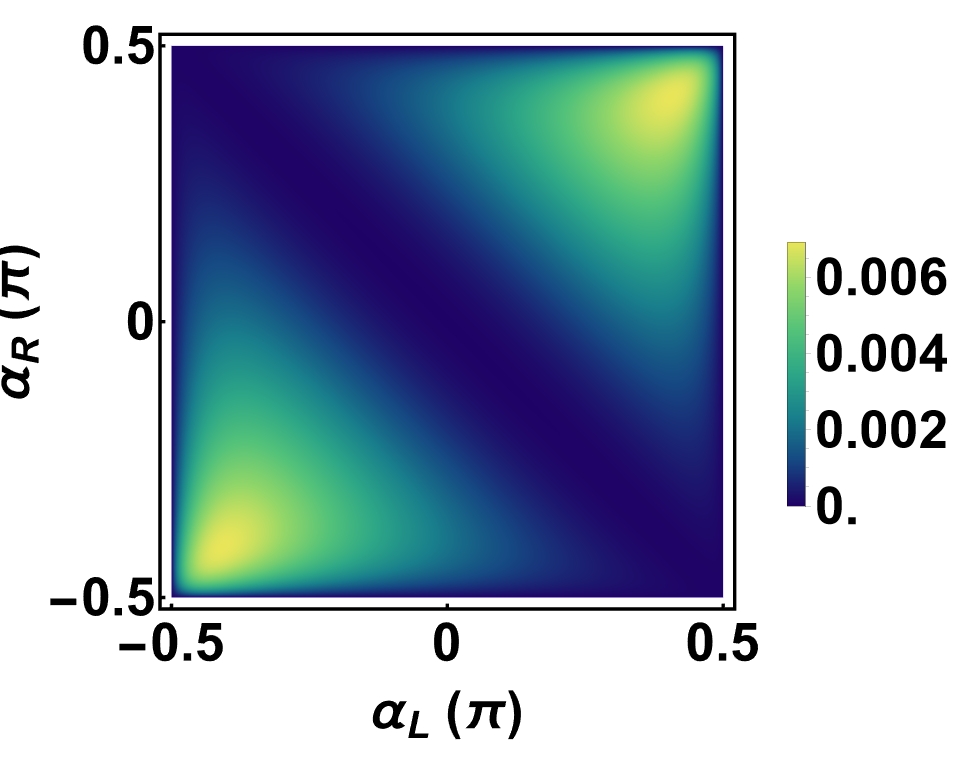}} \\
\textbf{SRP}\\
\subfloat[]{\includegraphics[width=.23\textwidth]{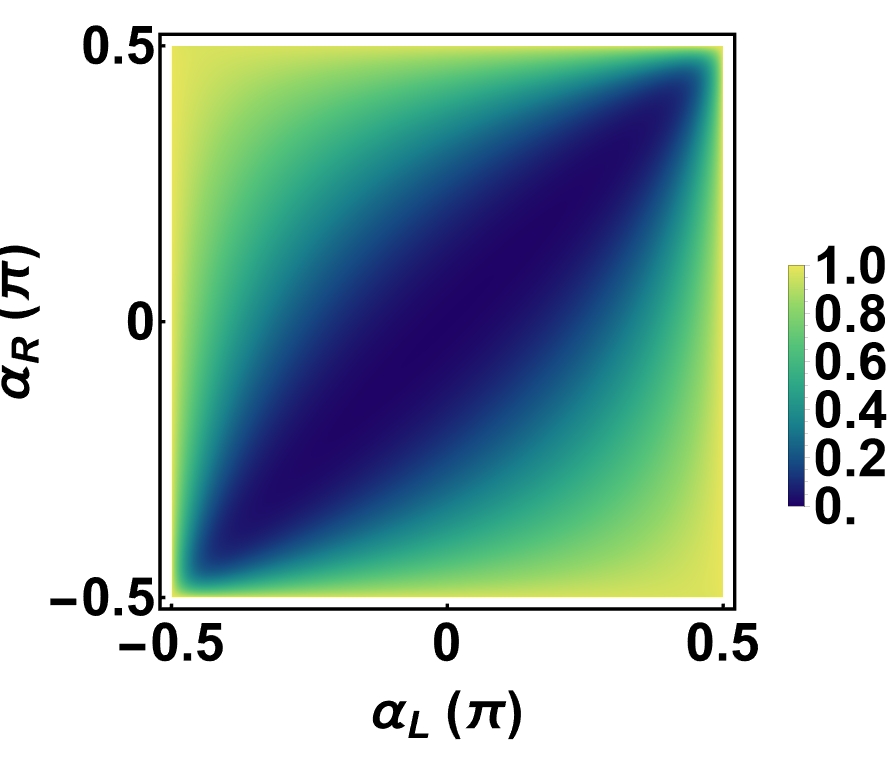}} 
\subfloat[]{\includegraphics[width=.24\textwidth]{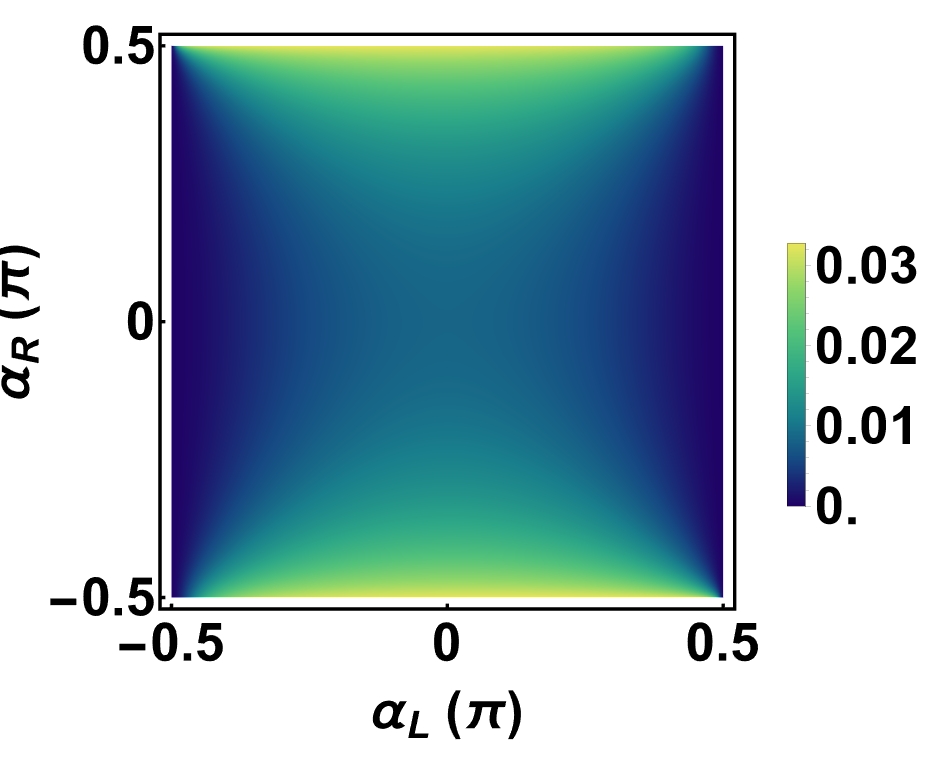}}
\subfloat[]{\includegraphics[width=.23\textwidth]{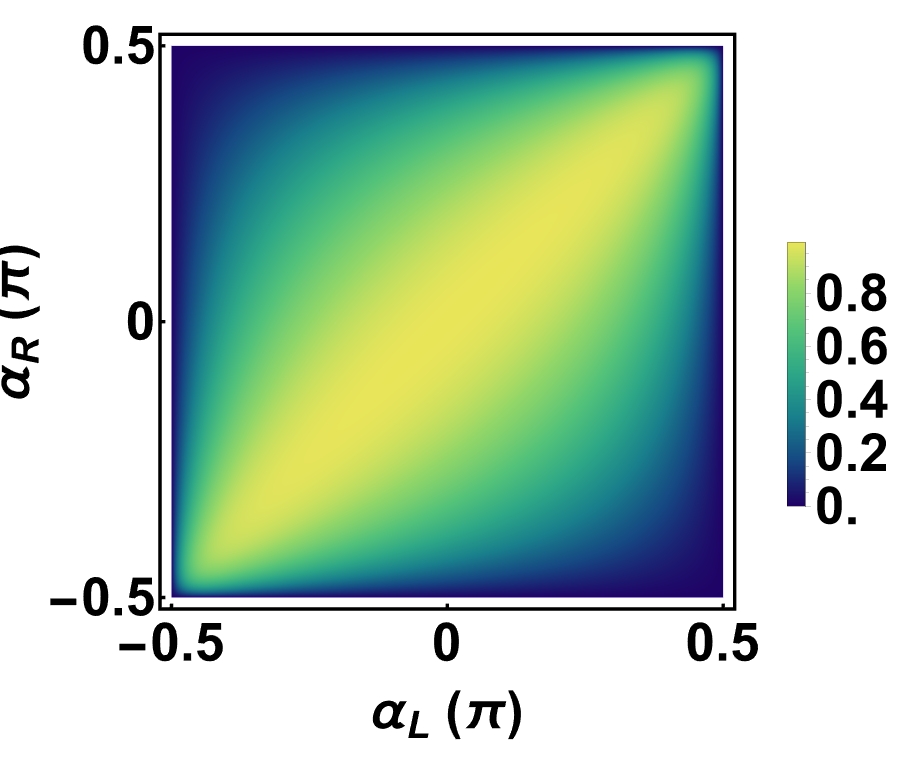}}
\subfloat[]{\includegraphics[width=.25\textwidth]{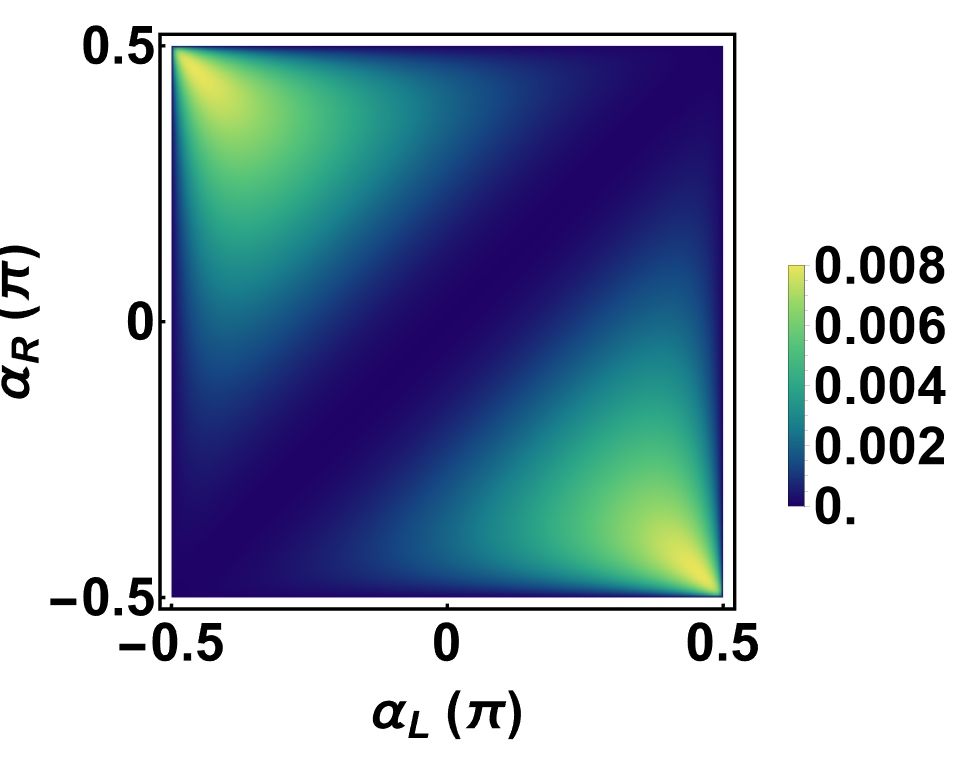}} \\

\caption{\textit{(color online)} Normalisation for asymmetric processes - the first two vertical columns represent reflection amplitudes for electrons and holes, respectively, while the last two represent the transmission amplitudes for the electrons and holes. (a-d) RAR: Variation of the reflection and the transmission amplitudes for RAR. (e-h) Variation of the reflection and the transmission amplitudes for SAR. (i-l) Variation of the reflection and the transmission amplitudes for the RSP. (m-p) Variation of the reflection and the transmission amplitudes for the SRP. The value of the superconducting gap is 1.2 meV. The potential $U$ in the superconducting region is 200$\Delta$. The width $d$ of the superconducting region is 200 nm.}\label{norm3}
\end{figure*}
\end{center}
\begin{center}
\begin{figure*}
\subfloat[]{\includegraphics[width=.23\textwidth]{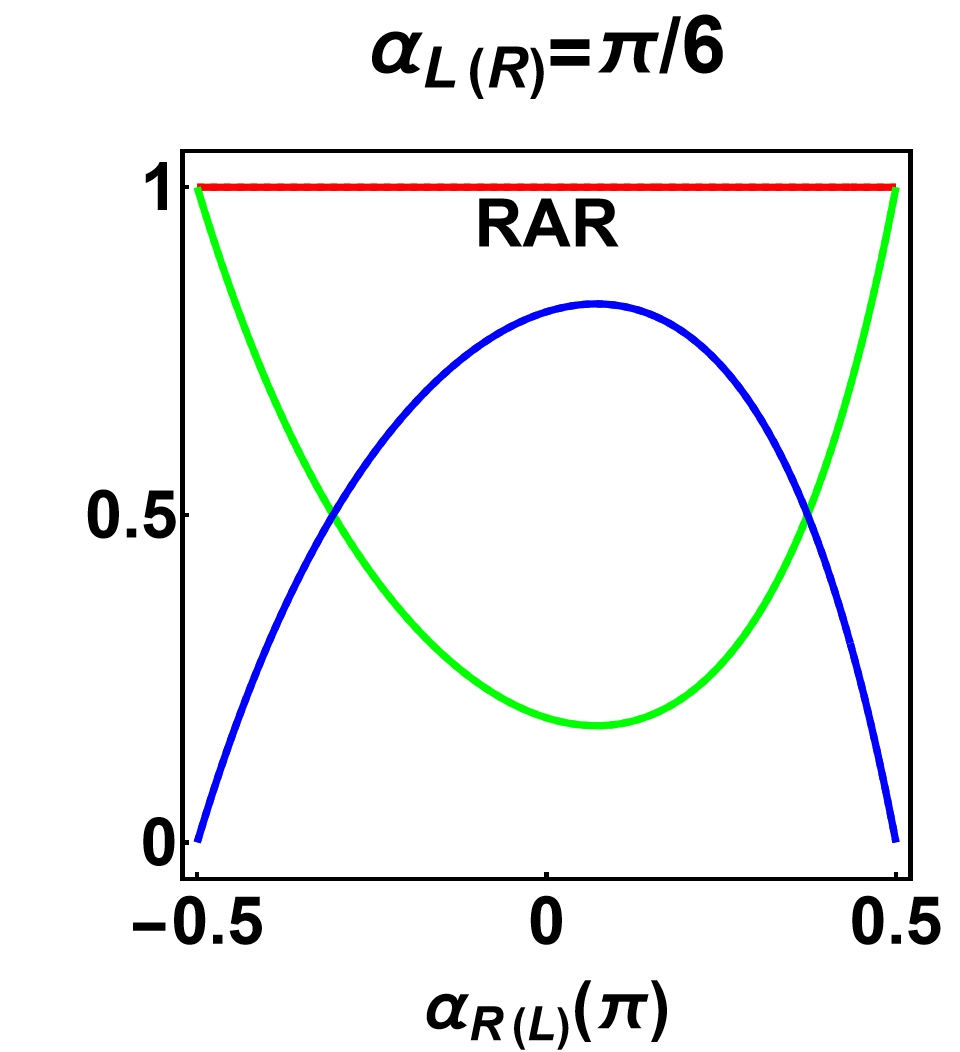}}
\subfloat[]{\includegraphics[width=.23\textwidth]{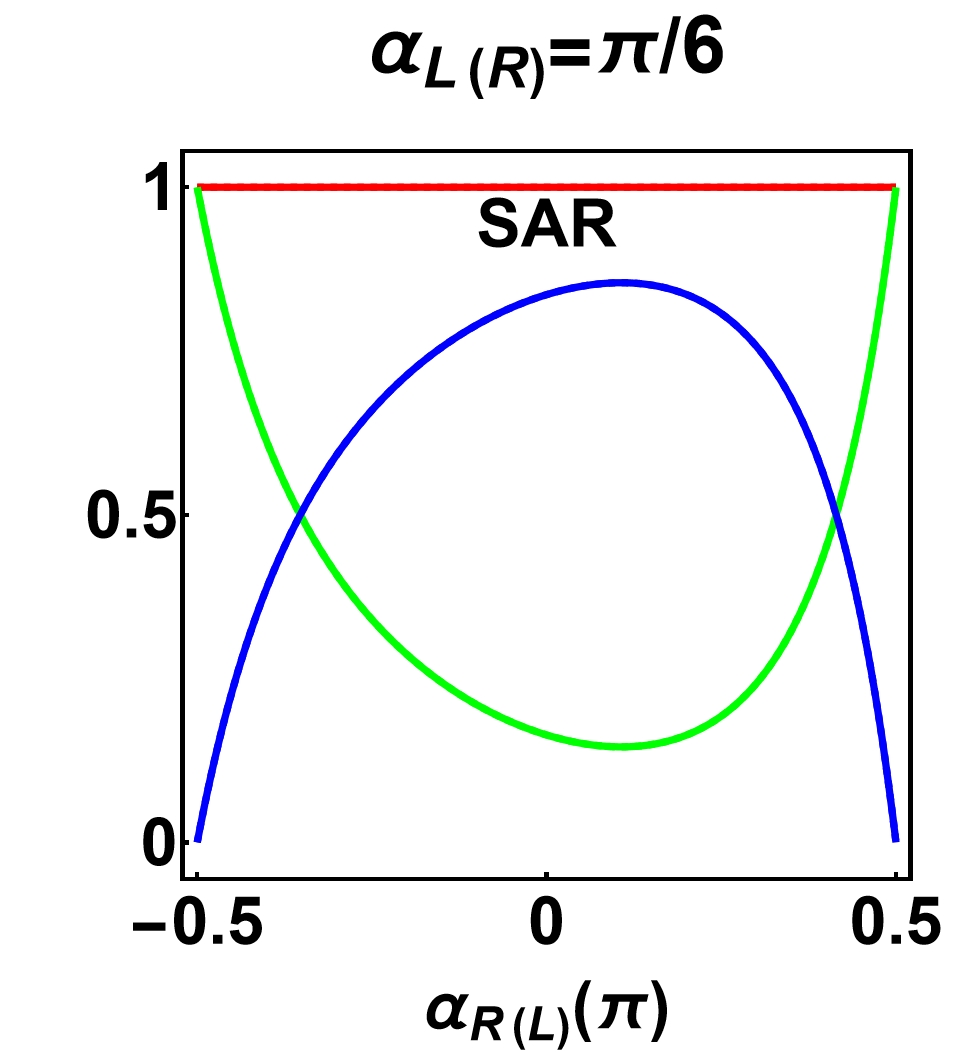}}
\subfloat[]{\includegraphics[width=.23\textwidth]{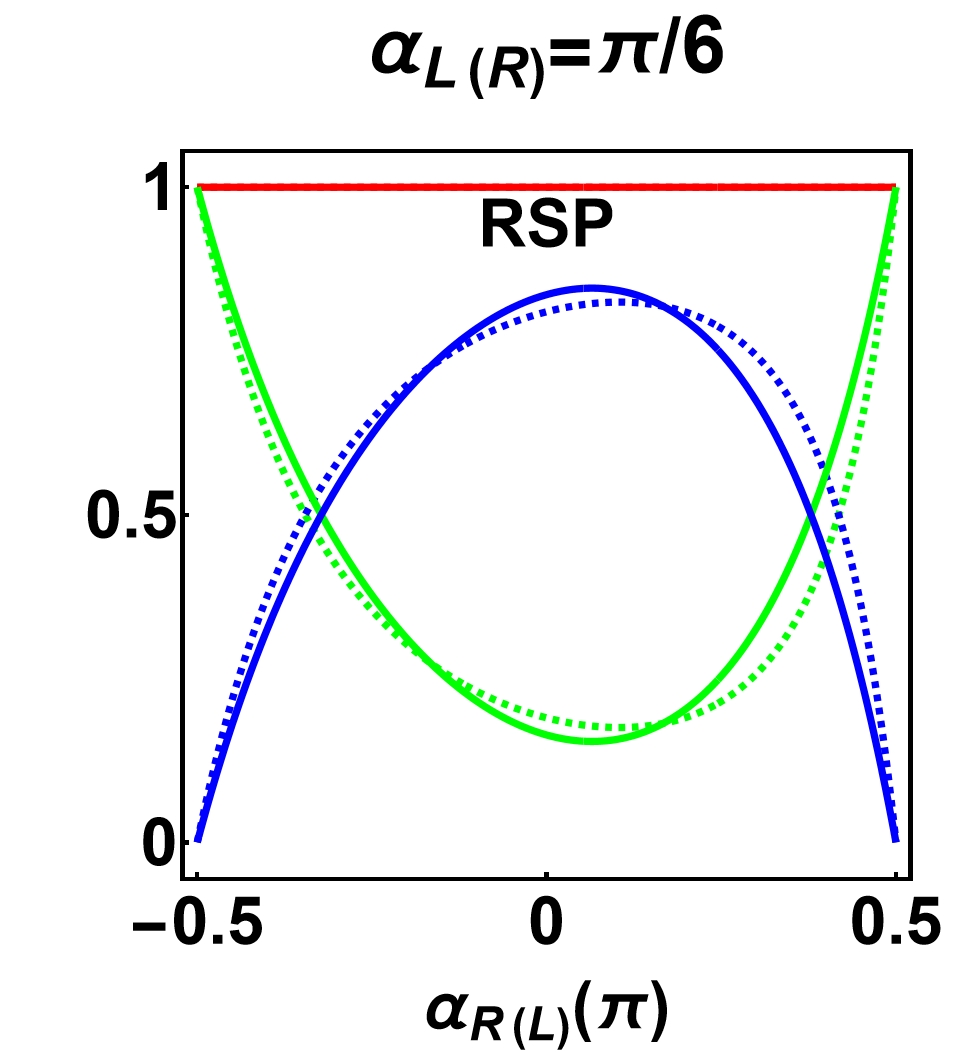}}
\subfloat[]{\includegraphics[width=.23\textwidth]{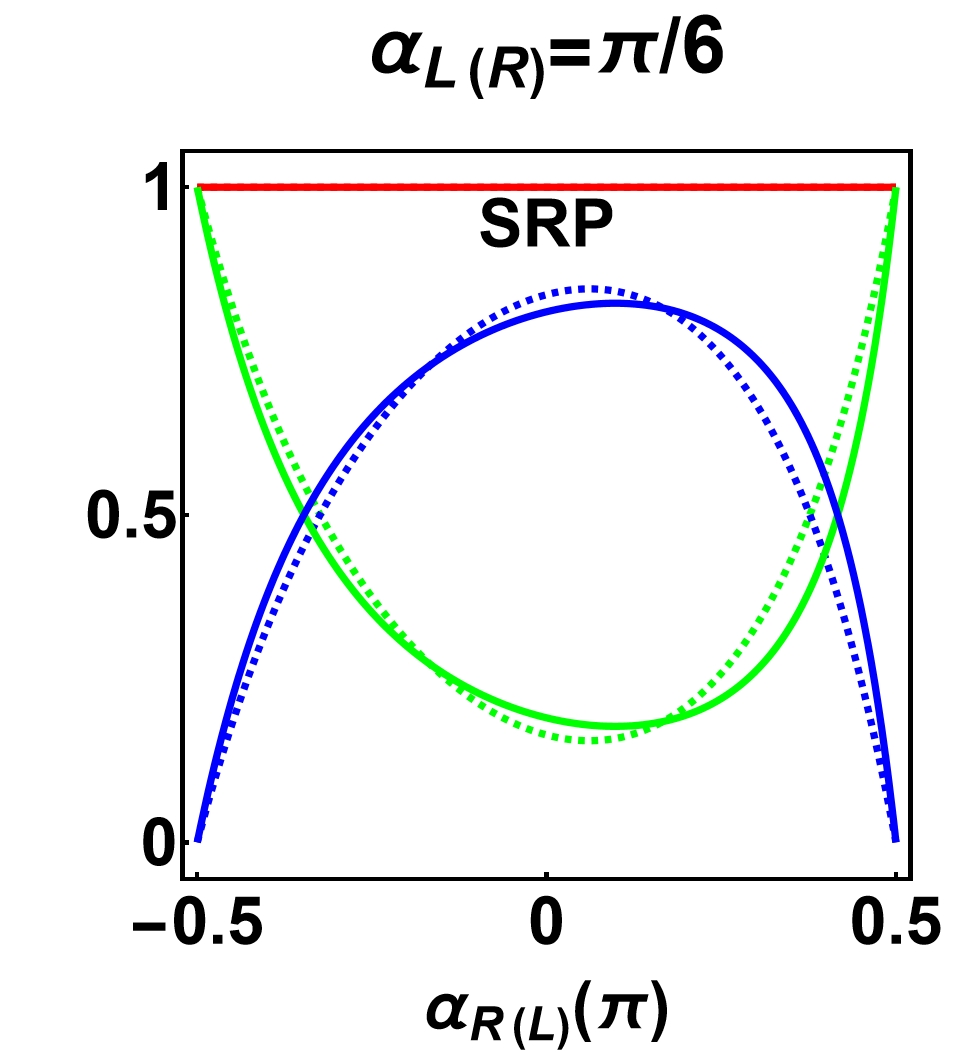}}
\caption{\textit{(color online)} {\color{black}Variation of the reflection and transmission amplitudes and total normalisation for various fixed incident angles for all four asymmetric cases - (a)  RAR, (b) SAR, (c) RSP and (d) SRP. 
The value of the superconducting gap is 1.2 meV. The potential $U$ in the superconducting region is 200$\Delta$. The width $d$ of the superconducting region is 200 nm. The $red$ line shows the normalisation in all the processes. The \textit{blue} line denotes the total transmission amplitude, and the \textit{green} line is the total reflection amplitude. The solid lines correspond to the case when the incident angle at the \textit{left} GS interface is varied, while the dashed lines correspond to the variation of the angle at the \textit{right} SG interface.}}\label{norm2}
\end{figure*}
\end{center}
For the asymmetric case, we obtain a general dispersion relation which involves a potential difference between the two graphene regions and, thus, corresponds to the symmetric case in the limit of zero potential difference between the two graphene regions. Based on the energy conditions given in Table \ref{tab:table1} and Eq.~(\ref{TR}), we get different dispersion relations for various processes. For example, the process can be retro on the \textit{left} interface and specular on the \textit{right} interface, which gives us the Retro-Specular process (RSP). One can also have the specular process at the \textit{left} interface and the retro process at the \textit{right} interface, giving us the Specular-Retro process (SRP). The dispersion relations for the four processes - RAR, SAR, RSP and SRP - are given below -
\begin{widetext}
\begin{subequations}
\begin{align}
&  \textbf{RAR}\nonumber \\ 
& 2\csc^{2}\beta -2\cos 2dk_{0}\tan\alpha_{L}\tan\alpha_{R}
-2\cosh 2d\kappa (\cot^{2}\beta - \sec\alpha_{L}\sec\alpha_{R})
- 2i\cot\beta(\sec\alpha_{L}+\sec\alpha_{R})\sinh 2d\kappa =0
\nonumber \\\\
& \textbf{SAR}\nonumber \\
& 2\csc^{2}\beta\sec\alpha_{L}\sec\alpha_{R}+2\cosh 2d\kappa\left(1-\frac{\cot ^{2}\beta}{\cos\alpha_{L}\cos\alpha_{R}}\right)  -2\cos 2dk_{0}\tan\alpha_{L}\tan\alpha_{R} - 2i\cot\beta(\sec\alpha_{L}+\sec\alpha_{R})\sinh2d\kappa= 0\\
& \textbf{RSP}\nonumber \\ 
& 2\csc ^{2}\beta \sec \alpha_{R} +2\cosh 2d\kappa(\sec\alpha_{L} - \sec\alpha_{R}\cot ^{2}\beta) - 2i\sin 2dk_{0}\tan\alpha_{L}\tan\alpha_{R}- 2i\cot\beta(1+\sec\alpha_{L}\sec\alpha_{R})\sinh 2d\kappa=0
\\
& \textbf{SRP}\nonumber \\ 
& 2\sec\alpha_{L}\csc ^{2}\beta - 2\cosh 2d\kappa(\sec\alpha_{L}\cot ^{2}\beta - \sec\alpha_{R}) -2i \sin 2dk_{0}\tan\alpha_{L}\tan\alpha_{R} - 2i(1+ \sec\alpha_{L}\sec\alpha_{R})\cot\beta\sinh 2d\kappa=0
\end{align} 
\label{dispasym}
\end{subequations}
\end{widetext}
The above four dispersion relations also imply that there is a mismatch between the Fermi levels of the graphene region on the two sides. These relations are transcendental and must be solved numerically to obtain real solutions for the energy $E$.

 We now plot the various reflection and transmission amplitudes associated with the four asymmetric processes in Fig.~\ref{norm3}. We have two different angles, $\alpha_L$ and $\alpha_R$, which can be controlled independently; we plot the four components separately to look at the complete normalisation. Some significant features observed in these plots are the hole transmission for RAR (Fig.~\ref{norm3}(d)) and SAR (Fig.~\ref{norm3}(h)) are complementary, as the usual retro and the specular processes are complementary. However, as we go on to more complex RSP (Fig.~\ref{norm3}(i-l)) and the SRP(Fig.~\ref{norm3}(m-p)) processes, we see that electron reflection and transmission are comparable while the hole transmission is complementary. In order to understand these normalisations in a better way, we make a cross-sectional plot by fixing one of the angles in Fig.~\ref{norm2}.

In Fig.~\ref{norm2}, we plot the total reflection and the transmission for each of the asymmetric processes. We plot these normalisations in two ways: first, by varying $\alpha_{L}$ and fixing the value of $\alpha_{R}$ (denoted by solid lines in Fig.~\ref{norm2}) and second, by varying $\alpha_{R}$, while keeping the value of $\alpha_{L}$ constant (denoted by dashed lines in Fig.~\ref{norm2}). We have chosen $\alpha_{L(R)}=\pi/6$ for all the cases. Now, for the RAR and the SAR cases, in Fig.~\ref{norm2}(a,b), as these processes at the \textit{left} and the \textit{right} interface are the same, the solid and the dashed lines overlap, and we see a perfect left-right symmetry. However, for the mixed processes (RSP and SRP) in Fig.~\ref{norm2}(c,d),  the processes at the two interfaces are different. Thus, we see that the solid and the dashed lines do not overlap completely, indicating the presence of an asymmetry in the system.
\begin{figure}
\subfloat[]{\includegraphics[width=.5\columnwidth]{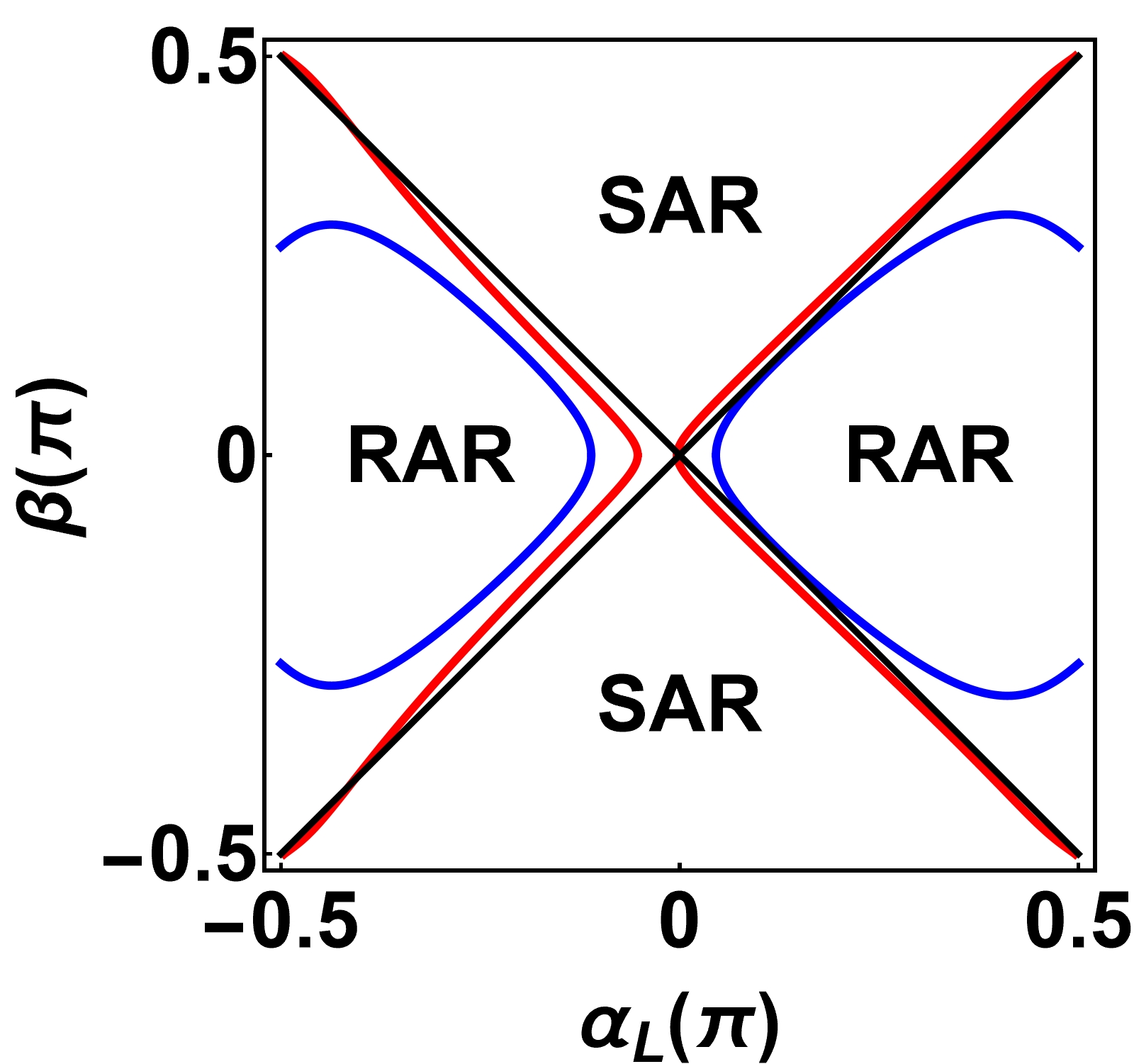}}
\subfloat[]{\includegraphics[width=.5\columnwidth]{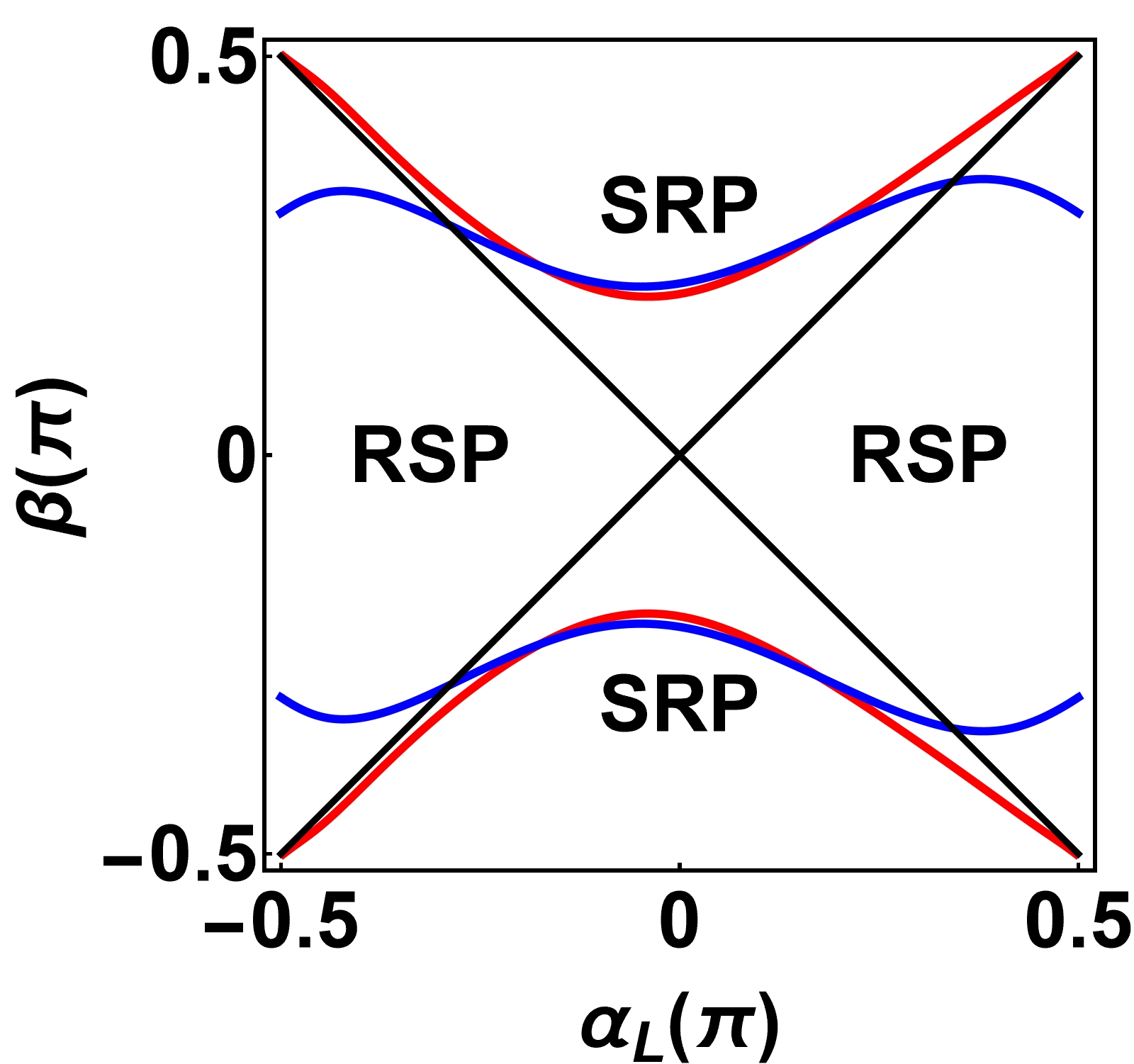}}\\
\caption{\textit{(color online)} Asymmetric processes - Variation of the energy-gap ratio, $\beta$, as a function of the incident angle, $\alpha_L$ for (a) RAR (\textit{blue} curves) and SAR (\textit{red} curves) and (b) for RSP (\textit{blue} curves) and SRP (\textit{red} curves) with $\alpha_R=\pi/6$. Parameters used are $\Delta$=1.2 meV, \textit{U}=200$\Delta$ and \textit{d}=200 nm.} \label{anorm}
\end{figure}

We now analyse the dispersion relations of Eq.~(\ref{dispasym}). Since these equations are transcendental in nature, we solve these numerically and obtain a graphical solution for the energy-gap ratio, $\beta$, as a function of the incident angle, $\alpha_L$, corresponding to RAR and SAR processes in Fig.~\ref{anorm}(a) and RSP and SRP in Fig.~\ref{anorm}(b) for a fixed value of $\alpha_R=\pi/6$, since we are only interested in the reflection processes at the first interface. We observe a shift along $\alpha_{L}$ in the energy dispersions in Fig.~\ref{anorm}(a,b), which indicates the presence of an asymmetry in the potentials applied to the \textit{left} and \textit{right} graphene regions. We see that the variations plotted in Fig.\ref{anorm}(a), show a qualitative similarity with the symmetric case plotted in Fig.\ref{norm}(c). The only difference is that the curves are now displaced from the centre, indicating that there is an asymmetry in the potential bias. In Fig.\ref{anorm}(b), we see that the curves for the RSP and the SRP processes have some overlap, indicating that these processes are mixed and cannot be separated in that region. 

The richness of the dispersion relation presented in Fig.~\ref{anorm} and the behaviour of transmission coefficients plotted in Fig.~\ref{norm}, Fig.~\ref{norm3} and Fig.~\ref{norm2}
promises interesting features in electron and hole transport in such GSG junction. To unnderstand this in the next section, we calculate the Goos-H\"anchen shift and the differential conductivity in such heterojunctions.

\section{Results and Discussion}\label{RD}
Using the transfer matrix framework developed in the preceding section \ref{hamwaven}, we shall calculate the Goos-H\"anchen shift and differential conductance over a representative range of energy $E$ and potential $V$ that can characterize such GSG junction.

\begin{center}
\begin{figure}
\subfloat[]{\includegraphics[width=.22\textwidth]{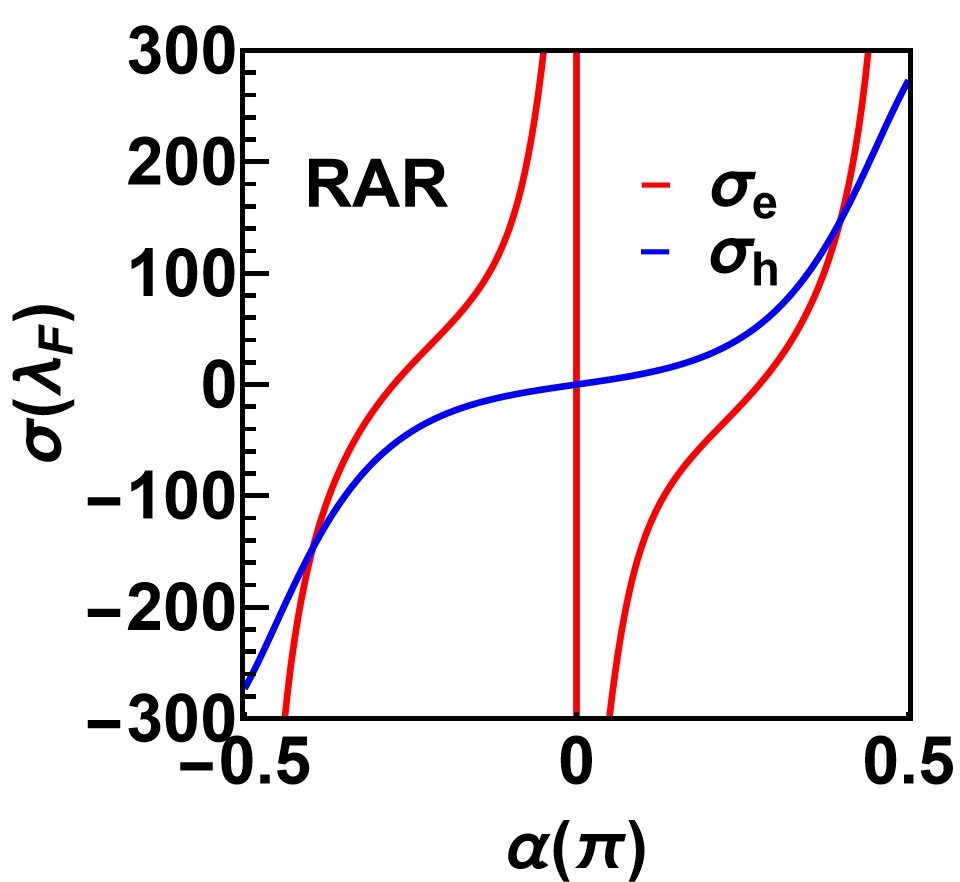}}
\subfloat[]{\includegraphics[width=.22\textwidth]{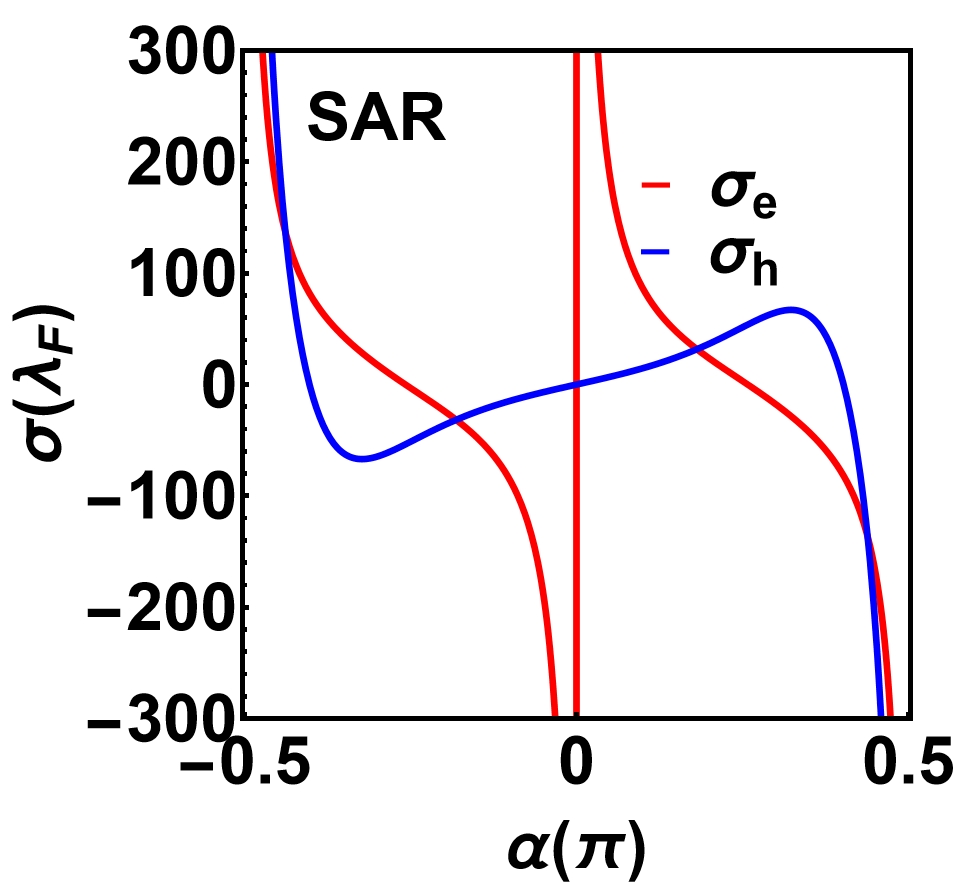}}\\
\subfloat[]{\includegraphics[width=.22\textwidth]{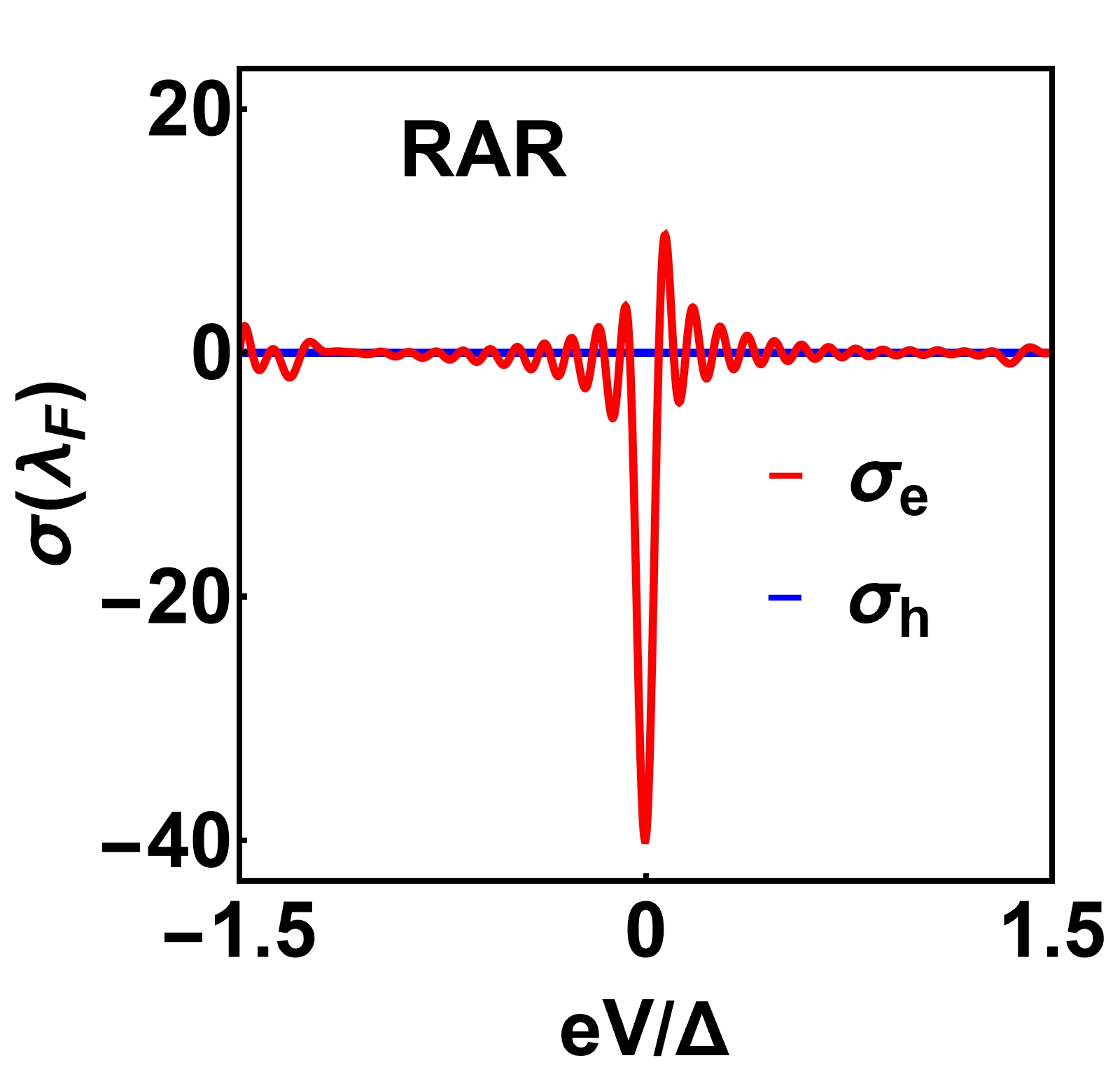}}
\subfloat[]{\includegraphics[width=.22\textwidth]{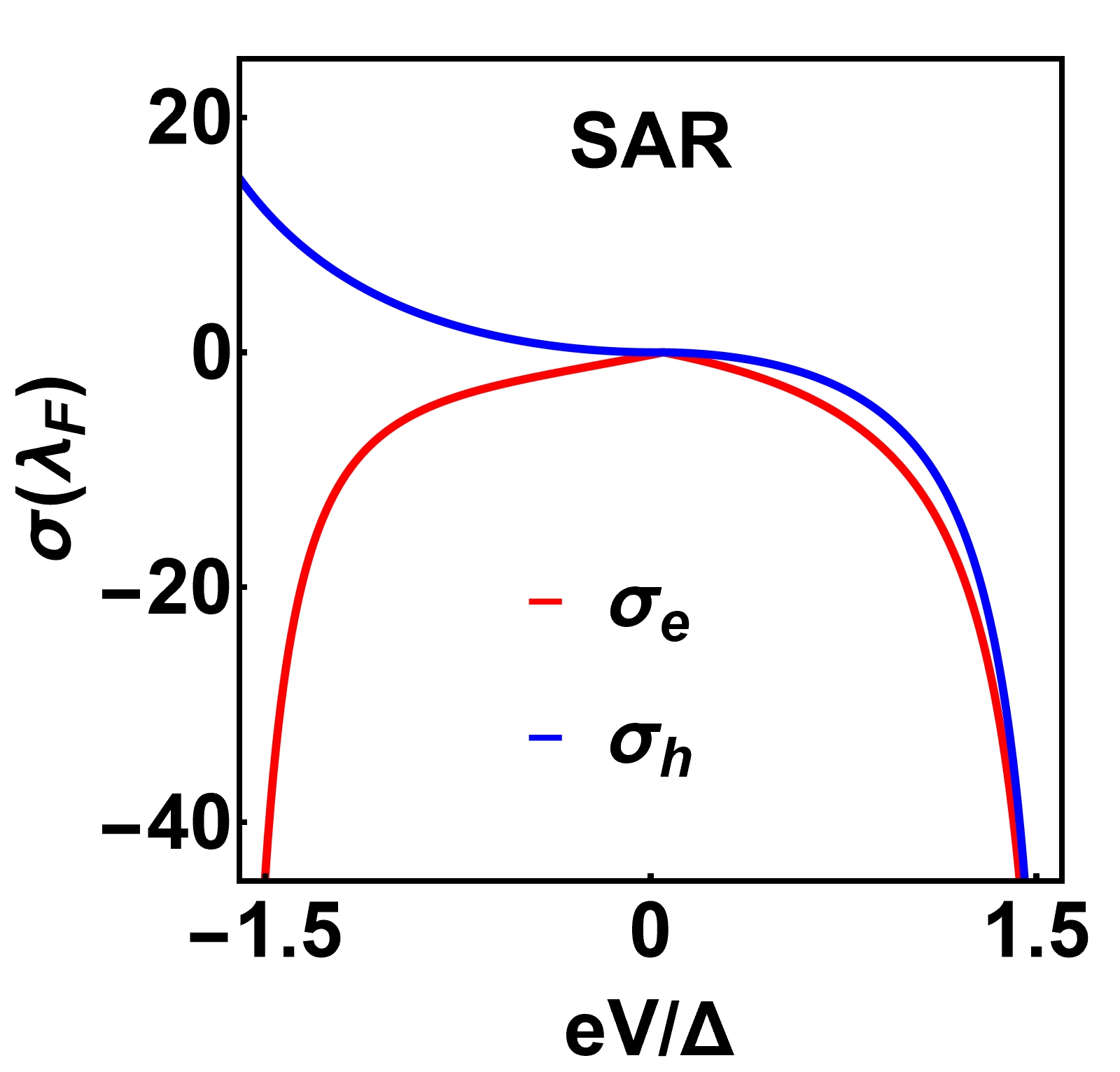}}\\
\subfloat[]{\includegraphics[width=.22\textwidth]{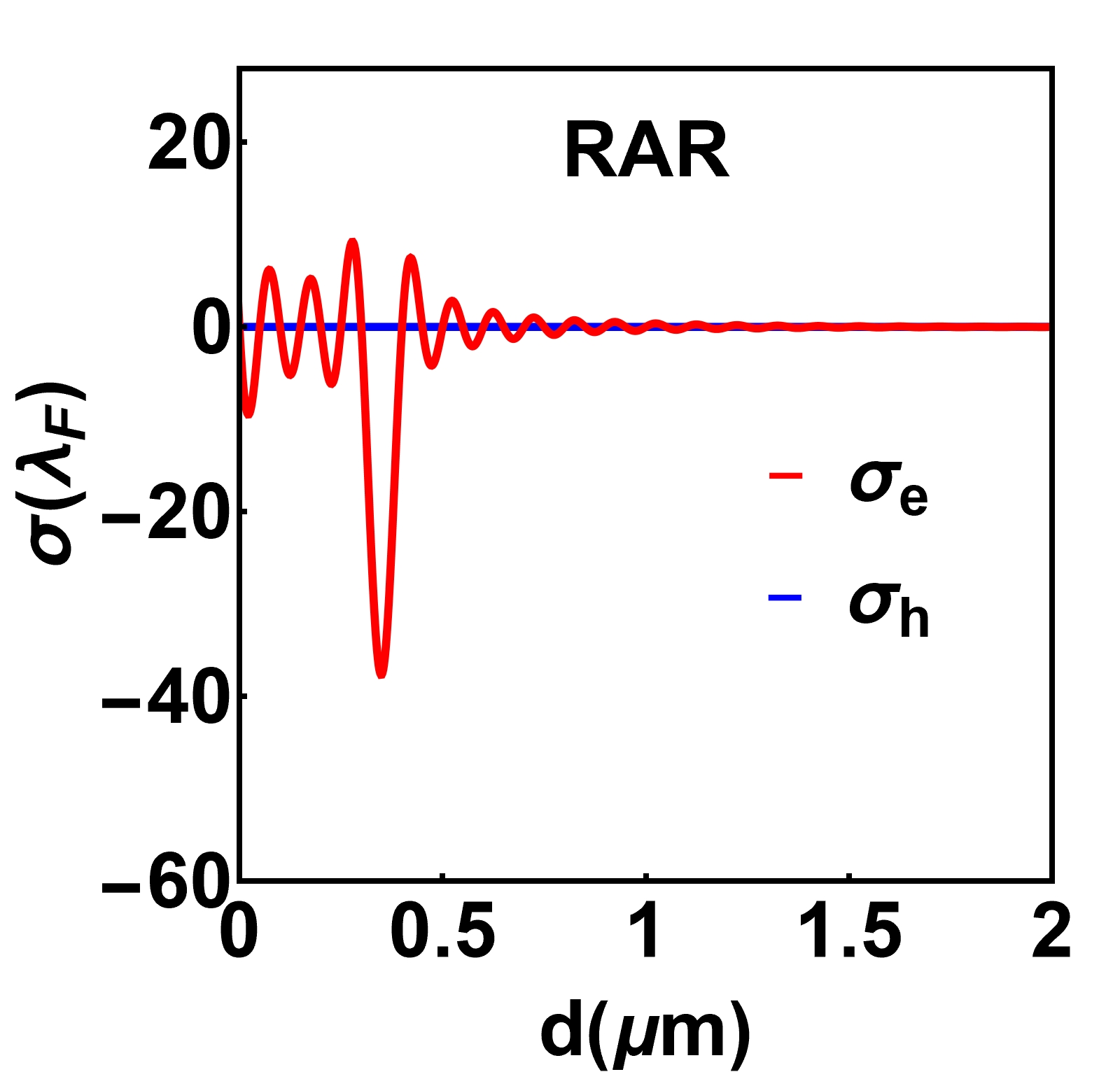}}
\subfloat[]{\includegraphics[width=.22\textwidth]{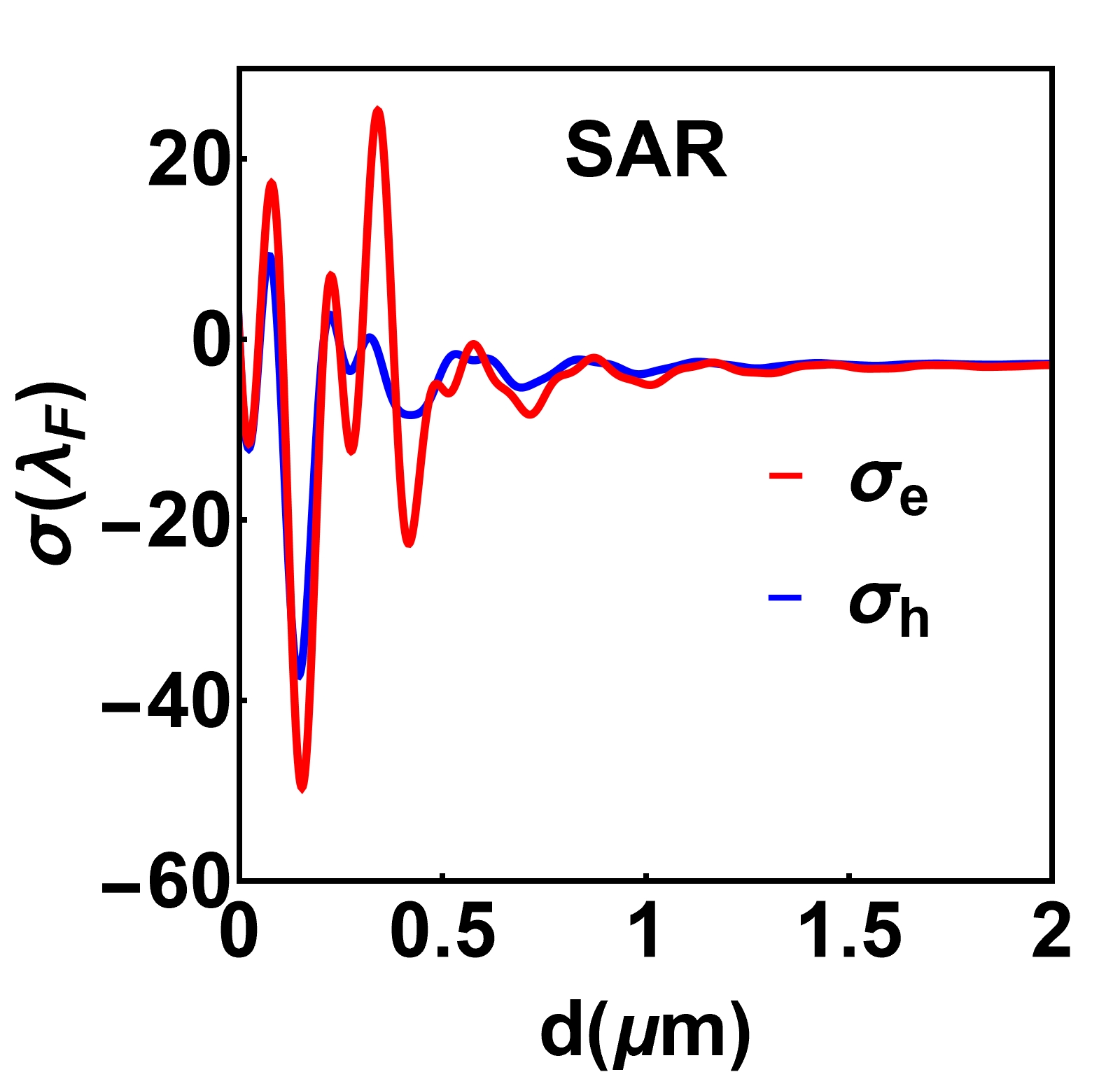}} 
\caption{\textit{(color online)} \color{black}{GH shift for symmetric processes - GH shift for the electron, $\sigma_e$ (\textit{red} curve) and the hole, $\sigma_h$ (\textit{blue} curve) for RAR (\textit{left} column) and SAR (\textit{right} column), (a-b) as a function of the incident angle, $\alpha$ ($\beta=0.1$ and $\Delta=1.2$ meV), (c-d) as a function of eV/$\Delta$ and (e-f) as a function of the width $d$ of the superconducting region. For (a-f), \textit{U}=200$\Delta$.}}\label{GHsym}
\end{figure}
\end{center}

\subsection{Goos-H\"anchen Shift} \label{GHshiftsec}
The transmission and reflection coefficients introduced in Eq.~(\ref{TR}) can be used to calculate
the Goos-H\"anchen shift at the GS interface using the following expression \cite{Beenakker2009, Sharma2011, Chen2013, Liu2018}
\begin{equation}
\sigma = -\delta'(q) = -\frac{\partial \delta}{\partial q} = -\text{Im }\left( \frac{\partial}{\partial q} \text{ln } r\right) \label{GHshift}
\end{equation}
where $\delta$ is the phase associated with the reflection amplitude $r$ that are respectively called $r_e$ for electron, and $r_h$ for hole and \textit{q} is the transverse wave vector which appears in the definition of $\alpha$ in Eq.~(\ref{param})(a,b). The corresponding expressions are given in Eq.~(\ref{reftranseq}) in Appendix \ref{append}. We find out the phase of $r_e$ and $r_h$ numerically and evaluate $\sigma_e$ for electron, $\sigma_h$ for hole using Eq.~(\ref{GHshift}).

\begin{figure*}
\subfloat[]{\includegraphics[width=.24\textwidth]{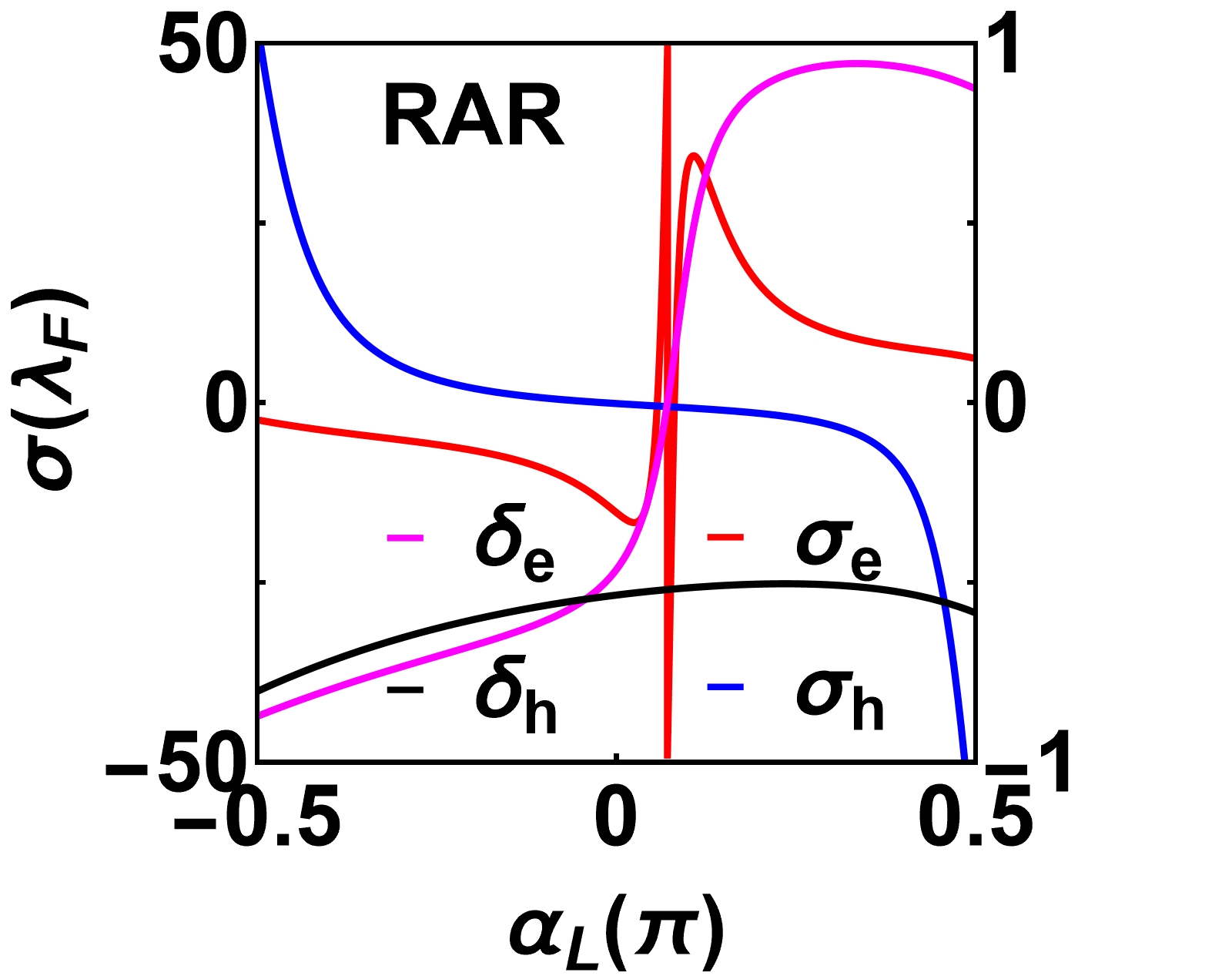}} 
\subfloat[]{\includegraphics[width=.24\textwidth]{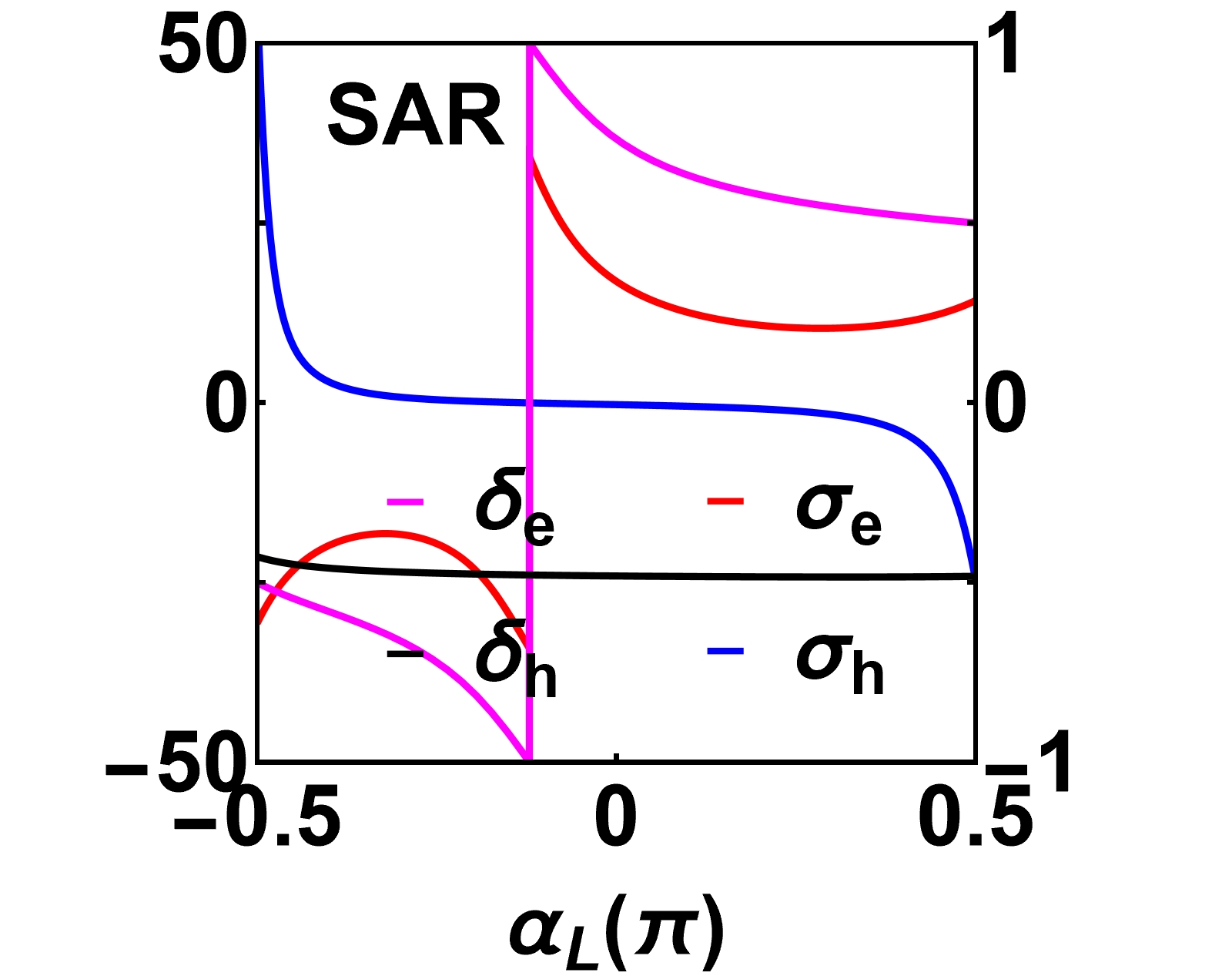}}
\subfloat[]{\includegraphics[width=.24\textwidth]{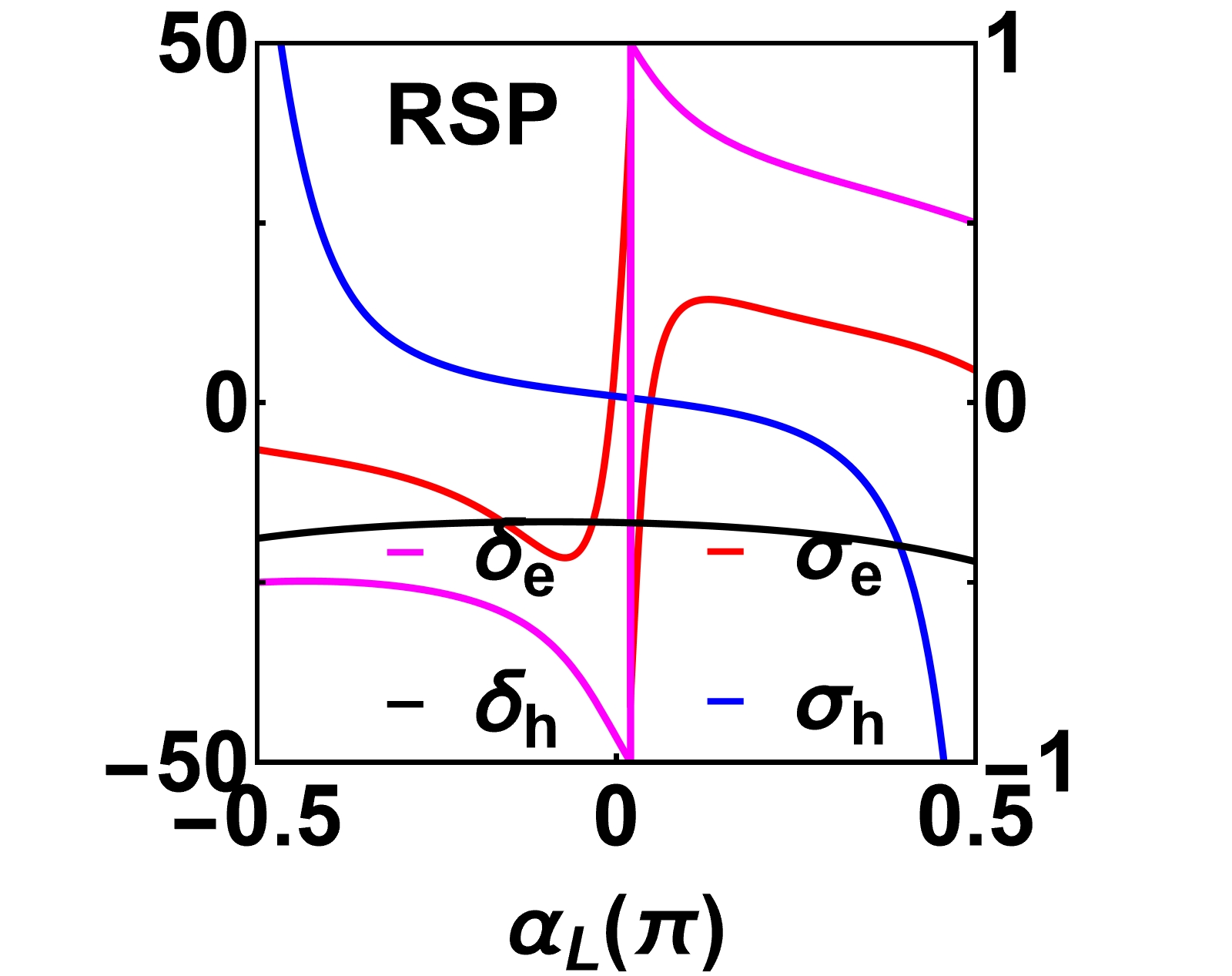}} 
\subfloat[]{\includegraphics[width=.24\textwidth]{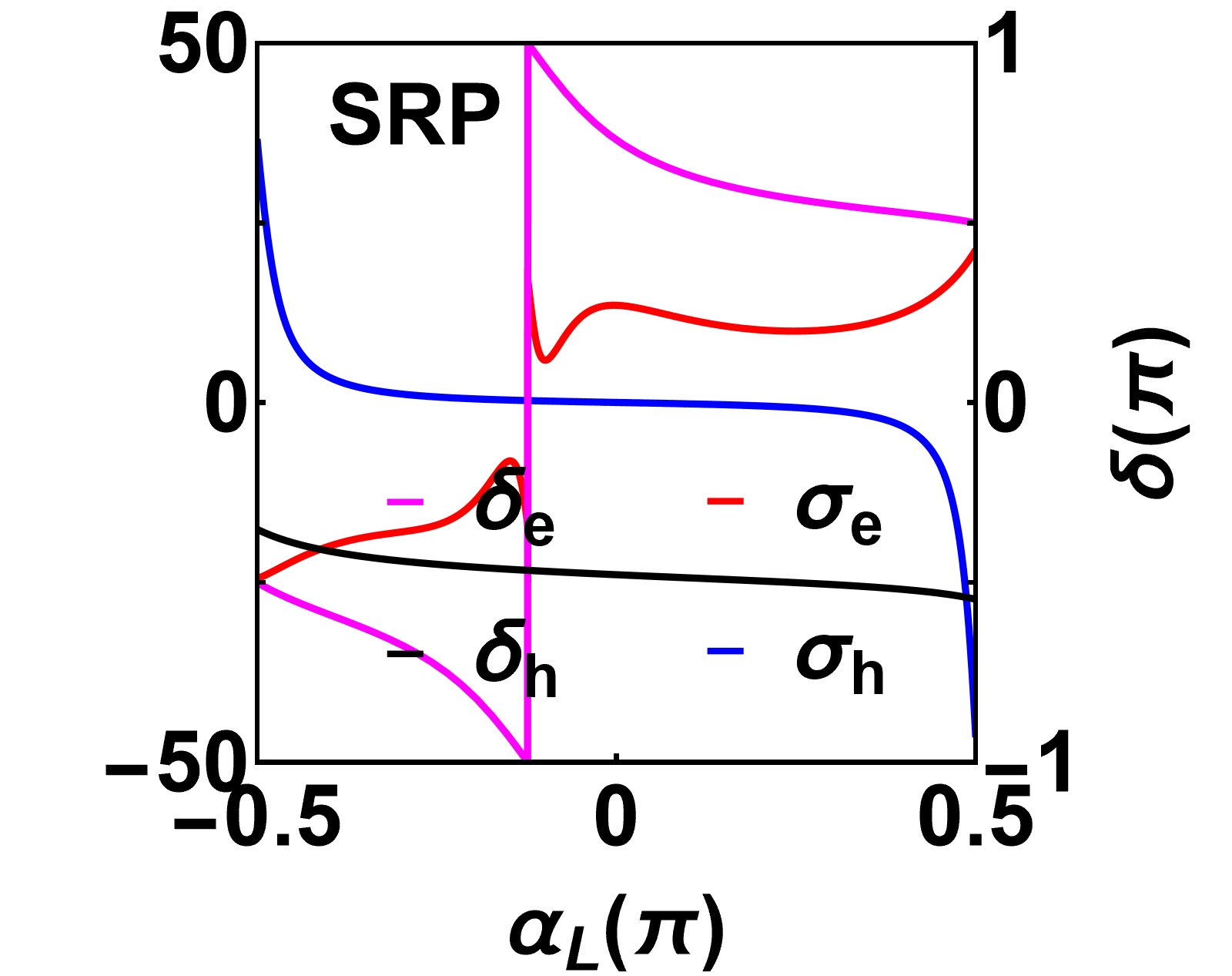}}\\
\subfloat[]{\includegraphics[width=.24\textwidth]{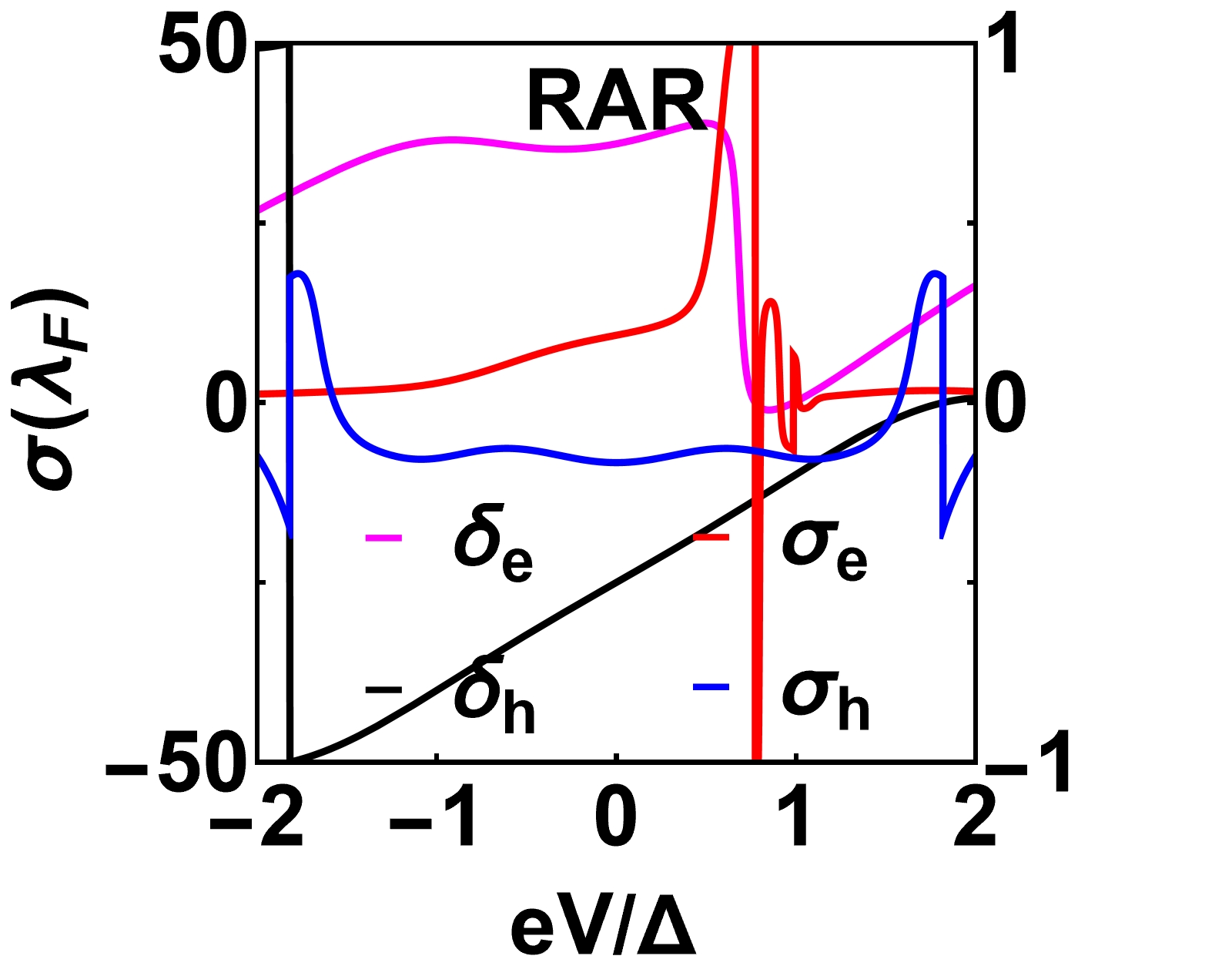}} 
\subfloat[]{\includegraphics[width=.24\textwidth]{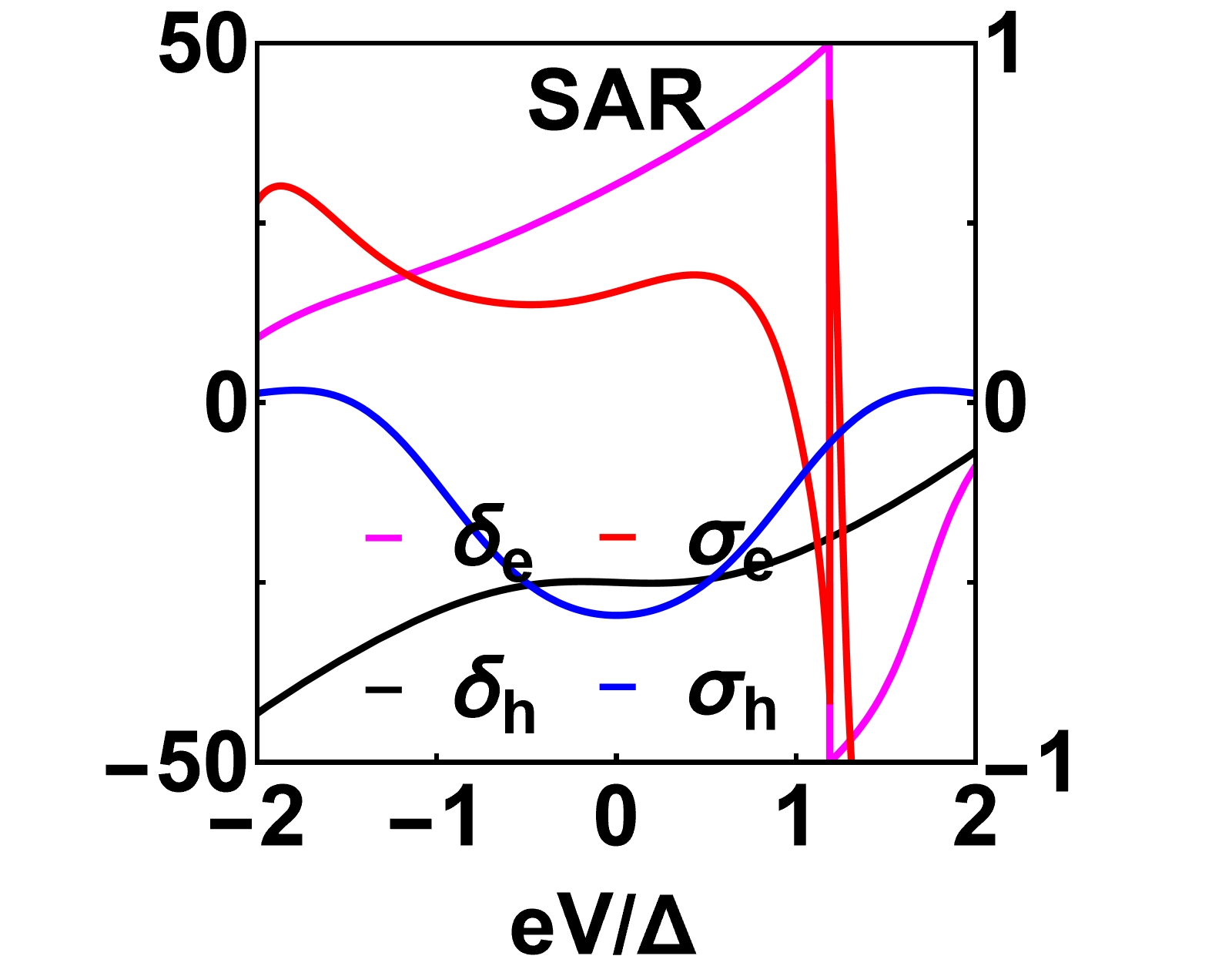}}
\subfloat[]{\includegraphics[width=.24\textwidth]{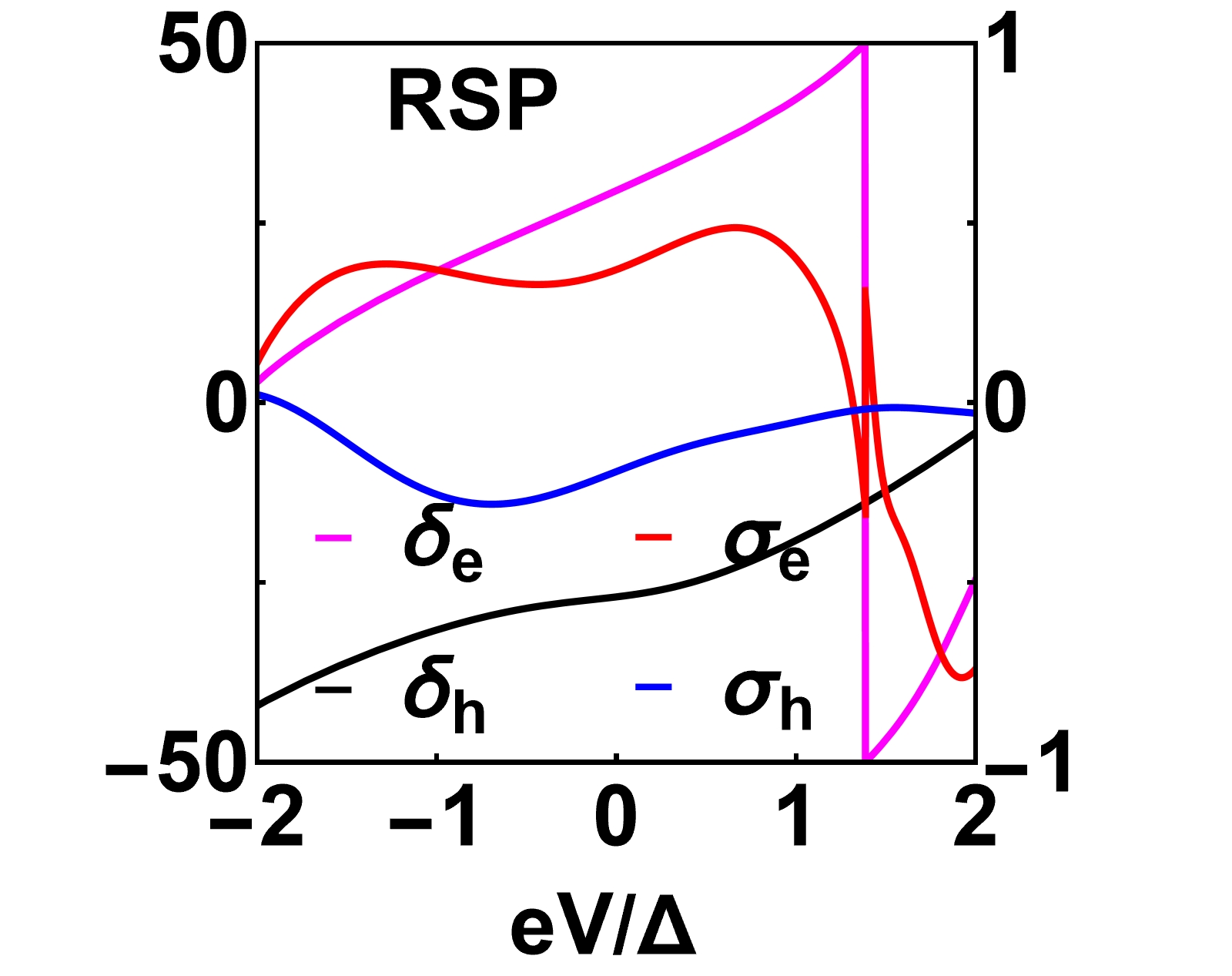}}
\subfloat[]{\includegraphics[width=.24\textwidth]{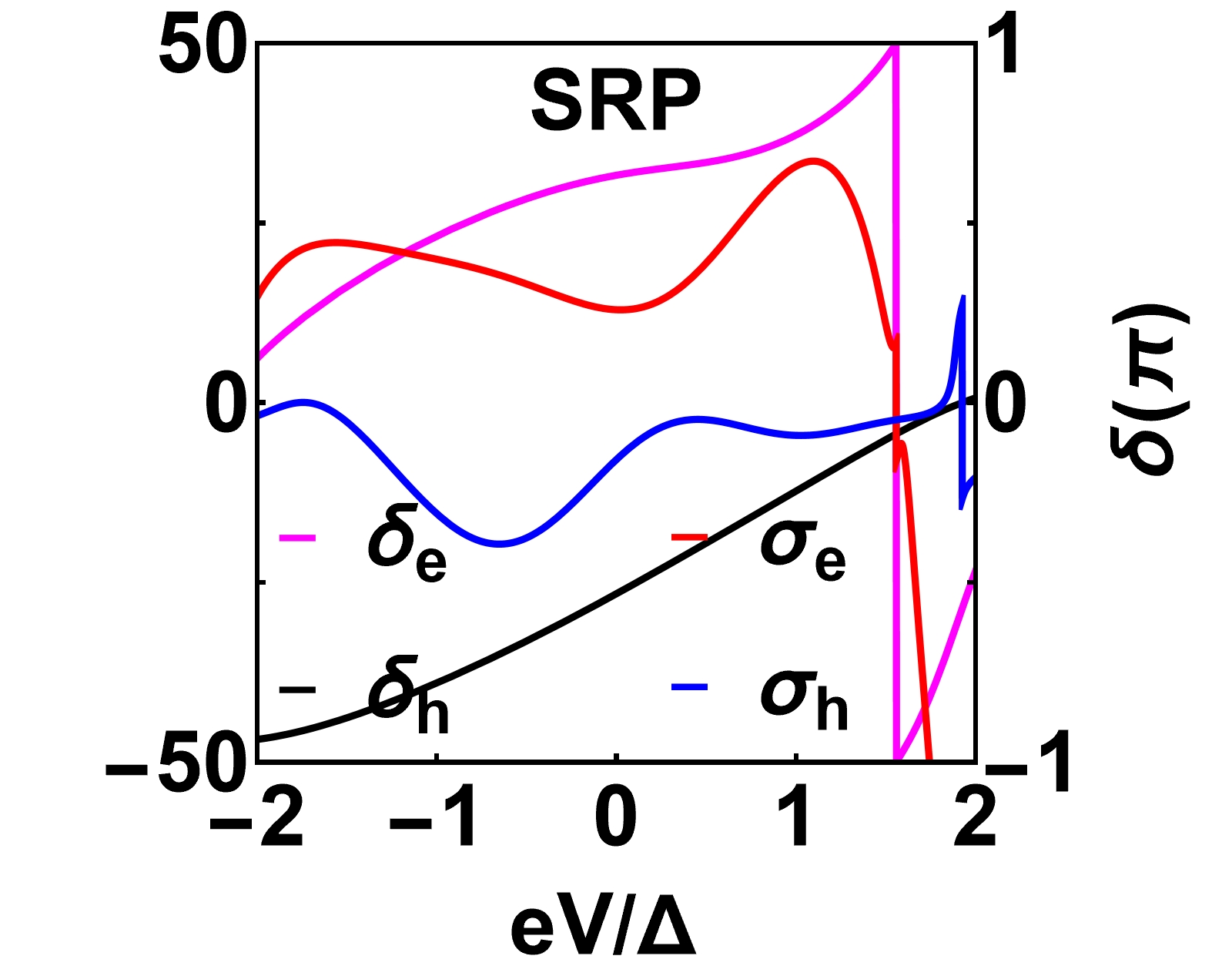}}  
\caption{\textit{(color online)} \color{black}{GH shift for asymmetric processes - (a-d) Variation of the GH shift, $\sigma$, and the reflection phases, $\delta$, as a function of $\alpha_L$ for a fixed value of $\alpha_R=\pi/6$. The \textit{red} curve denotes the GH shift for an electron, $\sigma_e$, while the \textit{blue} curve denotes the GH shift for a hole, $\sigma_h$. The \textit{magenta} curve denotes the reflection phase for an electron, $\delta_e$, while the \textit{black} curve denotes the reflection phase for a hole, $\delta_h$. (a) For RAR, $E_{F}=0.9\Delta$, $E = 0.4\Delta$, (b) For SAR, $E_{F}=0.01\Delta$, $E = 0.4\Delta$, (c) For RSP, $E_{F}=0.9\Delta$, $E = 0.8\Delta$, (d) For SRP, $E_{F}=0.05\Delta$, $E = 0.8\Delta$. (e-h) Variation of the GH shift, $\sigma$, and the reflection phases, $\delta$, as a function of $eV/\Delta$. The corresponding values of the Fermi energy, $E_F$, and incident particle energy, $E$, for RAR, SAR, RSP and SRP are the same as considered in (a-d). Other parameters used are  $\Delta=$1.2 meV, \textit{U}=200$\Delta$ and \textit{d}=200 nm. }}\label{ashift2}
\end{figure*}

In the optical case, the GH shift occurs at total internal reflection (TIR) and beyond \cite{Puri1986}. For the case of the GSG interface, the critical angle beyond which TIR takes place is given by Eq.~(\ref{param})(e) and remains the same for both the electron and the hole. The critical angle depends on the voltage $V$ through $E_{F}^{L,R}$ as given in Eq.~(\ref{param})(a,b). However, for the energy criteria considered for RAR and SAR as mentioned in Table \ref{tab:table1}, we see that we can get a finite GH shift for all possible values of the incident angle. Additionally, we see that at a certain value of the incident angle, the GH shift for the electron and the hole changes sign \cite{Beenakker2009, Chen2013, Liu2018}, as is observed in the case of symmetric (see Fig.~\ref{GHsym}), and asymmetric processes  (see Fig.~\ref{ashift2}). Such change of sign in GH shift was earlier observed in several other graphene-based structures\cite{Chen2011, Song2012, XiChen2013}.

\begin{figure*}[!t]
\subfloat[]{\includegraphics[width=.23\textwidth]{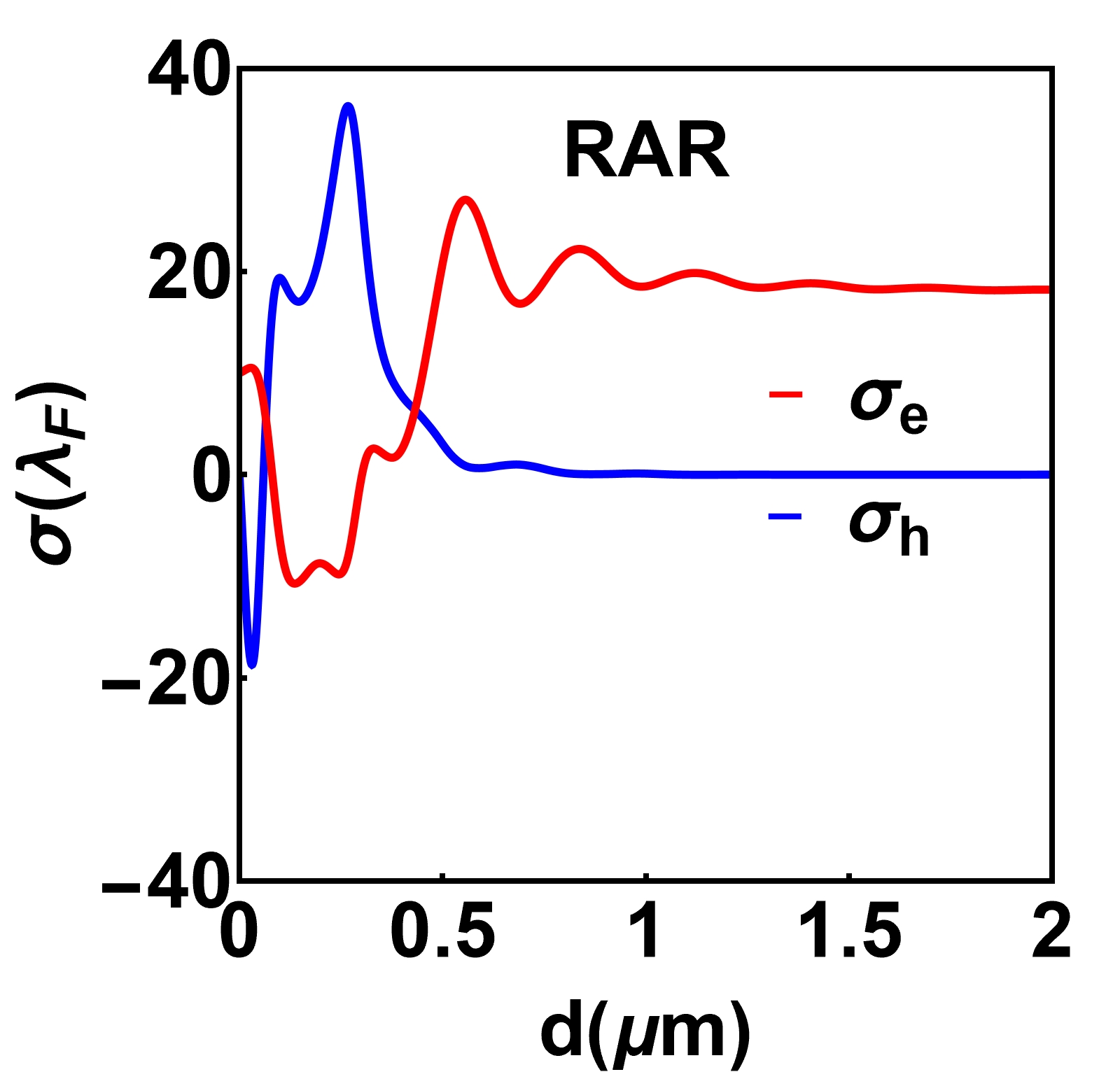}} 
\subfloat[]{\includegraphics[width=.23\textwidth]{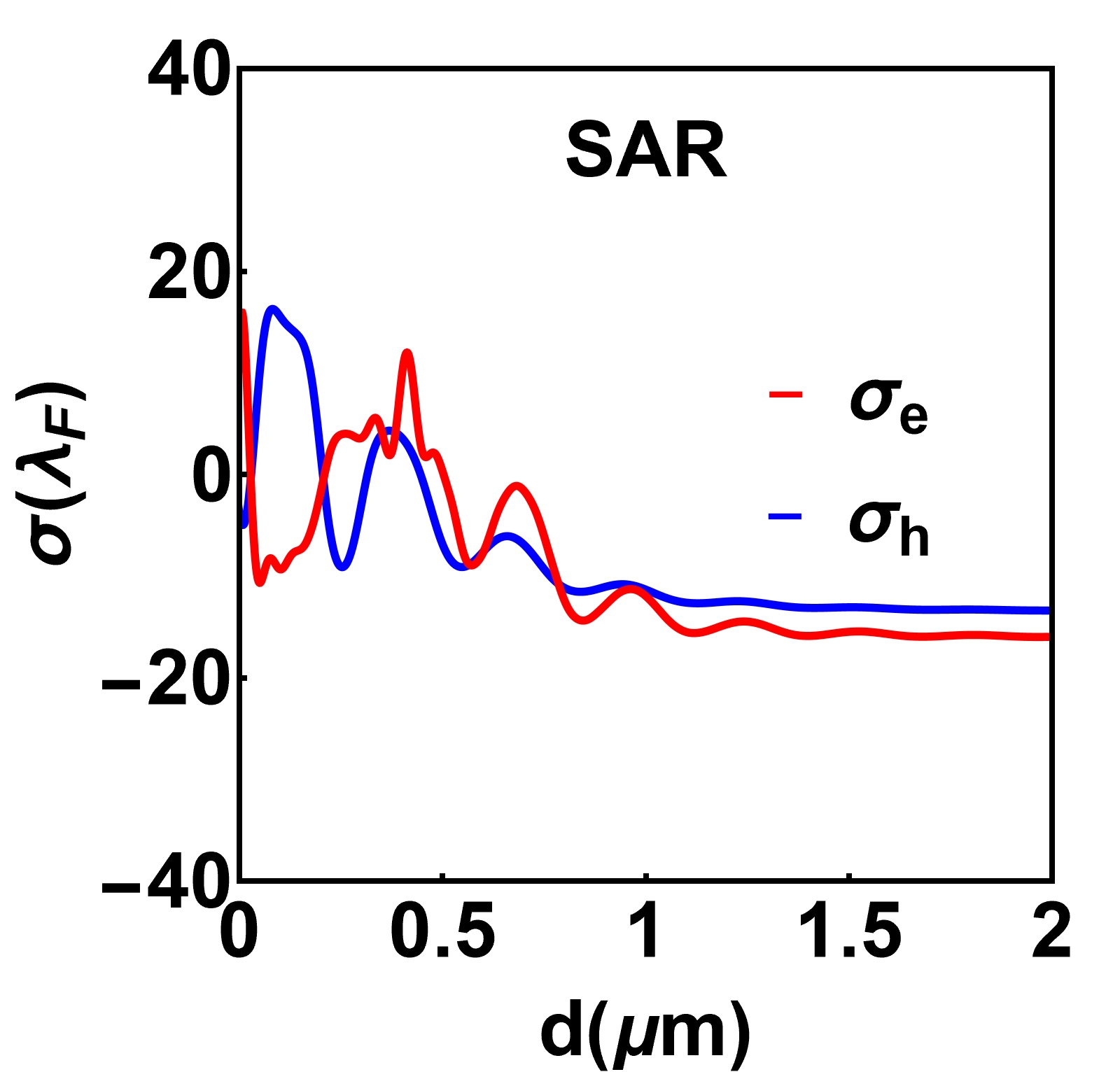}}
\subfloat[]{\includegraphics[width=.23\textwidth]{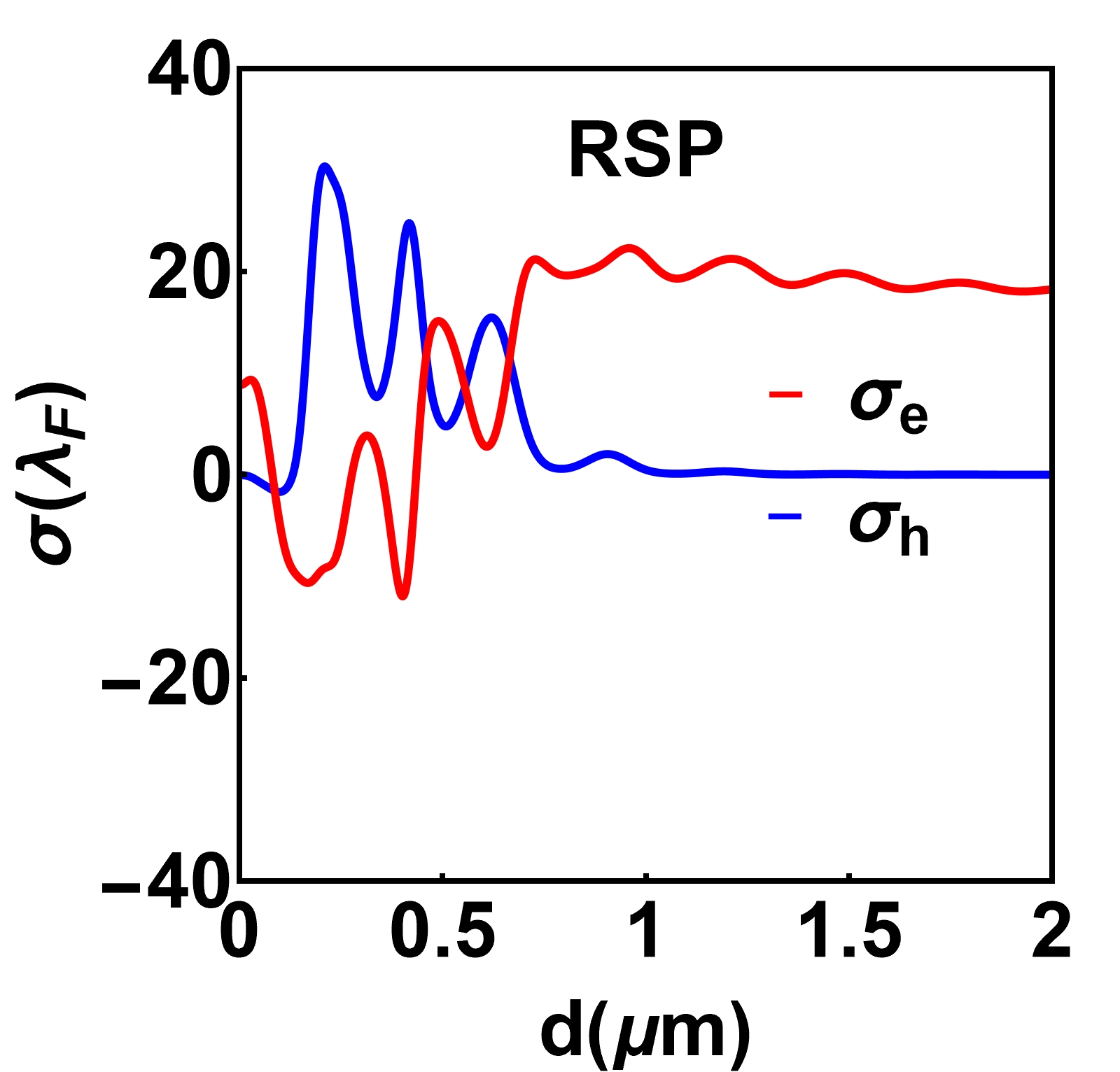}} 
\subfloat[]{\includegraphics[width=.23\textwidth]{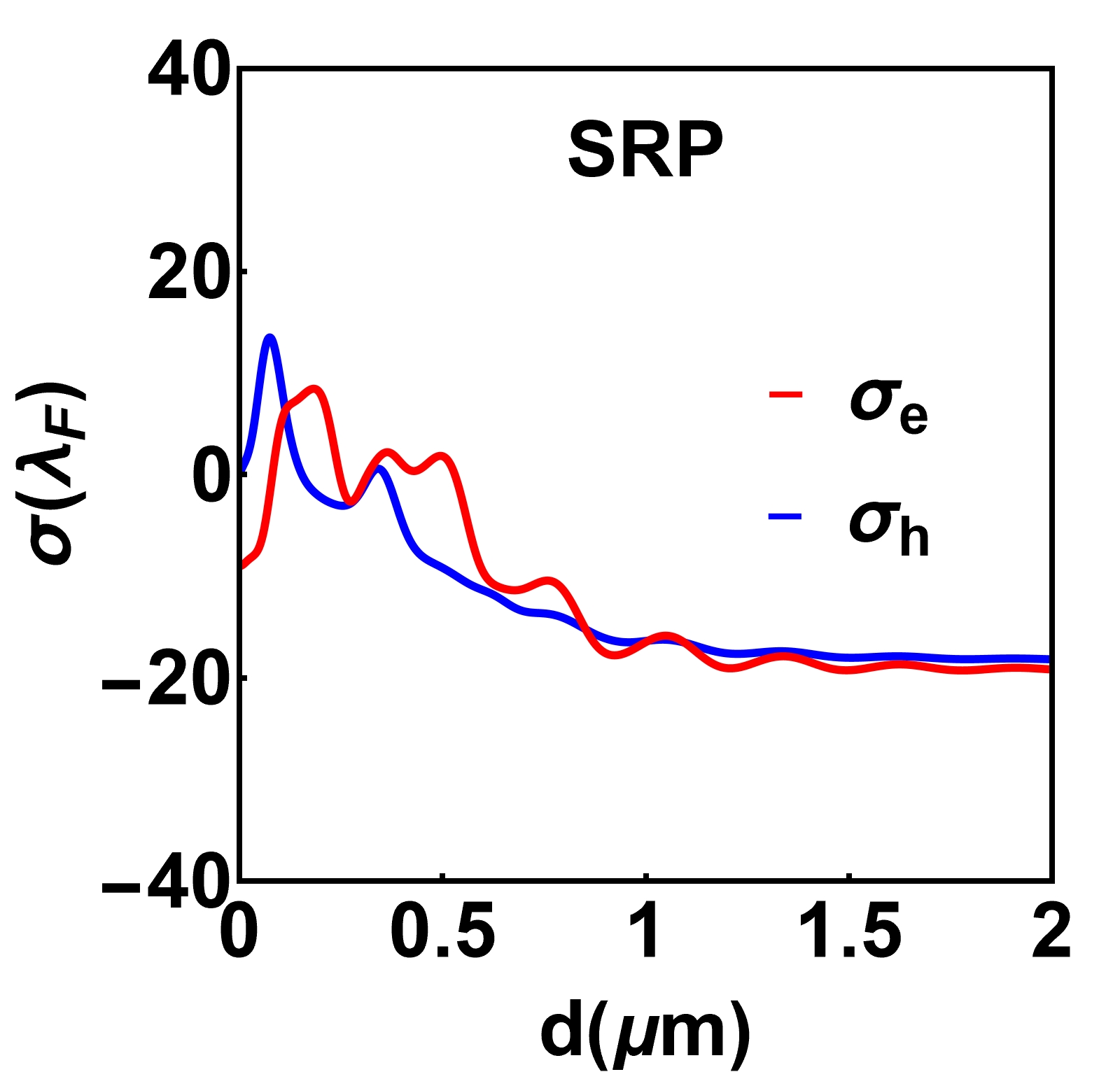}}
\caption{\textit{(color online)} \color{black}{GH shift for asymmetric processes - Variation of the GH shift, $\sigma$, as a function of the width \textit{d} of the superconducting region for (a) RAR, here $E_{F}=0.9\Delta$, $E = 0.4\Delta$, (b) SAR, here $E_{F}=0.01\Delta$, $E = 0.4\Delta$, (c) RSP, here $E_{F}=0.9\Delta$, $E = 0.8\Delta$, (d) SRP, here $E_{F}=0.05\Delta$, $E = 0.8\Delta$. The \textit{red} curves denote the GH shift for an electron, $\sigma_e$, while the \textit{blue} curves denote the GH shift for a hole, $\sigma_h$. The value of the superconducting gap, $\Delta$, is 1.2 meV. The potential $U$ in the superconducting region is 200$\Delta$ for (a-h). }}\label{ashift1}
\end{figure*}  
In Fig.~\ref{GHsym}(a,b), we plot the GH shift for the electron and the hole as a function of the incident angle, $\alpha$, defined in Eq.~(\ref{param})(a,b), for the symmetric RAR and SAR processes. The corresponding reflection coefficients are depicted in Fig.~\ref{norm}. We see that the GH shift changes signs for both the electron and the hole as depicted in  Fig.\ref{GHsym}(a,b). In Fig.~\ref{GHsym}(c,d), we plot the GH shift for the electron and the hole as a function of the parameter $eV/\Delta$. At $eV/\Delta$ close to zero, we see a sharp increase in the absolute value of the electron GH shift, $\sigma_e$, for the case of RAR. At this value, the critical angle for total internal reflection becomes $\frac{\pi}{2}$. In Fig.~\ref{GHsym}(e,f), we plot the GH shift as a function of the width \textit{d} of the superconducting region in a GSG junction both for RAR and SAR. We observe small oscillations in the GH shift with increasing $d$. These fluctuations or oscillations subside, leading to a stable behaviour with an almost vanishing $GH$ shift (within the accuracy of our numerical calculation) as $d$ is increased further.

After listing the features of GH-shift in a symmetric GSG junction, in Fig.~\ref{ashift1}, we plot the GH shift for the asymmetric junctions for the cases of  RAR, SAR, RSP and SRP as a function of the width \textit{d} of the superconducting region, both for the electron, $\sigma_e$ (\textit{red} curve), and the hole, $\sigma_h$ (\textit{blue} curve).
The variation of the reflection and transmission amplitudes for these four cases as a function of $\alpha_{L}$, and the corresponding dispersion  for these four cases were already depicted in Fig. \ref{norm2} and Fig. \ref{anorm}. 
We observe in Fig.~\ref{ashift1}(a-d) that both the electron GH-shift $\sigma_e$ (\textit{red} curve) and the hole GH-shift, $\sigma_h$ (\textit{blue} curve), show oscillatory behaviour as a function of $d$ for all the four processes for relatively narrower superconducting regions. These oscillations tend to subside with a further increase in the value of $d$, and the GH sift saturates to a given value.

To gain a better understanding of the GH-shift for asymmetric processes, in Fig.~\ref{ashift2}(a-d), we plot the corresponding $\sigma$ for both the electron, $\sigma_e$ (\textit{red} curve), and the hole, $\sigma_h$ (\textit{blue} curve) as a function of $\alpha_L$ for fixed $\alpha_R=\pi/6$ (defined in Eq.~(\ref{param})(a,b)). We observe that the electron GH shift shows an abrupt change at a value of $\alpha_{L}$ in the range $[-\frac{\pi}{2}, \frac{\pi}{2}]$ before showing a sign reversal. Also, the GH shift of the electron and the hole show diverging behaviour at $\alpha_{L}=\pm{\pi/2}$, angles corresponding to the grazing incidence.

In Fig.~\ref{ashift2}(e-h), we plot the GH shift $\sigma_e$ and $\sigma_h$, as a function of ($eV/\Delta$). For RAR in Fig.~\ref{ashift2}(e) and SAR in Fig.~\ref{ashift2}(f), $\sigma_{e}$ shows abrupt changes 
 at specific values of $(eV/\Delta)$ before showing a sign change. On the other hand, $\sigma_{H}$ shows abrupt changes at a given value of $(eV/\Delta)$ before showing sign change for RAR in Fig.~\ref{ashift2}(e), but varies smoothly in case of SAR as seen in Fig.~\ref{ashift2}(f).
The GH shifts $\sigma_{e,h}$ depicted in Fig.~\ref{ashift2}(g-h) for RSP and SRP, change  sign through a smooth variation.

Explanation of the detailed quantitative features of the variation of GH shift for the electron and hole as a function of different parameters as depicted in Fig.~\ref{ashift1}, and Fig.~\ref{ashift2} is difficult, as the expression of reflection and transmission coefficients given in Eq. (\ref{reftranseq}) in Appendix \ref{append} are very lengthy  and not amenable to a simple interpretation. However, some conclusions can be obtained by providing a comparison of the GH shift, $\sigma$, with the reflection phase, $\delta$, for the electron and the hole for asymmetric processes. This is also done in Fig.~\ref{ashift2}, by plotting it alongside the reflection phase and indicating the corresponding $y$-axis value on the right side of each plot. 

Particularly, in Fig.~\ref{ashift2}(a-h), we plot the corresponding reflection phases of the electron, denoted as $\delta_e$, and the reflection phase for the hole, denoted as $\delta_h$, for the asymmetric processes and so that the behaviour of the GH shift for every abrupt change in the reflection phase of the electron, $\delta_e$ (\textit{magenta} solid curve), and the hole, $\delta_h$ (\textit{black} solid curve) can be identified. The considered parameters and the energy constraints are provided in the figure caption. We see that whenever there is an abrupt change in the phase of the electron, the same is reflected in the GH shift leading to the formation of a spike. The reflection phase of the electron changes abruptly at certain values of the incident angle, $\alpha_L$, and the potential, $(eV/\Delta)$. Consequently, we see a spike in the GH shift at the same position, as expected from Eq. (\ref{GHshift}). The reflection phase for the hole on the other hand, varies smoothly, leading to a smooth variation of the GH shift for SAR and RSP, as can be verified from Fig.~\ref{ashift2}(a-d, f-g).  For $\sigma_{h}$ in the case of RAR and SRP, we again see a spike in Fig.~\ref{ashift2}(e,h) at a given value of $\frac{eV}{\Delta}$. This is again due to an abrupt change in the phase of the reflection amplitude at the same location. The sign reversals in $\sigma_{e,h}$, however, happen in all the cases.  We now look at the differential conductance of the junction.

\subsection{Conductance Oscillation}\label{conductsec}

In the preceding section \ref{RD},  we discussed in detail the GH shift of electrons and holes in GSG junctions in detail. To supplement the above discussion, we calculate another property that can characterise transport through such junctions, namely the conductivity of the GSG heterojunction within the  well-known BTK  framework\cite{Blonder1982}. To that purpose, we first obtain the reflection and the transmission coefficients for the electron(hole) from the transfer matrix and then substitute them into the expression to calculate the conductivity of the junction. The formula for the conductivity \cite{Blonder1982,Beenakker2006}  of the junction appears as

\begin{widetext}
\begin{equation}\label{acond}
\frac{\partial I}{\partial V} = g_{0}(V) \int_{0} ^{\frac{\pi}{2}}(1 - |r(eV,\alpha_{L},\beta)|^2 +|r_{A}(eV,\alpha_{L},\beta)|^2)\cos\alpha_{L} d\alpha_{L}
\end{equation}   
\end{widetext}
where $g_{0}(V)$ is defined as
\begin{equation*}
 g_{0}(V) =  \frac{4 e^2}{h}N(eV);~~N(eV) =\frac{(E_{F}+E)d}{\pi\hbar V}.
\end{equation*}
$r(eV,\alpha_{L})$ and $r_{A}(eV,\alpha_{L})$ respectively denote the reflectances due to normal and Andreev reflection. 

\begin{center}
\begin{figure}
\subfloat[]{\includegraphics[width=.22\textwidth]{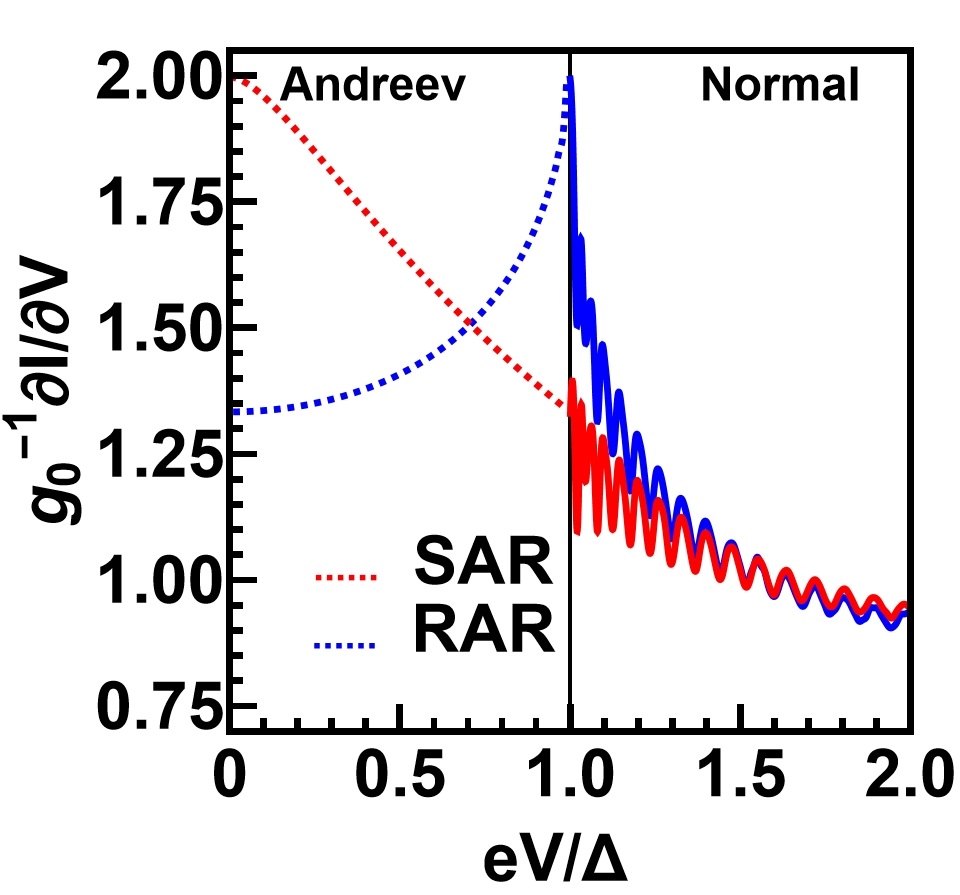}} 
\subfloat[]{\includegraphics[width=.22\textwidth]{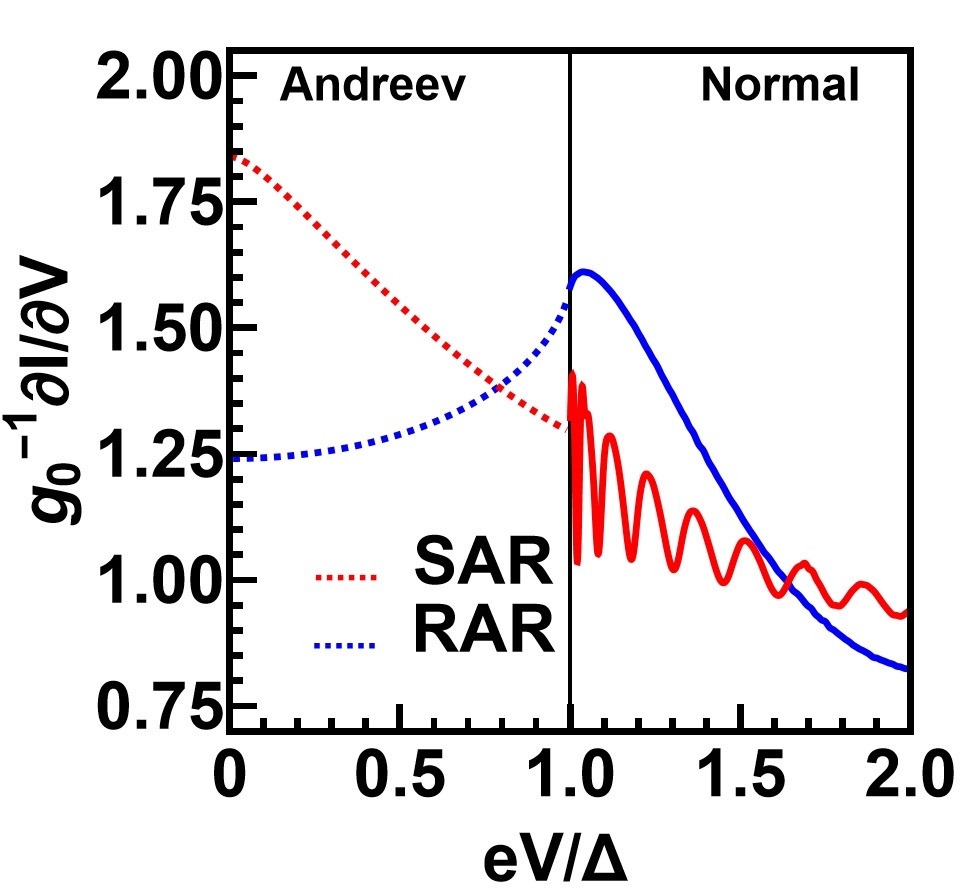}}
\caption{\textit{(color online)} \color{black}{Variation of the normalised conductance as a function of $(eV/\Delta)$ for the symmetric RAR and SAR processes for (a) \textit{U}=200$\Delta$ and (b) \textit{U}=50$\Delta$. The \textit{black} solid line at 1 demarcates the Andreev and the normal regions. The \textit{red} dotted(solid) curves correspond to RAR, while the \textit{blue} dotted(solid) curves correspond to SAR in the Andreev(normal) region.}}\label{cnd}
\end{figure}
\end{center}

We plot the normalised conductance as a function of $\frac{eV}{\Delta}$ for symmetric RAR (\textit{blue} curve) and SAR (\textit{red} curve) in Fig.~ \ref{cnd}(a) for \textit{U}=200$\Delta$ and in Fig.~\ref{cnd}(b) for \textit{U}=50$\Delta$. Depending on the ratio of the energy to the superconducting gap, we divide the $\frac{eV}{\Delta}$ axis into the Andreev and the normal part. The solid vertical line at 1 serves as the boundary between the two regions. On the left side of this vertical line, we only have the quasi-particle states in the gapped region that are contributing to the conductance as this region supports Andreev processes (dotted curves for SAR and RAR). Free particle states contribute to the conductance on the right side of this vertical line, as normal processes (solid curves for SAR and RAR) are supported. We do not see any oscillations in the Andreev region, but in the normal region, we see oscillations in the conductance for both RAR and SAR.

\begin{figure}[!t]
\subfloat[]{\includegraphics[width=.22\textwidth]{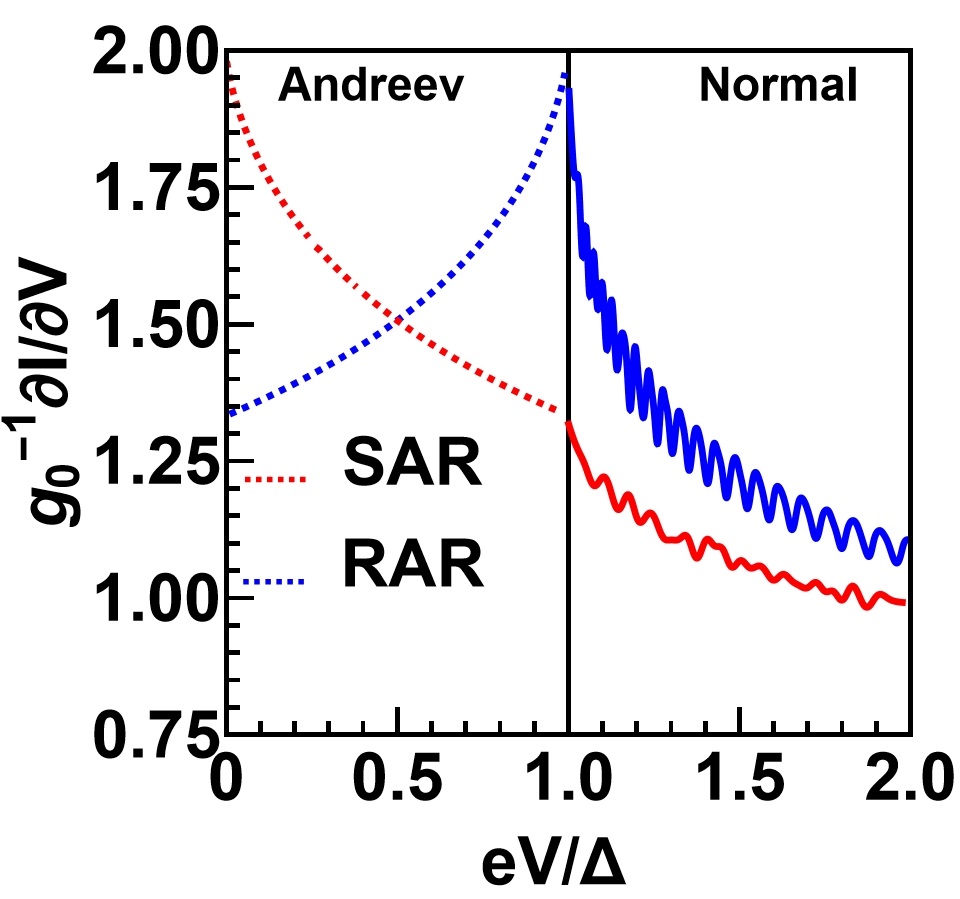}} 
\subfloat[]{\includegraphics[width=.22\textwidth]{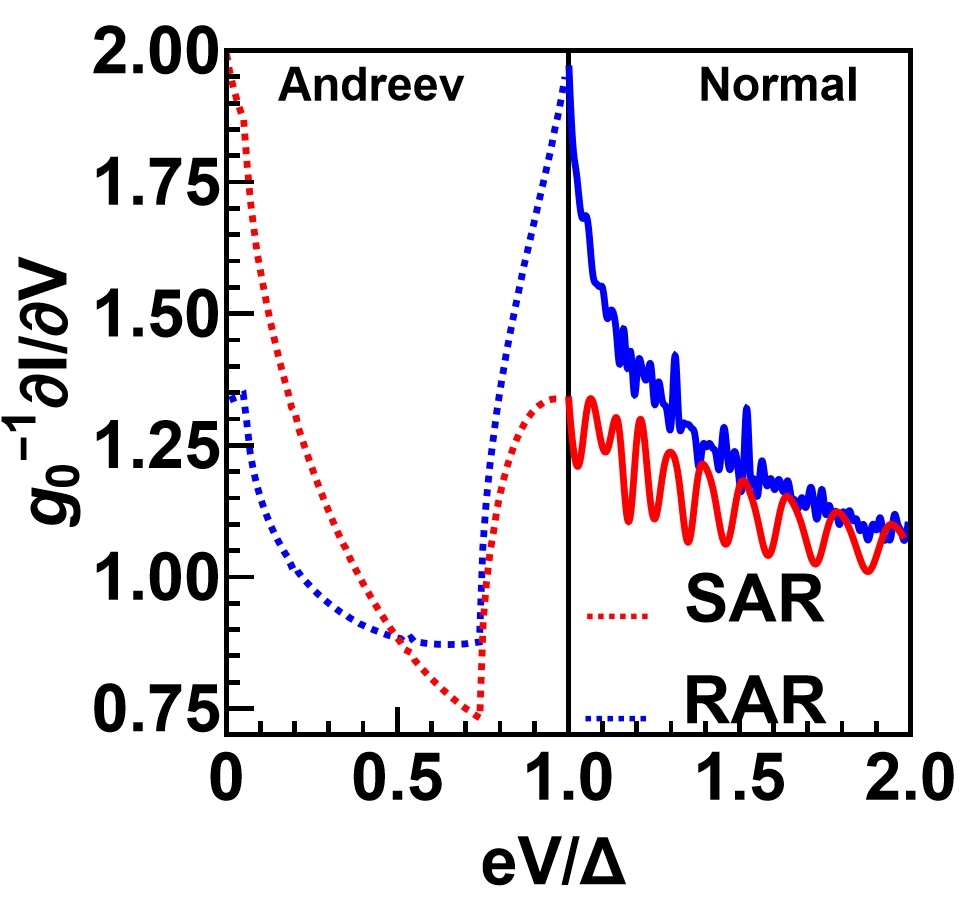}}\\
\subfloat[]{\includegraphics[width=.22\textwidth]{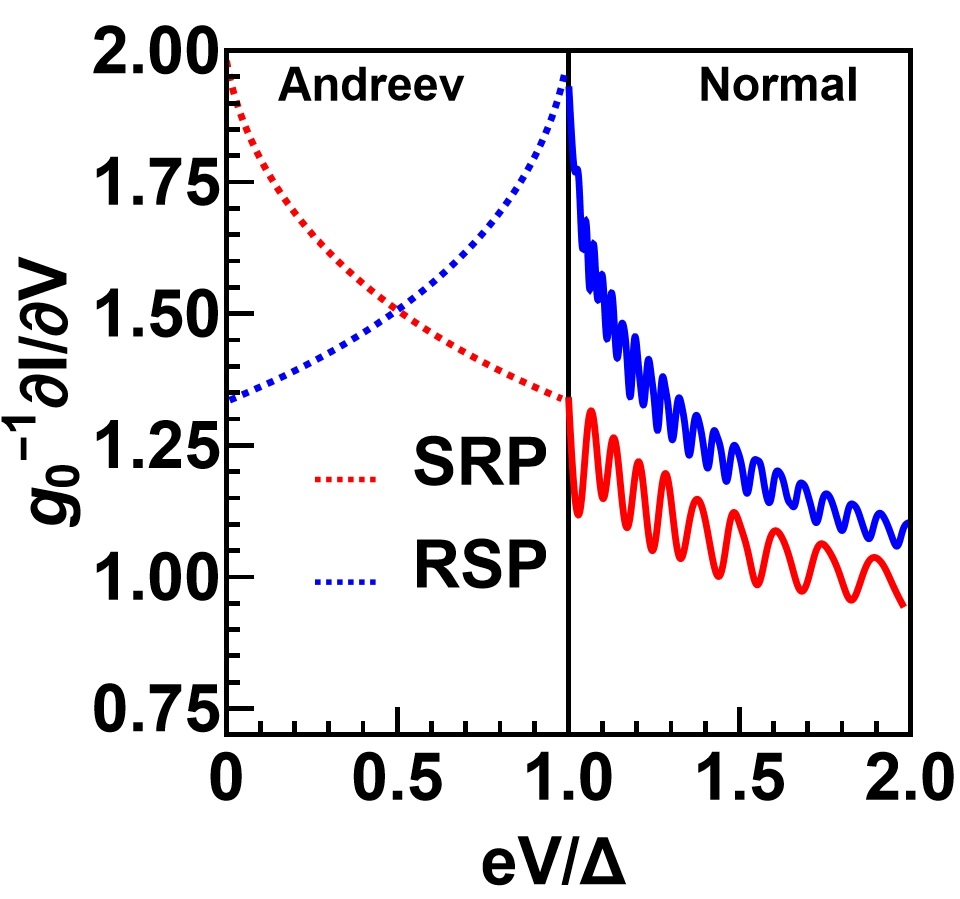}} 
\subfloat[]{\includegraphics[width=.22\textwidth]{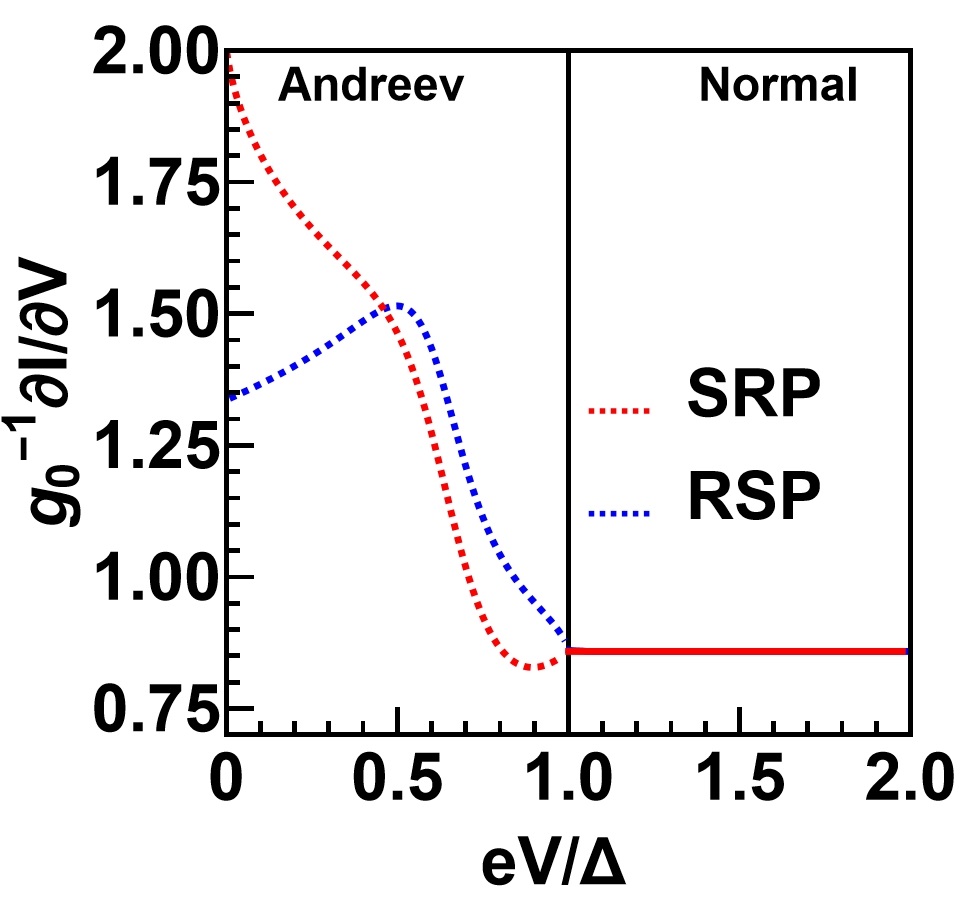}}
\caption{\textit{(color online)} Variation of the normalised conductance as a function of $(eV/\Delta)$ for the asymmetric RAR and SAR for (a) \textit{U}=200$\Delta$ and (b) \textit{U}=50$\Delta$, and for the asymmetric RSP and SRP for (c) \textit{U}=200$\Delta$ and (d) \textit{U}=50$\Delta$. The \textit{black} solid line at 1 demarcates the Andreev and the normal regions. The \textit{red} dotted(solid) curves correspond to RAR and RSP, while the \textit{blue} dotted(solid) curves correspond to SAR and SRP in the Andreev(normal) region.}\label{acnd}
\end{figure}

In the subsequent Fig.~\ref{acnd}, we plot the normalised conductance as a function of $\frac{eV}{\Delta}$ for the asymmetric processes RAR (\textit{blue} curve) and SAR (\textit{blue} curve) in Fig.~\ref{acnd}(a) for \textit{U}=200$\Delta$ and in Fig.~\ref{acnd}(b) for \textit{U}=50$\Delta$. Similarly, we show the variation of the normalized conductance for the asymmetric processes RSP (\textit{blue} curve) and SRP (\textit{blue} curve) in Fig.~\ref{acnd}(c) for \textit{U}=200$\Delta$ and in Fig.~\ref{acnd}(d) for \textit{U}=50$\Delta$. We again observe oscillatory behaviour in the conductance in the normal region preceded by a monotonic behaviour in the Andreev region as a function of $\frac{eV}{\Delta}$ in Fig.~\ref{acnd}(a). However, in Fig.~\ref{acnd}(b)
for $\textit{U}=50 \Delta$, that corresponds to a smaller potential in the superconducting regime as compared to the Fig.~\ref{acnd}(a)
the behaviour in the Andreev regime is non-monotonic and followed by the oscillatory behaviour of the corresponding plots in the normal region, which also partially overlap. In Fig.~\ref{acnd}(d) for $\textit{U}=50 \Delta$, in the normal regime, the normalised conductance for the SRP and RSP process coincides with the numerical accuracy of our calculation. This shows that the behaviour of the conductance is strongly dependent on $U$.

\section{Conclusion}\label{concludes}
In conclusion, we have studied the transport properties of a GSG junction in the ballistic regime in detail using a transfer matrix based approach. 
We have considered the effect of Andreev and normal reflection at the GS interface in such junctions, characterised 
the transport under various energy ranges by explicitly evaluating the reflectance and transmittance through such junctions both for electron and hole. We have particularly observed rich features such as the electronic analogue of GH shift in the junction for electrons as well as holes, and tried to correlate them with the corresponding features in reflection amplitude. Since the GH shift and the differential conductance, both are dependent on the reflection amplitude in such GSG junction, it is natural to ask a question whether there is any inter-relation between therm? However, given the complex nature of the expression for the transmittance and reflectance given in \ref{reftranseq} in appendix \ref{append} we are unable to provide any direct relation between these two quantities at this stage. Finally, we evaluated the normalised differential conductivity through such junction as a function of the bias voltage between the left and right terminal of junction. We hope our detailed analysis will augment further theoretical and and experimental studies to characterise electron transport through such junctions.

\appendix
\section{} \label{append}
\begin{widetext}
In this appendix, we provide the forms of the various matrices, $M$, and column vectors $G$ and $S$ used in the text.
\begin{align}\label{Mmat}
\begin{gathered}
M_{1}(x=0) = \begin{bmatrix}
\frac{e^{-i\frac{\alpha}{2}^{L}}}{\sqrt{\cos\alpha^{L}}} & \frac{e^{i\frac{\alpha^L}{2}}}{\sqrt{\cos\alpha^{L}}} & 0  & 0\\
\frac{e^{i\frac{\alpha}{2}^{L}}}{\sqrt{\cos\alpha^{L}}} & \frac{e^{-i\frac{\alpha}{2}^{L}}}{\sqrt{\cos\alpha^{L}}} & 0 & 0\\
0 & 0 & \frac{e^{-i\frac{\alpha'}{2}^{L}}}{\sqrt{\cos\alpha'^{L}}} &\frac{e^{i\frac{\alpha'}{2}^{L}}}{\sqrt{\cos\alpha'}}\\
0 & 0 & \frac{e^{i\frac{\alpha'}{2}^{L}}}{\sqrt{\cos\alpha'^{L}}} & \frac{e^{-i\frac{\alpha'}{2}^{L}}}{\sqrt{\cos\alpha'^{L}}}
\end{bmatrix} \\
M_{2}(x=0) = \begin{bmatrix}
e^{i\beta} & e^{ -i\beta} & e^{-i\beta} & e^{-i\beta}\\
e^{i\beta+i\gamma} & -e^{i\beta-i\gamma} & -e^{-i\beta-i\gamma} & e^{-i\beta + i\gamma}\\1 & 1 & 1 & 1\\e^{i\gamma} & -e^{-i\gamma} & -e^{-i\gamma} & e^{i\gamma}
\end{bmatrix} \\
M_{3}(x=d) = \begin{bmatrix}
e^{ik_{o}d-\kappa d+i\beta} & e^{-ik_{o}d+\kappa d -i\beta} & e^{-ik_{o}d-\kappa d-i\beta} & e^{ik_{o}d+\kappa d-i\beta}\\
e^{ik_{o}d-\kappa d+i\beta+i\gamma} & -e^{-ik_{o}d+\kappa d+i\beta-i\gamma} & -e^{-ik_{o}d-\kappa d-i\beta-i\gamma} & e^{ik_{o}d + \kappa d-i\beta + i\gamma}\\e^{ik_{o}d-\kappa d} & e^{-ik_{o}d+\kappa d} & e^{-ik_{o}d-\kappa d} & e^{ik_{o}d+\kappa d}\\e^{ik_{o}d-\kappa d+i\gamma} & -e^{-ik_{o}d+\kappa d-i\gamma} & -e^{-ik_{o}d-\kappa d-i\gamma} & e^{ik_{o}d+\kappa d+i\gamma}
\end{bmatrix} \\
M_{4}(x=d) = \begin{bmatrix}
\frac{e^{ik_{e}^{R}d}e^{-i\frac{\alpha}{2}^{R}}}{\sqrt{\cos\alpha^{R}}} & \frac{e^{-ik_{e}^{R}d}e^{i\frac{\alpha}{2}^{R}}}{\sqrt{\cos\alpha^{R}}} & 0  & 0\\
\frac{e^{ik_{e}^{R}d}e^{i\frac{\alpha}{2}^{R}}}{\sqrt{\cos\alpha^{R}}} & \frac{e^{-ik_{e}^{R}d}e^{-i\frac{\alpha}{2}^{R}}}{\sqrt{\cos\alpha^{R}}} & 0 & 0\\
0 & 0 & \frac{e^{ik_{h}^{R}d}e^{-i\frac{\alpha'}{2}^{R}}}{\sqrt{\cos\alpha'^{R}}} &\frac{e^{-ik_{h}^{R}d}e^{i\frac{\alpha'}{2}^{R}}}{\sqrt{\cos\alpha'^{R}}}\\
0 & 0 & \frac{e^{ik_{h}^{R}d}e^{i\frac{\alpha'}{2}^{R}}}{\sqrt{\cos\alpha'^{R}}} & \frac{e^{-ik_{h}^{R}d}e^{-i\frac{\alpha'}{2}^{R}}}{\sqrt{\cos\alpha'^{R}}}
\end{bmatrix}
\end{gathered}
\end{align}
The definitions of the G matrices are
\begin{align}\label{Gmat}
G(x=0)=\begin{bmatrix}
1\\r{e}\\r_{h}\\0
\end{bmatrix},~~
G(x=d)=\begin{bmatrix}
t{e}\\0\\0\\t{h}
\end{bmatrix}
\end{align}
The definitions of the various coefficients denoted by $a_{ij}$'s in Eq.~(\ref{aijs}) are as follows;
\begin{align}
  \label{acof1}
     a_{11} &= \frac{\sqrt{\cos \left(\alpha ^R\right)}\exp \left(-\frac{1}{2} i \left(2 d \left(k_e-i \kappa +k_0\right)+\alpha ^L-\alpha ^R\right)\right)}{2 \left(-1+e^{2 i \beta }\right) \left(1+e^{2 i \alpha ^R}\right) \sqrt{\cos \left(\alpha ^L\right)}}-e^{2 d \left(\kappa +i k_0\right)} \nonumber\\
     &-e^{2 d \left(\kappa +i k_0\right)+i \alpha ^L}+e^{2 i \beta } \left(e^{2 d \kappa }+e^{i \left(2 d k_0+\alpha ^L\right)}+e^{i \alpha ^R} \left(-e^{2 d \kappa }+e^{i \alpha ^L} \left(e^{2 d \kappa }+e^{2 i d k_0}\right)+e^{2 i d k_0}\right)+e^{2 i d k_0}-e^{2 d \kappa +i \alpha ^L}\right)\nonumber\\
     &-e^{i \alpha ^R} \left(e^{2 d \left(\kappa +i k_0\right)}+e^{i \alpha ^L} \left(1+e^{2 d \left(\kappa +i k_0\right)}\right)-1\right)+e^{i \alpha ^L}-1\nonumber\\
     a_{14} &= \frac{\csc (\beta ) e^{-i d k_{e}} \sinh (d \kappa ) \left(\sin \left(d k_0\right) \cos \left(\frac{1}{2} \left(\alpha '^L-\alpha ^R\right)\right)-i \cos \left(d k_0\right) \cos \left(\frac{1}{2} \left(\alpha '^L+\alpha ^R\right)\right)\right)}{\sqrt{\cos \left(\alpha ^R\right)} \sqrt{\cos \left(\alpha '^L\right)}}\\
     a_{21} &= \frac{\exp \left(-\frac{1}{2} i \left(2 d \left(-k_e-i \kappa +k_0\right)+\alpha ^L-\alpha ^R\right)\right)\sqrt{\cos \left(\alpha ^R\right)}}{2 \left(-1+e^{2 i \beta }\right) \left(1+e^{2 i \alpha ^R}\right) \sqrt{\cos \left(\alpha ^L\right)}} e^{2 d \left(\kappa +i k_0\right)}+e^{2 d \kappa +2 i d k_0+i \alpha ^L}+e^{2 i \beta }\nonumber \\
    &\left(e^{2 d \kappa }-e^{i \left(2 d k_0+\alpha ^L\right)}
    +e^{i \alpha ^R} \left(e^{2 d \kappa }+e^{i \alpha ^L} \left(-e^{2 d \kappa }+e^{2 i d k_0}\right)+e^{2 i d k_0}\right)-e^{2 i d k_0}-e^{2 d \kappa +i \alpha ^L}\right)-e^{i \left(2 d \left(k_0-i \kappa \right)+\alpha ^L+\alpha ^R\right)}\nonumber\\
    &-e^{2 d \left(\kappa +i k_0\right)+i \alpha ^R}+e^{i \alpha ^L}+e^{i \left(\alpha ^L+\alpha ^R\right)}-e^{i \alpha ^R}-1 \nonumber\\
    a_{24} & = \frac{\csc (\beta ) e^{-i d \left(k_0-k_e\right)} \sinh (d \kappa ) \left(-\cos \left(\frac{\alpha ^R}{2}\right) \sin \left(\frac{\alpha '^L}{2}\right)+e^{2 i d k_0} \sin \left(\frac{\alpha ^R}{2}\right) \cos \left(\frac{\alpha '^L}{2}\right)\right)}{\sqrt{\cos \left(\alpha ^R\right)} \sqrt{\cos \left(\alpha '^L\right)}} \nonumber
    \end{align}
\begin{align}
  \label{acof2}
    a_{31} & = \frac{-\csc (\beta ) e^{-i d \left(k_h+k_0\right)} \sinh (d \kappa ) \left(\sin \left(\frac{\alpha ^L}{2}\right) \cos \left(\frac{\alpha '^R}{2}\right)-e^{2 i d k_0} \cos \left(\frac{\alpha ^L}{2}\right) \sin \left(\frac{\alpha '^R}{2}\right)\right)}{\sqrt{\cos \left(\alpha ^L\right)} \sqrt{\cos \left(\alpha '^R\right)}} \nonumber\\
    a_{34} & = \frac{\sqrt{\cos \left(\alpha '^R\right)}\exp \left(-\frac{1}{2} i \left(2 d \left(k_h-i \kappa +k_0\right)+\alpha '^L-\alpha^R\right)\right)}{2 \left(-1+e^{2 i \beta }\right) \left(1+e^{2 i \alpha '^R}\right) \sqrt{\cos \left(\alpha '^L\right)}}e^{2 d \kappa }-e^{i \left(2 d k_0+\alpha '^L\right)}\nonumber\\
    &+e^{2 i \beta } \left(e^{2 d \left(\kappa k_0\right)}+e^{2 d \left(\kappa +i k_0\right)+i \alpha '^L}-e^{i \alpha '^R} \left(e^{2 d \left(\kappa +i k_0\right)}+\left(-1+e^{2 d \left(\kappa +i k_0\right)}\right) e^{i \alpha '^L}+1\right)+e^{i \alpha '^L}-1\right)\nonumber\\
    &+e^{i \left(2 d k_0+\alpha '^L+\alpha '^R\right)}+e^{i \left(2 d k_0+\alpha '^R\right)}-e^{2 i d k_0}-e^{2 d \kappa +i \alpha '^L}-e^{2 d \kappa +i \left(\alpha '^L+\alpha '^R\right)}+e^{2 d \kappa +i \alpha '^R}\nonumber\\
    a_{41} & = \frac{\csc (\beta ) e^{i d k_h} \sinh (d \kappa ) \left(-\sin \left(d k_0\right) \cos \left(\frac{1}{2} \left(\alpha ^L-\alpha '^R\right)\right)+i \cos \left(d k_0\right) \cos \left(\frac{1}{2} \left(\alpha ^L+\alpha '^R\right)\right)\right)}{\sqrt{\cos \left(\alpha ^L\right)} \sqrt{\cos \left(\alpha '^R\right)}}\\
    a_{44} & = \frac{\sqrt{\cos \left(\alpha '^R\right)}\exp \left(-\frac{1}{2} i \left(2 d \left(-k_h-i \kappa +k_0\right)+\alpha '^L-\alpha '^R\right)\right)}{2 \left(-1+e^{2 i \beta }\right) \left(1+e^{2 i \alpha '^R}\right) \sqrt{\cos \left(\alpha '^L\right)}}-e^{2 d \kappa }-e^{i \left(2 d k_0+\alpha '^L\right)}\nonumber\\
    &+e^{2 i \beta } \left(e^{2 d \left(\kappa +i k_0\right)} \left(1+e^{i \alpha '^L} \left(1+e^{i \alpha '^R}\right)+\left(-1+e^{i \alpha '^L}\right) \left(-1+e^{i \alpha '^R}\right)\right)\right)\nonumber\\
    &-e^{i \alpha '^R} \left(e^{2 i d k_0} \left(1+e^{i \alpha '^L}\right)+e^{2 d \kappa } \left(-1+e^{i \alpha '^L}\right)\right)-e^{2 i d k_0}+e^{2 d \kappa +i \alpha '^L}\nonumber
\end{align}

\begin{align}
  \label{reftranseq}
    r_{e} = \frac{a_{32} a_{44}-a_{34} a_{42}}{a_{31} a_{44}-a_{34} a_{41}} ;~~~
    r_{h} = \frac{a_{33} a_{44}-a_{34} a_{43}}{a_{31} a_{44}-a_{34} a_{41}} ;~~~
    t_{e} = \frac{a_{44}}{a_{31} a_{44}-a_{34} a_{41}} ;~~~ 
    t_{h} = \frac{a_{34}}{a_{31} a_{44}-a_{34} a_{41}}
\end{align}

\end{widetext}

\end{document}